\documentclass[usenatbib]{mnras}

\usepackage{newtxtext,newtxmath}
\usepackage{multirow}
\usepackage{amsmath}
\usepackage{tablefootnote}
\usepackage{threeparttable}

\usepackage[T1]{fontenc}
\usepackage{ae,aecompl}

\usepackage{graphicx}	
\usepackage{amsmath}	
\usepackage{amssymb}	
\usepackage[labelfont=bf]{subcaption}
\usepackage[]{hyperref}

\title[ALMA observations of local radio galaxies]{\centering{The AGN fuelling/feedback cycle in nearby radio galaxies \\
  I. ALMA observations and early results.}}

\author[I. Ruffa et al.]{Ilaria Ruffa,$^{1,2}$\thanks{E-mail: i.ruffa@ira.inaf.it}
Isabella Prandoni,$^{2}$
Robert A. Laing,$^{3}$
Rosita Paladino,$^{2}$
\newauthor
Paola Parma,$^{2}$
Hans de Ruiter,$^{2}$
Arturo Mignano,$^{2}$
Timothy A. Davis,$^{4}$
\newauthor
Martin Bureau,$^{5,6}$
and Joshua Warren$^{5}$
\\
$^{1}$Dipartimento di Fisica e Astronomia, Universit\`{a} degli Studi di Bologna, via P.\ Gobetti 93/2, 40129 Bologna, Italy\\
$^{2}$INAF - Istituto di Radioastronomia, via P.\ Gobetti 101, 40129 Bologna, Italy\\
$^{3}$Square Kilometre Array Organisation, Jodrell Bank Observatory, Lower Withington, Macclesfield, Cheshire SK11 9DL, UK\\
$^{4}$School of Physics \& Astronomy, Cardiff University, Queens Buildings, The Parade, Cardiff CF24 3AA, UK\\
$^{5}$Sub-dept.\ of Astrophysics, Dept.\ of Physics, University of Oxford, Denys Wilkinson Building, Keble Road, Oxford OX1 3RH, UK\\
$^{6}$Yonsei Frontier Lab and Department of Astronomy, Yonsei University, 50 Yonsei-ro, Seodaemun-gu, Seoul 03722, Republic of Korea
}

\date{Accepted XXX. Received YYY; in original form ZZZ}

\pubyear{2018}

\def\kms{{{km\,s$^{-1}$}}}
\def\co{ $^{12}$CO(2-1)}
\def\cotoh2{CO-to-H$_{2}$}

\hypersetup{draft}

\begin{document}
\label{firstpage}
\pagerange{\pageref{firstpage}--\pageref{lastpage}}
\maketitle

\begin{abstract}
This is the first paper of a series exploring the multi-frequency properties of a sample of eleven nearby low excitation radio galaxies (LERGs) in the southern sky. We are conducting an extensive study of different galaxy components (stars, warm and cold gas, radio jets) with the aim of improving our understanding of the AGN fuelling/feedback cycle in LERGs.
We present ALMA Band 6\co\ and continuum observations of nine sources. Continuum emission from the radio cores was detected in all objects. Six sources also show mm emission from jets on kpc/sub-kpc scales. The jet structures are very similar at mm and cm wavelengths. We conclude that synchrotron emission associated with the radio jets dominates the continuum spectra up to 230~GHz. 
The\co\ line was detected in emission in six out of nine objects, with molecular gas masses ranging from $2 \times 10^{7}$ to $2 \times 10^{10}$~M$_{\rm \odot}$. The CO detections show disc-like structures on scales from $\approx$0.2 to $\approx$10~kpc. In one case (NGC\,3100) the CO disc presents some asymmetries and is disrupted in the direction of the northern radio jet, indicating a possible jet/disc interaction. In IC\,4296, CO is detected in absorption against the radio core as well as in emission. In four of the six galaxies with CO detections, the gas rotation axes are roughly parallel to the radio jets in projection; the remaining two cases show large misalignments. 
In those objects where optical imaging is available, dust and CO appear to be co-spatial.
\end{abstract}




\section{Introduction}\label{sec:intro}
It is widely believed that feedback processes associated with active galactic nuclei (AGN) can potentially play a fundamental role in shaping galaxies over cosmic time \citep[e.g.][]{Ciotti09,Ciotti10,Debuhr12,King15,Harrison18}. AGN feedback can change the physical conditions of the surrounding interstellar medium (ISM), preventing gas cooling or even expelling the gas from the nuclear regions, thus impacting star formation processes and the subsequent evolution of the host galaxy \citep[e.g.][]{Combes17,Harrison17}. Two main variants of feedback are commonly discussed: radiative and kinetic. The former is typically associated with a radiatively efficient (quasar-like) AGN; the latter with an energetic outflow or jet \citep[e.g.][]{Best12,Alexander12,Heckman14}. Radio jets produce some of the clearest manifestations of AGN feedback such as cavities in the hot intracluster gas and jet-driven outflows in the ISM \citep[see e.g.][for a review]{Fabian12}.

In the local Universe, radio galaxies (RGs), which by definition show strong kinetic (jet-induced) feedback, are typically hosted by the most massive early type galaxies (ETGs). They can be divided into two classes  according to their optical spectra (e.g.\ \citealt{Heckman14}).  High-excitation radio galaxies (HERGs) have spectra showing strong, quasar/Seyfert-like emission lines, accrete at $\ga 0.01~\dot{M}_{\rm Edd}$ (where $\dot{M}_{\rm Edd}$ is the Eddington accretion rate\footnote{$\dot{M}_{\rm Edd} = \dfrac{4\pi \, G \, M_{\rm SMBH} \, m_{\rm p}}{\varepsilon \, c \, \sigma_{\rm T}}$, where G is the gravitational constant, $M_{\rm SMBH}$ is the mass of the central super-massive black hole (SMBH), $m_{\rm p}$ is the mass of the proton, $\varepsilon$ is the accretion efficiency, $c$ is the speed of light and $\sigma_{\rm T}$ is the cross-section for Thomson scattering.}) and are radiatively efficient, thereby producing radiative as well as kinetic feedback. Low-excitation radio galaxies (LERGs) have spectra with weak, LINER-like emission lines, accrete at $\ll 0.01 \dot{M}_{\rm Edd}$ and their feedback is almost entirely kinetic. It has been proposed that the dichotomy in accretion rate is a consequence of different trigger mechanisms and fuelling sources \citep[e.g.][]{Hardcastle07}. In this scenario, HERGs are fuelled by cold gas transported to their nuclei through merging or collisions with gas-rich galaxies, whereas LERGs are powered by direct accretion from the hot phase of the inter-galactic medium (IGM). \citet{Allen06} supported the hot accretion scenario, finding a correlation between the jet power of LERGs and the Bondi accretion rate (i.e. the rate of spherical accretion of the hot, X-ray emitting medium; \citealp{Bondi52}). \citet{Russell13} later reported a lower significance level for that correlation, however. More realistic models for LERGs, requiring chaotic accretion, were first proposed by \citet{Sanders81} and then elaborated by several authors, based on results from numerical simulations \citep[see e.g.][]{King07,Wada09,Naya12,Gaspari13,King15,Gaspari15,Gaspari17}. In this mechanism (now refereed to as \textit{chaotic cold accretion}) the hot gas from the galaxy halo cools to temperatures lower than 10$^{3}$~K and forms dense clouds of cold gas. Within a few Bondi radii\footnote{The Bondi radius is given by $r_{\rm B} = \dfrac{2 G \, M_{\rm SMBH}}{c_{\rm s}^{2}}$, where $G$ is the gravitational constant, $M_{\rm SMBH}$ is the mass of the SMBH, and $c_{\rm s}$ is the speed of sound.} chaotic inelastic collisions between the clouds are frequent enough to cancel their angular momentum, allowing them to accrete onto the SMBH.

The most compelling evidence that cold gas can play a role in fuelling LERGs is that it is frequently detected in these sources, with masses that are potentially capable of powering the jets by accretion  \citep[$M_{\rm H_2} \sim 10^{7} - 10^{10}$~M$_{\rm \odot}$, e.g. ][]{Prandoni07,Prandoni10,Ocana10}. Hints of cold gas clouds falling towards active nuclei have recently been observed in some objects \citep[e.g.][]{Tremblay16,Maccagni18}, providing support for this hypothesis. 
The presence of cold gas alone is not direct evidence of fuelling, however. For example, the cold gas in 3C\,31 is found to be in ordered rotation and stable orbits \citep[][]{Okuda05}: in cases like this the accretion rate may be relatively low. In other galaxies, the molecular gas appears to be outflowing or interacting with the radio jets, rather than infalling \citep[e.g.][]{Alatalo11,Combes13,Oosterloo17}. The origin of the observed gas also remains unclear: it may cool from the hot gas phase or come from stellar mass loss, interactions or minor mergers.

Our project aims at exploring the multi-frequency properties of a complete flux and volume-limited sample of eleven local LERGs in the southern sky. The purpose is to carry out an extensive study of the various  galaxy components (stars, warm and cold gas, radio jets) to get a better understanding of the AGN fuelling/feedback cycle in LERGs, using the ATLAS\textsuperscript{3D} galaxies as a radio-quiet control sample \citep{Cappellari11}. In this framework we have acquired Atacama Pathfinder EXperiment (APEX) \co\ integrated spectra \citep[][Laing et al.\ in preparation, hereafter Paper~II]{Prandoni10}, Very Large Telescope (VLT) Visible Multi Object Spectrograph (VIMOS) integral-field-unit spectroscopy for the entire sample (Warren et al., in preparation) and Karl G. Jansky Very Large Array (JVLA) high-resolution continuum observations at 10~GHz for five sources (Ruffa et al., in preparation). Archival Hubble Space Telescope (HST) or ground-based optical/near-IR and VLA images are also used, when available. 

In this paper, we present Atacama Large Millimeter/submillimeter Array (ALMA) Cycle 3\co\ and 230~GHz continuum observations of nine objects. The paper is structured as follows. In Section~\ref{sec:sample} we present the sample, describing the selection criteria and available data. In Section~\ref{sec:ALMA observations} we describe the ALMA observations and data reduction; the analysis performed on the data products is detailed in Section~\ref{sec:analysis}. In Section~\ref{sec:results} we present the results for individual galaxies. General properties of the sample are discussed in the context of earlier work  in Section~\ref{sec:discussion}. We summarise our conclusions in Section~\ref{sec:conclusion}.
In Appendix~A, included as supplementary material in the on-line version of the paper, we present a brief description of the re-analysed archival VLA data and optical/IR imaging used in our analysis.

Throughout this work we assume a standard $\Lambda$CDM cosmology with H$_{\rm 0}=70$\,km\,s$^{-1}$\,Mpc$^{\rm -1}$, $\Omega_{\rm \Lambda}=0.7$ and $\Omega_{\rm M}=0.3$. All of the velocities in this paper are given in the optical convention.

\section{The southern radio galaxy sample}\label{sec:sample}
We have defined a complete volume- and flux-limited sample of eleven RGs in the southern sky. This sample was selected from \citet{Ekers89}, who presented a complete sample of 91 radio galaxies from the Parkes 2.7-GHz survey, all located in the declination range $-17^{\circ}<\delta<-40^{\circ}$, with a radio flux-density limit of 0.25~Jy at 2.7~GHz and an optical magnitude limit of m\textsubscript{v}$=17.0$. From this sample, we selected those sources satisfying the following criteria:
\begin{enumerate}
\item elliptical/S0 galaxy optical counterpart;
\item host galaxy redshift $z<0.03$.
\end{enumerate}
This resulted in a sample of eleven radio galaxies, all with low or intermediate 1.4 GHz radio powers: P$_{\rm 1.4}$\textsubscript{GHz}$\leq10^{25.5}$~W~Hz$^{-1}$. The majority of the sources have FR\,I radio morphologies \citep{Fanaroff74}; one is classified as intermediate between FR\,I and FR\,II and one as FR\,II. 
Based on the available optical spectroscopy (\citealt{Tad93,Smith00,Colless03,Coll06,Jones09}), all of the radio galaxies in our southern sample have [OIII] line luminosities below the relation shown in Fig. 2 of \citet{Best12} and, as argued in that paper, can be securely classified as LERGs. 
The main characteristics of our southern sample galaxies are listed in Table~\ref{tab:Southern Sample}. 

\begin{table*}
\centering
\caption{General properties of the southern radio galaxy sample.}\label{tab:Southern Sample}
\begin{tabular}{l l c c c c c c c}
\hline
\multicolumn{1}{c}{ Radio} &
\multicolumn{1}{c}{ Host } &
\multicolumn{1}{c}{ z } & 
\multicolumn{1}{c}{ S$_{\rm 1.4}$ } &
\multicolumn{1}{c}{ log~P$_{\rm 1.4}$ } &
\multicolumn{1}{c}{ FR } &
\multicolumn{1}{c}{ D$_{\rm L}$ } &
\multicolumn{1}{c}{ v\textsubscript{opt} } \\
\multicolumn{1}{c}{ source } & 
\multicolumn{1}{c}{ galaxy } &
\multicolumn{1}{c}{ }  &  
\multicolumn{1}{c}{ } &
\multicolumn{1}{c}{ }  &
\multicolumn{1}{c}{ class } &
\multicolumn{1}{c}{ }   &
\multicolumn{1}{c}{ }   \\
\multicolumn{1}{c}{ } &
\multicolumn{1}{c}{ } &
\multicolumn{1}{c}{  } &
\multicolumn{1}{c}{ (Jy) } &
\multicolumn{1}{c}{ (W\,Hz$^{-1}$)} &
\multicolumn{1}{c}{ } &
\multicolumn{1}{c}{ (Mpc)}  &
\multicolumn{1}{c}{  (\kms)}   \\
\multicolumn{1}{c}{(1)} &
\multicolumn{1}{c}{ (2)} &
\multicolumn{1}{c}{ (3)} &
\multicolumn{1}{c}{ (4)} &
\multicolumn{1}{c}{ (5) } &
\multicolumn{1}{c}{ (6) } &
\multicolumn{1}{c}{ (7)} &
\multicolumn{1}{c}{ (8)} \\
\hline
PKS 0007$-$325&  IC\,1531 & 0.0256 & 0.5 & 23.9 &  I & 112.0 & 7681$\pm$25 \\
PKS 0131$-$31 & NGC\,612 & 0.0298 & 5.6 & 25.1 & I/II & 130.4 & 8913$\pm$29 \\
PKS 0320$-$37 & NGC\,1316 & 0.0058 & 150 & 25.1 &  I & 25.3 & 1734$\pm$10\\
PKS 0336$-$35 & NGC\,1399 & 0.0047 & 2.2 & 23.0 &  I & 20.4 & 1408$\pm$4\\
PKS 0718$-$34 &  $-$ & 0.0284 & 2.1 & 24.6 &  II & 124.1 & 8900$\pm$128 \\
PKS 0958$-$314& NGC\,3100 & 0.0088 & 0.5 & 23.0 &  I & 38.0 & 2629$\pm$20 \\
PKS 1107$-$372& NGC\,3557 & 0.0103 & 0.8 & 23.3 &  I & 44.5 & 3079$\pm$18 \\
PKS 1258$-$321& ESO 443-G 024 & 0.0170 & 1.2 & 23.9 &  I & 73.9 & 1689$\pm$18 \\
PKS 1333$-$33 &  IC\,4296 & 0.0125 & 4.5 & 24.2 &  I & 53.9 &  3737$\pm$10 \\
PKS 2128$-$388& NGC\,7075 & 0.0185 & 0.9 & 23.8 & I & 80.3 & 5466$\pm$20 \\
PKS 2254$-$367& IC\,1459 & 0.0060 & 1.2 & 23.9& I$^{*}$ & 25.9 & 1689$\pm$18\\
\hline
\end{tabular}
\parbox[t]{1\textwidth}{ \textit{Notes.} Columns: (1) Name of the radio source. (2) Host galaxy name. (3) Galaxy redshift taken from the NASA/IPAC extragalactic database (NED). (4) Radio flux density at 1.4~GHz; this is the most accurate value given in NED and includes all the radio emission associated with the source. (5) Logarithmic-scale radio power at 1.4~GHz derived from S$_{\rm 1.4}$ and D$_{\rm L}$. (6) Fanaroff-Riley class \citep{Fanaroff74}.
  (7) Luminosity distance derived from the redshift given in column (3) and assuming the cosmology in Section~\ref{sec:intro}. (8) Best estimate of the optical stellar velocity from NED, given in the LSRK system, for comparison with CO velocities in Table~\ref{tab:line parameters}.\\
$^{\ast}$FR\, I structure on sub-arcsecond scale \citep[see][]{Tingay15}.}
\end{table*}

\section{ALMA observations and data reduction}\label{sec:ALMA observations}
We used ALMA Band 6 ($\approx$230~GHz) to observe nine of the sample members. Two sources were excluded because the estimated integration times were unreasonably long. In NGC\,1316 (PKS 0320$-$37, Fornax A) the CO  is known to be distributed over an area much larger than the  single-pointing field-of-view (FOV) of the ALMA main array at 230~GHz \citep{Horellou01}; a large mosaic and short-spacing information would be required to image it adequately. Based on measurements with APEX (Paper\,II), NGC\,1399 (PKS 0336-35) appeared to be too faint for a reliable detection.

ALMA observations were taken during Cycle 3, between March and July 2016 (PI: I.\ Prandoni). Table~\ref{tab:ALMA observations summary} summarises the details of the observations. The total time on-source ranged from 3 to 30 minutes. The spectral configuration consisted of four spectral windows: one centred on the redshifted frequency ($\nu_{\rm sky}$) of the \textsuperscript{12}CO(J=2-1) line (rest frequency 230.5380~GHz) and divided into 1920 1.129~MHz-wide channels; the other three, used to map the continuum emission, had 128 31.25-MHz-wide channels. Between 36 and 43 12-m antennas were used, with maximum baseline lengths ranging from 460~m to 1.1~km. The maximum recoverable spatial scale (MRS), together with the major and minor axis full width half maxima (FWHM) and position angle of the synthesized beam for each observation are reported in Table~\ref{tab:ALMA observations summary}. Titan and Pallas were used as primary flux calibrators; J1037$-$2934, J1107$-$4449, J2537$-$5311, J0538$-$4405 and J1427$-$4206 were observed as secondary standards if no solar-system object was available.

We reduced the data using the Common Astronomy Software Application \citep[{\sc casa};][]{McMullin07} package, version 4.7.2, calibrating each dataset  separately using customized \textsc{python} data reduction scripts.

\subsection{Continuum imaging}
The three continuum spectral windows and the line-free channels in the line spectral window were used to produce the continuum maps, using the \textsc{clean} task in multi-frequency synthesis (MFS) mode with one Taylor series term \citep{Rau11}. All the continuum maps were made using natural weighting in order to maximise the sensitivity, with the goal of imaging emission from the jets. Since the cores are detected at high signal-to-noise in all of the targets, multiple cycles of phase-only self calibration were performed in all cases. Additional amplitude and phase self-calibration was performed for the brightest cores only. This allowed us to obtain root-mean square (rms) noise levels ranging from 0.02 to 0.06~mJy~beam$^{-1}$ for synthesized beams of 0.3 -- 1 arcsec FWHM.

Two-dimensional Gaussian fits were performed within the regions covered by the continuum emission in order to estimate the spatial extent of each observed component. Table~\ref{tab:Continuum images} summarises the main properties of the millimetre continuum maps, which are shown in Fig.~\ref{fig:continuum}. For each object, the upper panel shows the ALMA millimetre continuum image, while the lower panel shows an archival VLA continuum map (at 4.9, 8.5 or 14.9~GHz), chosen to match as closely as possible the angular resolution of the corresponding ALMA image. All of the archival VLA datasets used in this work have been re-analysed using  self-calibration and multi-scale {\sc clean} as appropriate. Details of the VLA observations and data reduction are given in Appendix A (provided as on-line only supplementary material), together with additional VLA images showing the large-scale radio structures of four sources. Although a detailed comparison of the mm and cm continuum properties of our sources is not possible with the available archival cm-wave data, a brief analysis is provided in Sections~\ref{sec:results} and \ref{sec:continuum_analysis}.

\subsection{Line imaging} 
After applying the continuum self-calibration, line emission was isolated in the visibility plane using the {\sc casa} task {\sc uvcontsub} to form a continuum model from linear fits in frequency to line-free channels and to subtract it from the visibilities. We then produced a data cube of CO channel maps using the {\sc clean} task with natural weighting.
The channel velocities were initially computed in the source frame with zero-points corresponding to the redshifted frequency of the CO(2-1) line ($\nu$\textsubscript{sky}, Table~\ref{tab:ALMA observations summary}).  The continuum-subtracted dirty cubes were cleaned in regions of line emission (identified interactively) to a threshold equal to 1.5 times the rms noise level, determined in line-free channels. Several channel widths (i.e.\ spectral bins) were tested to find a good compromise between signal-to-noise ratio (S/N) and resolution of the line profiles; the final channel widths range from 10 to 40\,\kms.

We clearly detect\co\ emission in six out of nine sources, with S/N ranging from 8 to 45. The cleaned CO data cubes are characterised by rms noise levels (determined in line-free channels) between 0.2 and 1.3~mJy~beam$^{-1}$.

Three targets (PKS~0718$-$34, ESO~443$-$G~024 and IC\,1459) are undetected in CO. In these cases, line emission (and consequently line-free channels) could not be identified in the spectral window centred on the redshifted CO emission; the continuum was thus modelled using the three continuum spectral windows only. The continuum-subtracted dirty cubes for these targets were cleaned down to 1.5 times the expected rms noise level, with conservative spectral channel widths between 75 and 80~km~s$^{-1}$. The $1\sigma$  noise levels measured in the cleaned channel maps ranged from 0.2 to 0.6~mJy~beam$^{-1}$; in these cases 3$\sigma$ upper limits are tabulated.

Table~\ref{tab:line images} summarises the properties of the CO data cubes.

\begin{table*}
\centering
\caption{ALMA Cycle 3 observations.}
\label{tab:ALMA observations summary}
\begin{tabular}{l r c c c c c c c c}
\hline
\multicolumn{1}{c}{ Target } &
\multicolumn{1}{c}{ Date } & 
\multicolumn{1}{c}{ $\nu$\textsubscript{sky}  (v\textsubscript{cen})} &
\multicolumn{1}{c}{ Time } & 
\multicolumn{1}{c}{   MRS } & 
\multicolumn{1}{c}{   $\theta$\textsubscript{maj} } & 
\multicolumn{1}{c}{   $\theta$\textsubscript{min}  } & 
\multicolumn{1}{c}{   PA} & 
\multicolumn{1}{c}{   Scale }\\        
\multicolumn{1}{c}{  } &       
\multicolumn{1}{c}{  } &   
\multicolumn{1}{c}{   (GHz) (km s$^{-1}$) } &          
\multicolumn{1}{c}{   (min)} &          
\multicolumn{1}{c}{   (kpc, arcsec)} &     
\multicolumn{2}{c}{ (arcsec) } &   
\multicolumn{1}{c}{   (deg) } &
\multicolumn{1}{c}{   (pc)} \\       
\multicolumn{1}{c}{   (1) } &   
\multicolumn{1}{c}{   (2) } &
\multicolumn{1}{c}{   (3) } &
\multicolumn{1}{c}{   (4) } & 
\multicolumn{1}{c}{   (5) } &   
\multicolumn{1}{c}{   (6) } &
\multicolumn{1}{c}{   (7) } &               
\multicolumn{1}{c}{   (8) } &
\multicolumn{1}{c}{   (9) } \\
\hline
 IC\,1531 &  2016-06-02 & 224.7774 (7702) & 12.0 & 5.6, 10.9 &  0.7  &  0.6  & 87  & 360  \\ 
  NGC\,612  &    2016-07-30   &    223.8426  (8974) &    3.0   &  6.6, 11.0 &   0.3 & 0.3 &    -75 &    180 \\  
 PKS 0718-34 &  \begin{tabular}[c]{@{}c@{}}  2016-05-02 \\ 2016-05-03 \end{tabular} &    223.8622  (8904)  &    32.2    &    6.3, 11.0    &   0.7    &     0.6    &    -80    &    400    \\
 NGC\,3100 &    2016-03-22    &    228.6299 (2484)   &    28.5     &    1.9, 10.6    &    0.9    &    0.7    &    -87    &    160   \\
  NGC\,3557  &    2016-06-03/04    &    228.2319 (2999)   &    22.5    &    2.0, 9.7    &    0.6    &    0.5    &    -70     &    130    \\ 
ESO 443-G 024  &    2016-05-01     &    226.6839  (5089)  &   24.2     &      3.7, 10.8    &     0.7    &     0.6    &    -63     &    240    \\
 IC\,4296  &  \begin{tabular}[c]{@{}c@{}}    2016-06-04 \\ 2016-06-11 \end{tabular}   &    227.7110  (3705)  &    25.5    &    2.8, 10.8    &    0.6    &     0.6    &    -84    &    150   \\  
 NGC\,7075 &    2016-05-03     &    226.4196  (5483)   &     24.5    &    4.1, 10.8    &    0.6    &     0.6    &    -76     &    230    \\
 IC\,1459  &    2016-04-11     &    229.1614 (1819)   &    11.4    &    1.3, 10.7    &     1.0    &     0.8    &    -71     &    120    \\   
\hline
\end{tabular}
\parbox[t]{1\textwidth}{ \textit{Notes.} $-$ Columns: (1) Target name. (2) Observation dates. (3)\co\ redshifted (sky) centre frequency estimated using the redshift listed in column (3) of Table~\ref{tab:Southern Sample}; the corresponding velocity (v\textsubscript{cen}; LSRK system, optical convention) is reported in parentheses. (4) Total integration time on-source. (5) Maximum recoverable scale in kiloparsec for the array configuration, and corresponding scale in arcseconds. (6) Major axis FWHM of the synthesized beam. (7) Minor axis FWHM of the synthesized beam. (8) Position angle of the synthesized beam. (9) Spatial scale corresponding to the major axis FWHM of the synthesized beam.}
\end{table*}

\begin{table*}
\centering
\caption{Properties of the ALMA continuum images.} 
\label{tab:Continuum images}
\begin{tabular}{l c c c c c c c c c c c }
\hline
\multicolumn{1}{c}{ Target } & 
\multicolumn{1}{c}{ rms }&
\multicolumn{1}{c}{ S$_{\rm 230}$ } &
\multicolumn{2}{c}{ Size FWHM } &
\multicolumn{1}{c}{ PA }\\
\multicolumn{1}{c}{  } & 
\multicolumn{1}{c}{ (mJy~beam$^{-1}$)  } &
\multicolumn{1}{c}{ (mJy) } &
\multicolumn{1}{c}{ (arcsec$^{2}$) } &
\multicolumn{1}{c}{  (pc$^{2}$) } &
\multicolumn{1}{c}{ (deg) } \\
\multicolumn{1}{c}{  (1) } &
\multicolumn{1}{c}{  (2) } &
\multicolumn{1}{c}{  (3) } &
\multicolumn{1}{c}{  (4) } &
\multicolumn{1}{c}{  (5) } &
\multicolumn{1}{c}{  (6) } \\
\hline
IC\,1531 & 0.05 &    108$\pm$10.5 &     &   &    \\
core &     &     105$\pm$10.5 &  (0.3 $\times$ 0.1)  &    (65 $\times$ 50)  &  141$\pm$19  \\
 SE jet$^{1}$ &    &    3.0$\pm$0.3 &   $-$  &  $-$  &  $-$  \\
 \hline
NGC\,612 &  0.06  &  29$\pm$2.9  &    &   &    \\
core &    &  29$\pm$2.9 &   (0.03 $\times$ 0.01)  &  (20 $\times$ 10)  &  137$\pm$44  \\
\hline
PKS 0718-34 &  0.02  &   17$\pm$1.4  &     &  &    \\
core &    &    14$\pm$1.4 &   (0.1$\times$0.08)  &  (60$\times$50)  &  107$\pm$38  \\
 NE jet &    &   1.0$\pm$0.1  &   (3.1$\times$1.5)  &  (1800$\times$860)  &  33$\pm$5   \\
SW jet &    &   1.5$\pm$0.1  &   (5.6$\times$1.0)   &  (3200$\times$570)  &  49$\pm$2   \\
\hline
NGC\,3100 & 0.02  &  50$\pm$4.3  &     &   &   \\
core &    &    43$\pm$4.3 &   (0.11$\times$0.05)  &  (20$\times$10)  &   170$\pm$89  \\
 N jet &   &  2.2$\pm$0.2  &   (2.4$\times$1.0)  &  (440$\times$180)  &  167$\pm$2  \\
S jet &   &  4.3$\pm$0.4  &   (1.6$\times$0.7)  &  (290$\times$130)  &  164$\pm$4  \\
\hline
NGC\,3557 &  0.03  &  30$\pm$2.5  &     &    &     \\
core &    &    25$\pm$2.5 &   (0.11$\times$0.1)  &  (23$\times$20)  &  62$\pm$22  \\
E jet &    &  2.0$\pm$0.2  &   (4.1$\times$0.7)  &  (870$\times$150)  &  77$\pm$2   \\
W jet &    &  2.5$\pm$0.2  &    (5.4$\times$0.6)  &  (1145$\times$130)  &  77$\pm$3   \\
\hline
ESO 443-G 024 &    0.02   &    61$\pm$5.3  &     &      \\
core &    &    53$\pm$5.3 &   (0.1$\times$0.08)  &  (35$\times$30)  &  111$\pm$19  \\
SE jet &       &    2.3$\pm$0.2  &    (2.5$\times$1.3)  &   (870$\times$450)   &  115$\pm$4   \\
NW jet &      &    5.3$\pm$0.5  &   (4.5$\times$1.0)  &   (1570$\times$350)  &   115$\pm2$   \\
\hline
IC\,4296 &  0.02  &   190$\pm$19.0  &     &    &   \\
core &    &    190$\pm$19.0 &   (0.1$\times$0.1)  &  (30$\times$30)  &  61$\pm$35  \\
\hline
NGC\,7075 &  0.02  &     19$\pm$1.7  &     &    &    \\
core &    &  17$\pm$1.7   &   (0.1$\times$0.07)  &  (40$\times$30)  &  70$\pm$40  \\
E jet$^{1}$ &    &     1.8$\pm$0.1  &   $-$  &  $-$  &  $-$  \\
\hline
IC\,1459$^{2}$ &   0.03  &    217$\pm$21 &   (0.08$\times$0.06)  &  (10$\times$7)  &  124$\pm$12  \\
\hline
\end{tabular}
\parbox[t]{1\textwidth}{ \textit{Notes.} $-$ Columns: (1) Target name. (2) 1$\sigma$ rms noise level measured in emission-free regions of the cleaned continuum map. (3) 230~GHz continuum flux density; the total, core and jet flux densities are quoted separately. The uncertainties are estimated as $\sqrt{{\rm rms}^{2} + (0.1 \times S_{\rm 230})^{2}}$, and the second term dominates in all cases. Errors on total flux densities are obtained through error propagation. (4) Size (FWHM) deconvolved from the synthesized beam. The sizes were estimated by performing 2$-$D Gaussian fits to identifiable continuum components. (5) Spatial extent of each component corresponding to the angular sizes in column (4). (6) Position angle of the corresponding component, defined North through East.\\
$^{1}$Unresolved component. \\
$^{2}$The FR\,I structure of this source is on milli-arcsecond (mas) scales \citep[$\leq40$~mas;][]{Tingay15} and is unresolved in our images.}
\end{table*}

\begin{table*}
\centering
\caption{Properties of the $^{12}$CO(2-1) line images.}
\label{tab:line images}
\begin{tabular}{ l c c c c c c}
\hline
\multicolumn{1}{c}{ Target } &
\multicolumn{1}{c}{ rms } &
\multicolumn{1}{c}{ Peak flux } &
\multicolumn{1}{c}{ S/N  } &
\multicolumn{1}{c}{ $\Delta$v\textsubscript{chan}  } \\
\multicolumn{1}{c}{  } &
\multicolumn{1}{c}{ (mJy~beam$^{-1}$) } &
\multicolumn{1}{c}{ (mJy~beam$^{-1}$) } &
\multicolumn{1}{c}{  } &
\multicolumn{1}{c}{ (km~s$^{-1}$) } \\
\multicolumn{1}{c}{ (1) } &
\multicolumn{1}{c}{ (2)} &
\multicolumn{1}{c}{ (3) } &
\multicolumn{1}{c}{ (4) } &
\multicolumn{1}{c}{ (5) } \\
\hline
 IC\,1531  &  0.7  &    12.4  &   18  &   20  \\
 NGC\,612    &  1.3  &  18.3   &   14  &  20  \\
  PKS 0718-34  &  0.2  &  $<0.6$  &   $-$  &   80  \\
  NGC\,3100  &  0.6  &  28.3 &  45 &  10  \\
  NGC\,3557   &  0.4  &  16.3 &   38  &  22 \\
 ESO 443-G 024  &  0.2   &  $<0.6$  &   $-$  &   75  \\
 IC\,4296  &  0.2  &  2.0  &  8   &   40  \\
 NGC\,7075 &  0.4  &  4.0  &  10  &   40  \\
  IC\,1459   &  0.6  &  $<1.8$  &   $-$  &    80  \\
\hline
\end{tabular}
\parbox[t]{1\textwidth}{ \textit{Notes.} $-$ Columns: (1) Target name. (2) 1$\sigma$ rms noise level measured in line-free channels at the channel width listed in column (5). (3) Peak flux density of the line emission. (4) Peak signal-to-noise ratio of the detection. (5) Final channel width of the data cube (km\,s$^{-1}$ in the source frame).}
\end{table*}

\section{Image Cube analysis}\label{sec:analysis}
\subsection{CO moment maps}
Integrated intensity (moment 0), mean velocity (moment 1) and line velocity width (moment 2) maps of the detected lines were created from the cleaned, continuum-subtracted CO data cubes using the masked moment technique as described by \citet[][see also \citealt{Bosma81a,Bosma81b,Kruit82,Rupen99}]{Dame11}. In this technique, a copy of the cleaned data cube is first Gaussian-smoothed spatially (with a FWHM equal to that of the synthesised beam) and then Hanning-smoothed in velocity. A three-dimensional mask is then defined by selecting all the pixels above a fixed flux-density threshold; this threshold is chosen so as to recover as much flux as possible while minimising the noise. We used thresholds varying from 1.2 to 2$\sigma$, depending on the significance of the CO detection (higher threshold for noisier maps). The moment maps were then produced from the un-smoothed cubes using the masked regions only \citep[e.g.][]{Davis17}. The resulting maps are shown in Figs~\ref{IC1531} $-$ \ref{NGC7075}, where the velocity zero-points are defined to be the intensity-weighted centroids of the observed CO emission. Based on their moment 0 and moment 1 maps, all detections show clear evidence of gas rotation, associated with discs or ring-like structures. There is evidence for asymmetries in the molecular gas distribution in a number of cases, particularly NGC\,3100 and IC\,4296 (see Sect.~\ref{sec:results} for more details).

For completeness, we show the moment 2 maps of all the targets detected in CO. It is worth noting however that they represent intrinsic velocity broadening only for those sources detected with a high S/N and/or whose CO emission is well resolved by our ALMA observations: this is certainly the case for NGC\,612 and NGC\,3100. Otherwise, the velocity dispersion (i.e. line-of-sight velocity width) is likely to be dominated by partially resolved velocity gradients within the host galaxy \citep[beam smearing; e.g.][]{Davis17}.

The extent of the molecular gas was estimated by performing 2D Gaussian fits to the moment 0 maps within the regions covered by the CO emission.
Table~\ref{tab:line parameters} summarises the estimated sizes, which are given as deconvolved major and minor axis FWHM. The detected discs are typically confined to kpc or sub-kpc scales except for NGC\,612, whose CO disc extends for at least 9.6~kpc along its major axis.

\subsection{Line widths and profiles}\label{sec:profiles}
The integrated spectral profiles of the six galaxies detected in CO(2-1) were extracted from the observed data cubes within boxes including all of the CO emission. The spectral profiles are shown in Figures~\ref{IC1531}$-$\ref{NGC7075} (panel d). The dimensions of the boxes used to extract the spectra are indicated in the figure captions.

All of the integrated spectral profiles exhibit the classic double-horned shape expected from a rotating disc. In one case (IC\,4296) a strong absorption feature was also detected (Fig.~\ref{fig:ic4296_spectrum}). Line widths were measured as full-width at zero intensity (FWZI) as well as FWHM. The former was defined as the full velocity range covered by spectral channels (identified interactively in the channel map) with CO intensities $\geq$3$\sigma$. These channels are highlighted in grey in Figures~\ref{fig:ic1531_spectrum}$-$\ref{fig:ngc7075_spectrum} and we also tabulate the flux integrated over this range. FWHM was defined directly from the integrated spectra as the velocity difference between the two most distant channels from the line centre with intensities exceeding half of the line peak. We did not attempt to fit models to the spectra at this stage. 

For non-detections, 3$\sigma$ upper limits on the integrated flux densities were calculated from the relation \citep[e.g.][]{Koay15}:
\begin{eqnarray}\label{eq:upper_limits}
  \Sigma S_{\rm CO} \Delta {\rm v}~({\rm Jy~beam}^{-1}~{\rm km~s}^{-1}) < 3\sigma \Delta {\rm v}_{\rm FWHM} \sqrt{\dfrac{\Delta {\rm v}}{\Delta {\rm v}_{\rm FWHM}}}
\end{eqnarray}   
where $\Delta$v is the channel width of the data cube in which the rms noise level ($\sigma$) is measured (see Table~\ref{tab:line images}) and $\Delta$v\textsubscript{FWHM} is the expected line FWHM. We assumed the line FWHM measured from APEX CO(2-1) observations (Paper II; see also Table~\ref{tab:line parameters}). The factor $\sqrt{\Delta {\rm v} / \Delta {\rm v}_{\rm FWHM}}$ accounts for the expected decrease in noise level with increasing bandwidth \citep{Wrobel99}. Equation~\ref{eq:upper_limits} is only valid if all of the molecular gas is concentrated within the synthesized beam (a few hundred parsec). If, more realistically, the molecular gas is distributed on larger scales, this assumption leads to a significant underestimation of the total flux limit. In this case, an estimate of the gas surface density upper limit is more meaningful (see Section~\ref{sec:mol_masses}).

The CO(2-1) line parameters are listed in Table~\ref{tab:line parameters}. v$_{\rm CO}$, the intensity-weighted velocity centroids, are our best estimates of the systemic velocity of the line emission. The errors given for v$_{\rm CO}$ are assumed to be equal to the channel widths of the corresponding integrated spectrum. The values of v$_{\rm CO}$ are all consistent (within the combined errors) with the stellar velocities v$_{\rm opt}$ listed in Table~\ref{tab:Southern Sample}.

\subsection{Molecular gas masses}\label{sec:mol_masses}
We adopted the following relation to estimate the total molecular gas masses, including contributions from heavy elements \citep[M\textsubscript{mol};][]{Bolatto13}:
\begin{footnotesize}
\begin{multline}\label{eq:gas mass}
M_{\rm mol}=\dfrac{1.05\times10^{4}}{R_{\rm 21}}~\left(\dfrac{X_{\rm CO}}{2\times10^{20}~\dfrac{{\rm cm}^{-2}}{{\rm K~km~s}^{-1}}}\right) \\ \times \left(\dfrac{1}{1+z}\right)~\left(\dfrac{\Sigma S_{\rm CO}\Delta\nu}{{\rm Jy~km~s}^{-1}}\right)~\left(\dfrac{D_{{\rm L}}}{{\rm Mpc}}\right)^{2},
\end{multline}
\end{footnotesize}
where $\Sigma S$\textsubscript{CO}$\Delta$v is the CO(2-1) flux integrated over velocity, $R$\textsubscript{21} is the CO(2-1) to CO(1-0) flux ratio, $z$ is the galaxy redshift, $D$\textsubscript{L} is the luminosity distance, and $X$\textsubscript{CO} is the \cotoh2\ conversion factor. $X$\textsubscript{CO} depends on the molecular gas conditions (e.g.\ excitation, dynamics, geometry) and the properties of the environment \citep[e.g.\ metallicity; see][]{Bolatto13} and is therefore likely to vary systematically between different galaxy types.
Little is known about $X$\textsubscript{CO} in nearby ETGs, because most studies are focused on gas-rich (and high gas metallicity) populations, typically late-type galaxies. Following \citet{Bolatto13} and \citet{Tremblay16}, we assume the average Milky Way value of  $X$\textsubscript{CO}$=2 \times 10^{20}$\,cm$^{-2}$K\,km\,s$^{-1}$.

Another important uncertainty concerns the CO(2-1) to CO(1-0) flux ratio, $R$\textsubscript{21}, which depends on optical depth and excitation conditions of the molecular gas \citep[e.g.][]{Braine92}. $R$\textsubscript{21} can vary significantly between objects.
Measurements of  $R$\textsubscript{21} have been made for gas-rich disc galaxies \citep[e.g.][]{Sandstrom13} and for radio-quiet ETGs \citep[e.g.][]{Young11}, but little is known about local radio-loud ETGs. The presence of a radio-loud AGN can significantly affect the conditions of the molecular gas in the surrounding regions \citep[e.g.][]{Oosterloo17}, but it is not clear whether such phenomena are common.

To date, the best estimate of $R$\textsubscript{21} for radio galaxies comes from the CO(2-1) to CO(1-0) brightness temperature ratios measured by \citet{Ocana10}. The mean ratio for 15 radio-loud ETGs observed in both CO transitions with the Institute de Radioastronomie Millimetrique (IRAM) 30m telescope is 2.32. This is the same as the flux density ratio provided that the CO emission is unresolved, since the ratios of the square of wavelength and the beam size cancel precisely \citep[e.g.][]{David14,Temi18}. We therefore take $R$\textsubscript{21}$=2.32$. We plan to test this assumption directly through observations of different CO transitions for our sample.

The estimated molecular masses range from $2.0\times10^{7}$ to $2.0\times10^{10}$~M\textsubscript{$\odot$}; upper limits for the non-detections are in the range $1.0-6.7\times10^{6}$~M$_{\rm \odot}$.  As discussed in the previous section, these upper limits will be underestimated if the CO emission is resolved. We have therefore also measured values or limits for the CO surface density ($\Sigma$\textsubscript{CO}).  For the detected sources, values averaged over the area covered by the CO emission range from 900 to few thousands of M$_{\rm \odot}$ pc$^{-2}$.  Limits for the undetected sources are typically a few hundred M$_{\rm \odot}$ pc$^{-2}$.  Values and limits for  M\textsubscript{mol} and $\Sigma$\textsubscript{CO} are listed in Table~\ref{tab:line parameters}.

\begin{table*}
\centering
\caption{Main\co\ line integrated parameters.}
\label{tab:line parameters}
\begin{tabular}{l c c  r  r  c c r r}
\hline
\multicolumn{1}{c}{ Target} &
\multicolumn{1}{c}{ Line FWHM }&
\multicolumn{1}{c}{ Line FWZI  }&
\multicolumn{1}{c}{ $\Sigma S\textsubscript{CO}\Delta{\rm v}$}&
\multicolumn{1}{c}{ M\textsubscript{mol} } &
\multicolumn{1}{c}{ v\textsubscript{CO}  } &
\multicolumn{1}{c}{ Size FWHM } &
\multicolumn{1}{c}{ $\Sigma$\textsubscript{CO}  } &\\
\multicolumn{1}{c}{  } &
\multicolumn{1}{c}{ (km~s$^{-1}$) } &
\multicolumn{1}{c}{ (km~s$^{-1}$) } &
\multicolumn{1}{c}{ (Jy km~s$^{-1}$) } &
\multicolumn{1}{c}{ (M$_{\rm \odot}$) } &
\multicolumn{1}{c}{ (km~s$^{-1}$) } &
\multicolumn{1}{c}{ (kpc$^{2}$) } &
\multicolumn{1}{c}{  (M$_{\rm \odot}$~pc$^{-2}$)} &\\
\multicolumn{1}{c}{ (1) } &
\multicolumn{1}{c}{ (2) } &
\multicolumn{1}{c}{ (3) } &
\multicolumn{1}{c}{ (4) } &
\multicolumn{1}{c}{ (5) } &
\multicolumn{1}{c}{ (6) } &
\multicolumn{1}{c}{ (7) } &
\multicolumn{1}{c}{ (8) } \\
\hline
 IC\,1531  &  260  &  260  &  2.0$\pm$0.2  &  $(1.1\pm0.1)\times10^{8}$  &   7686$\pm$20 & (0.25$\pm$0.05)$\times$(0.22$\pm$0.06)  &   3.8$\times$10$^{3}$ \\
 NGC\,612   &   760   &  840  &  273$\pm$27  &  $(2.0\pm0.2)\times10^{10}$  &  8924$\pm$20 & (9.6$\pm$0.3)$\times$(1.2$\pm$0.05)  &   1.0$\times$10$^{4}$ \\
  PKS 0718-34  &  334$^{1}$   &  480$^{1}$  &  $<0.1$  &  $<6.7\times10^{6}$  & $-$ & $-$  &   $<2.7\times10^{2}$ \\
   NGC\,3100 &  340 &  440  &  18$\pm$1.8 &  $(1.2\pm0.1)\times10^{8}$ &  2600$\pm$10 &  (1.6$\pm$0.3)$\times$(0.5$\pm$0.08)  &   9.1$\times10^{2}$ \\
 NGC\,3557  &  440  &  484  &  7.0$\pm$0.7  &  $(6.2\pm0.6)\times10^{7}$  &  3089$\pm$22 & (0.3$\pm$0.02)$\times$(0.2$\pm$0.01) &   3.7$\times10^{3}$  \\
 ESO 443-G 024  &  786$^{1}$   &  1000$^{1}$  &  $<0.1$  &  $<3.5\times10^{6}$  &  $-$ & $-$  &   $<3.9\times10^{2}$   \\
  IC\,4296$^{2}$  &  760  &  760  &  1.6$\pm$0.1  &  $(2.0\pm0.2)\times10^{7}$  &  3760$\pm$40 & (0.2$\pm$0.02)$\times$(0.04$\pm$0.02)  &   $3.0\times10^{3}$  \\
 NGC\,7075 &  560   &  600  &  1.0$\pm$0.1  &  $(2.9\pm0.2)\times10^{7}$  &  5510$\pm$40 &  $<0.2$  &   $3.3\times10^{3}$ \\
 IC\,1459  &  492$^{1}$   &  640$^{1}$  &   $<0.4$  &   $<1.0\times10^{6}$  &  $-$ & $-$  &   $<7.6\times10^{2}$ \\ 
\hline
\end{tabular}
\parbox[t]{1\textwidth}{ \textit{Notes.} $-$ Columns: (1) Target name. (2) Line FWHM defined directly from the integrated spectra as the velocity difference between the two most distant channels from the line centre with intensities $\geq$50\% of the line peak. (3) Full velocity range covered by spectral channels (identified interactively in the channel map) with CO intensities $\geq$3$\sigma$ (FWZI, grey shaded region in Figs.~\ref{IC1531}$-$\ref{NGC7075}). (4) Integrated CO flux density measured integrating numerically over all the channels in the range defined by the FWZI. Upper limits of the undetected sources are in units of Jy~beam$^{-1}$ km~s$^{-1}$. The velocity ranges in columns (2) -- (4) are measured in the source frame. (5) Molecular gas mass derived using Equation~\ref{eq:gas mass}. (6) CO systemic velocity in the LSRK frame (optical convention), determined numerically as the intensity-weighted centroid of the mean velocity map. (7) Size (FWHM, deconvolved from the beam) of the CO emission. (8) CO surface density over the area covered by the CO emission. The surface densities of the undetected sources were estimated over the beam area. \\
$^{1}$The line FWHM and FWZI of undetected sources are those estimated from APEX spectra (Paper II).\\
$^{2}$The molecular gas mass of IC\,4296 is measured considering CO emission only. If the channels of the absorption feature are included (and integrated as CO emission), we obtain $\Sigma S_{\rm CO}\Delta{\rm v}$ = (1.9$\pm$0.1)~Jy~km~s$^{-1}$, M$_{\rm mol}$=$(2.3\pm0.8)\times10^{7}$~M$_{\rm \odot}$.}
\end{table*}

\begin{figure*}
\centering
\begin{subfigure}[t]{.30\textwidth}
\centering
\caption{\textbf{IC\,1531}}\label{fig:ic1531_cont}
\includegraphics[width=\linewidth]{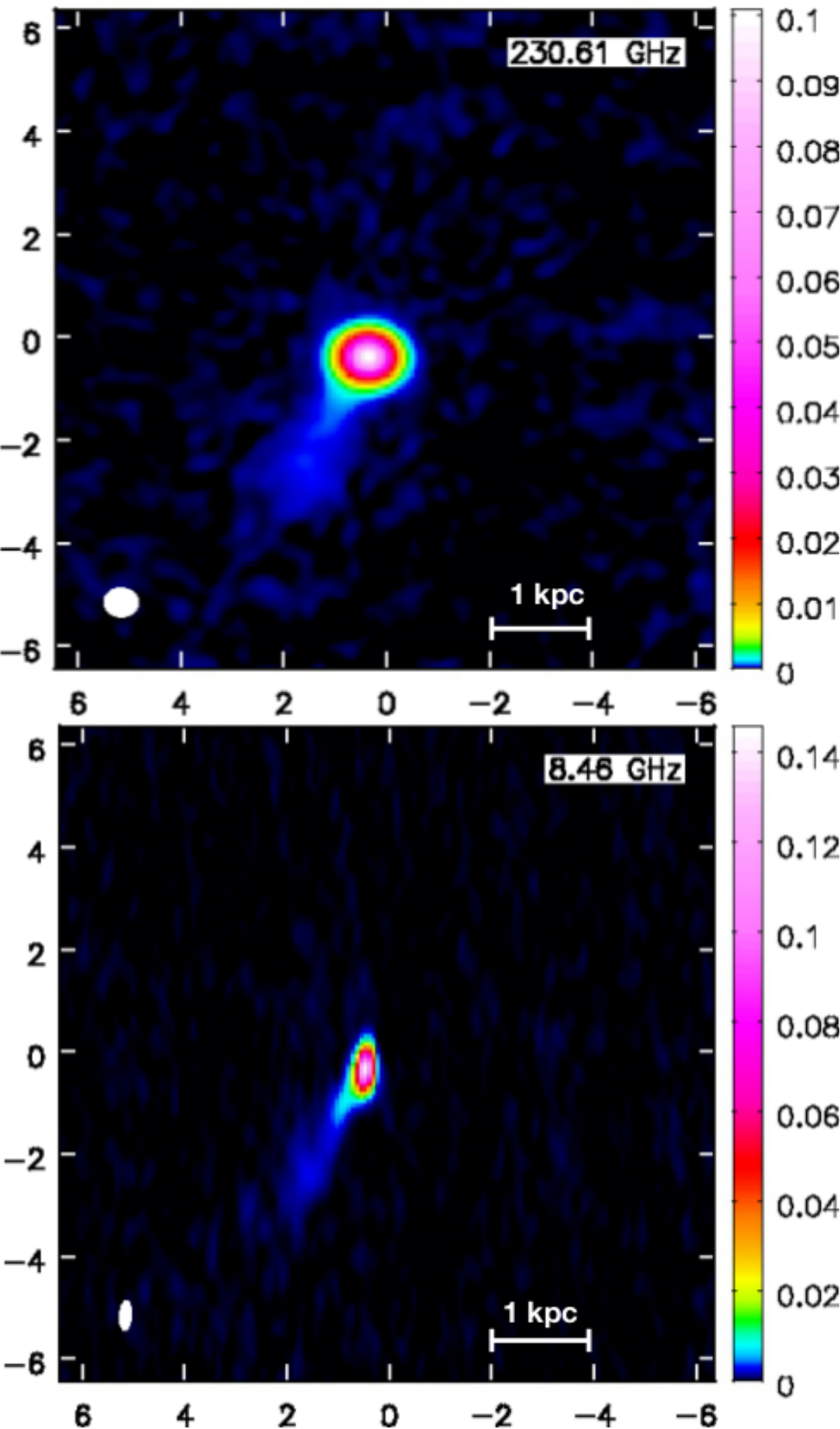}
\end{subfigure}
\hspace{3mm}
\begin{subfigure}[t]{.31\textwidth}
\centering
\caption{\textbf{NGC\,612}}\label{fig:ngc612_cont}
\includegraphics[width=\linewidth]{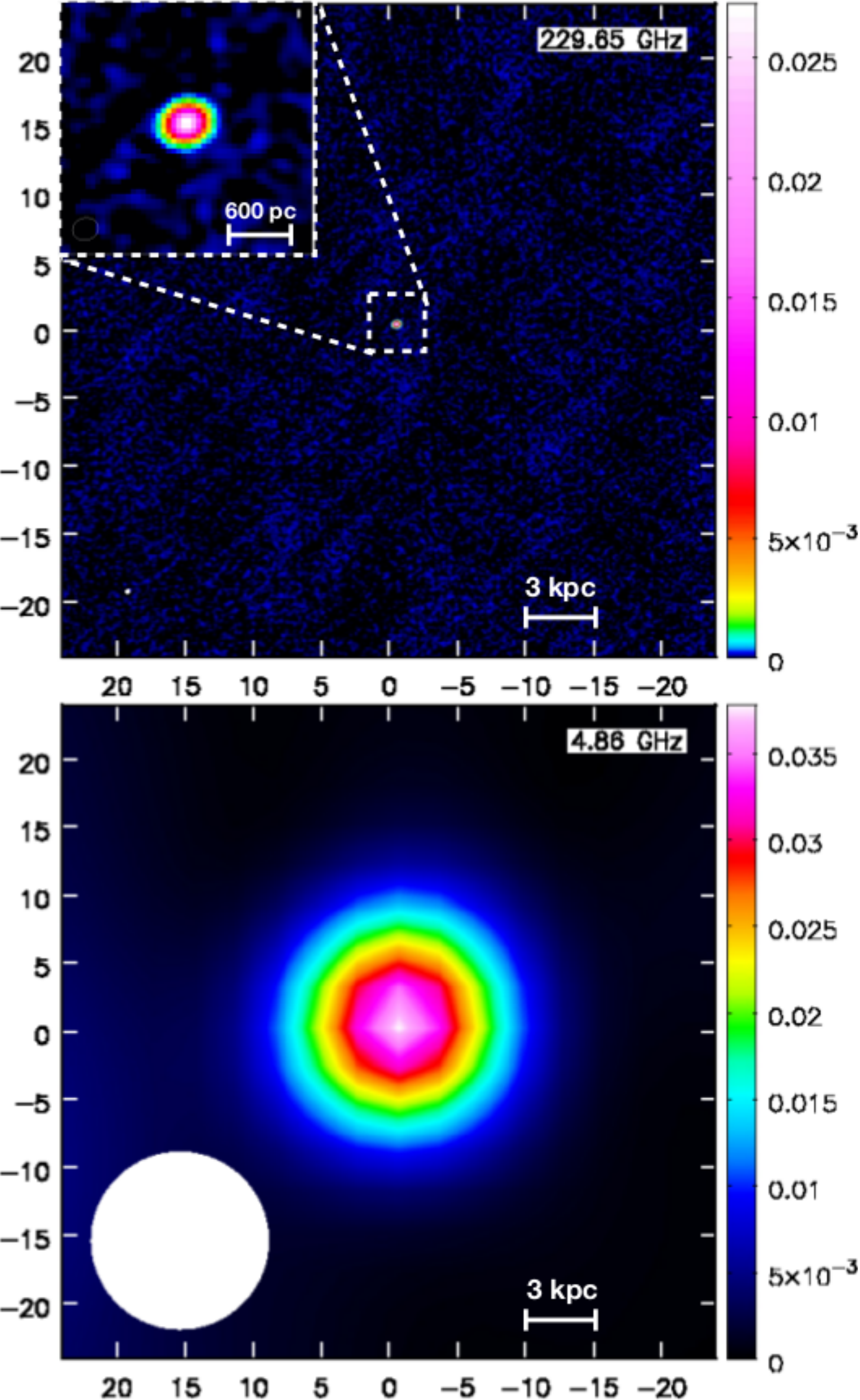}
\end{subfigure}
\hspace{3mm}
\begin{subfigure}[t]{.32\textwidth}
\centering
\caption{\textbf{PKS\,0178-34}}\label{fig:pks0718_cont}
\includegraphics[width=\linewidth]{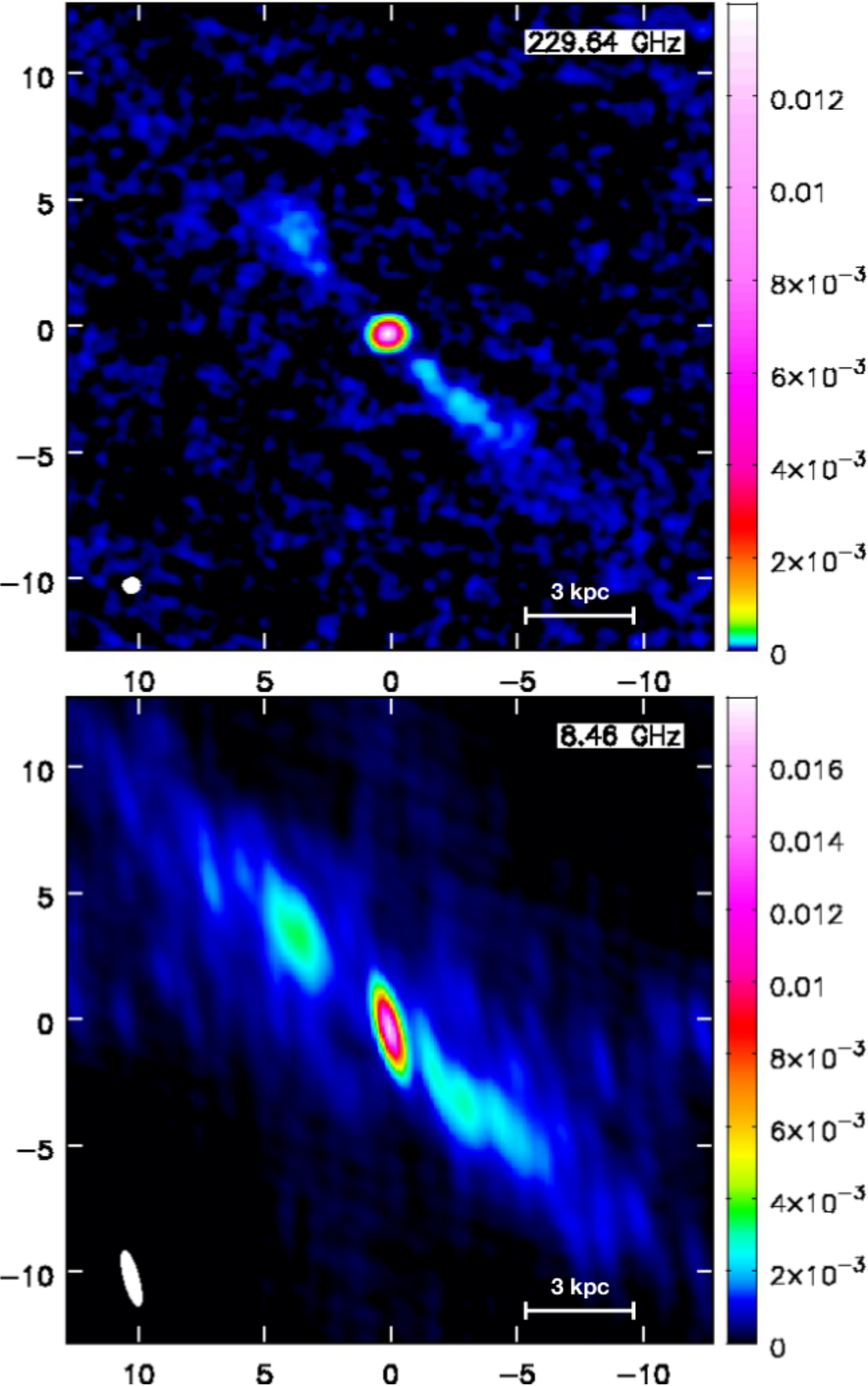}
\end{subfigure}

\medskip
\begin{subfigure}[t]{.31\textwidth}
\centering
 \caption{\textbf{NGC\,3100}}\label{fig:ngc3100_cont}
\includegraphics[width=\linewidth]{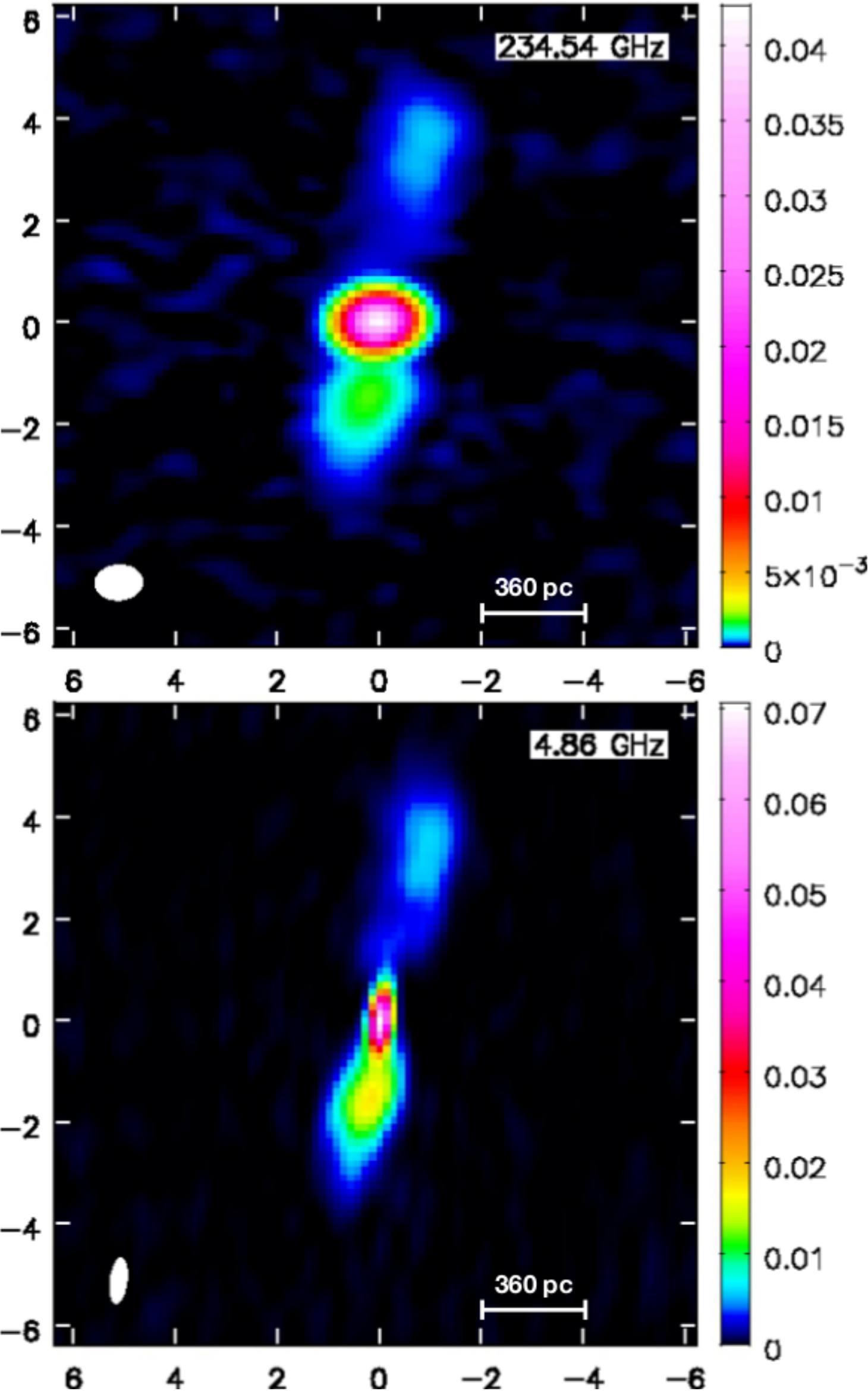}
\end{subfigure}
\hspace{3mm}
\begin{subfigure}[t]{.31\textwidth}
\centering
\caption{\textbf{NGC\,3557}}\label{fig:ngc3557_cont}
\includegraphics[width=\linewidth]{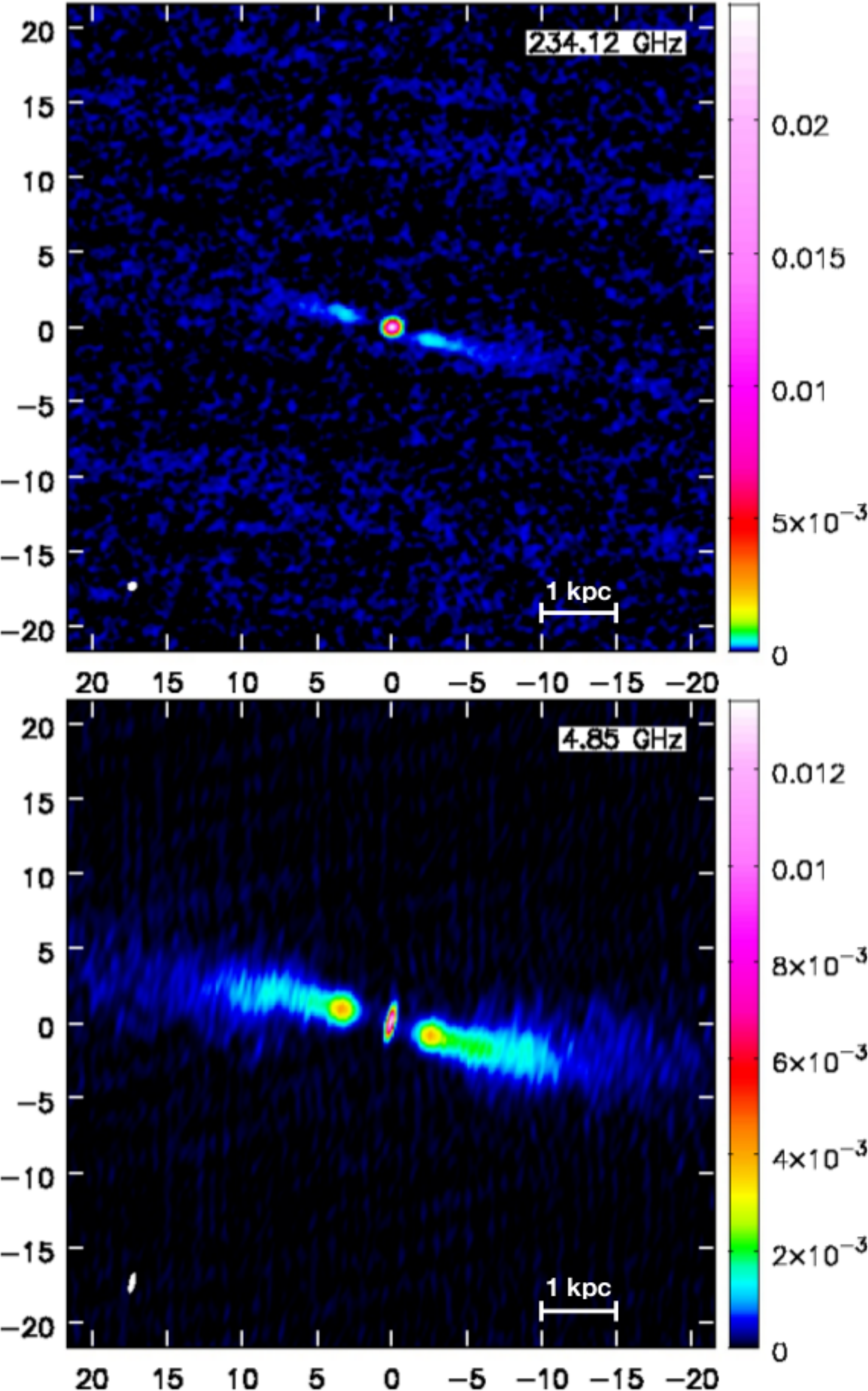}
\end{subfigure}
\hspace{3mm}
\begin{subfigure}[t]{.31\textwidth}
\centering
\caption{\textbf{ESO\,443-G 024}}\label{fig:eso443_cont}
\includegraphics[width=\linewidth]{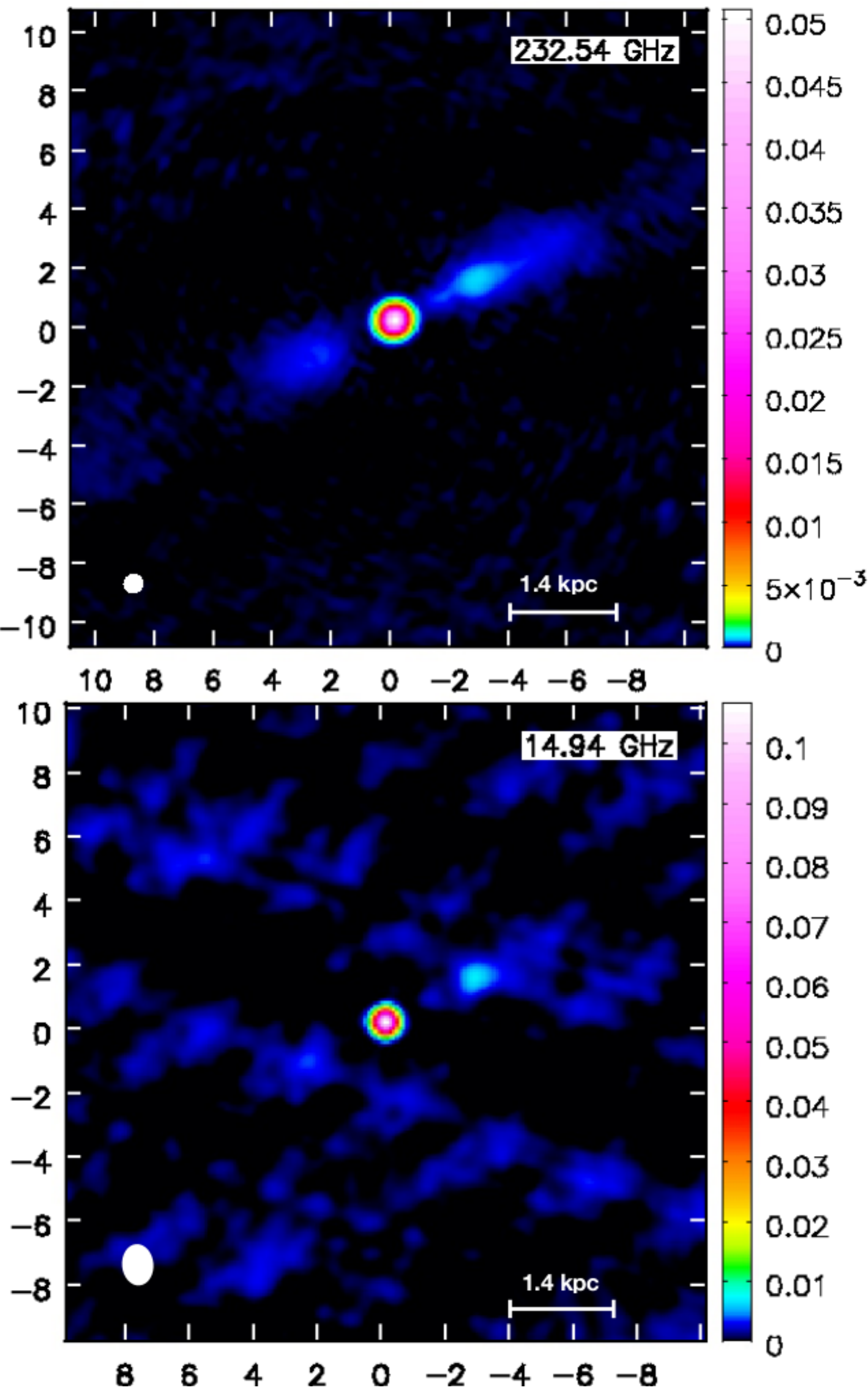}
\end{subfigure}
\caption{Naturally weighted ALMA Band 6 (upper panels) and archival VLA (lower panels) continuum maps of each target. The reference frequency of each observation is indicated in the top-right corner of the panel. The wedge on the right of each map shows the colour scale in Jy~beam$^{-1}$. Coordinates are given as relative positions with respect to the image phase centre in arcseconds; East is to the left and North to the top. The synthesised beam and the scale bar are shown in the bottom-left and bottom-right corner, respectively, of each panel. The properties of the ALMA continuum images are summarised in Table~\ref{tab:Continuum images}. Information about the archival radio images is provided in Appendix A.\label{fig:con230GHz}}
\end{figure*}

\begin{figure*}\ContinuedFloat
\begin{subfigure}[t]{.31\textwidth}
\centering
\caption{\textbf{IC\,4296}}\label{fig:ic4296_cont}
\includegraphics[width=\linewidth]{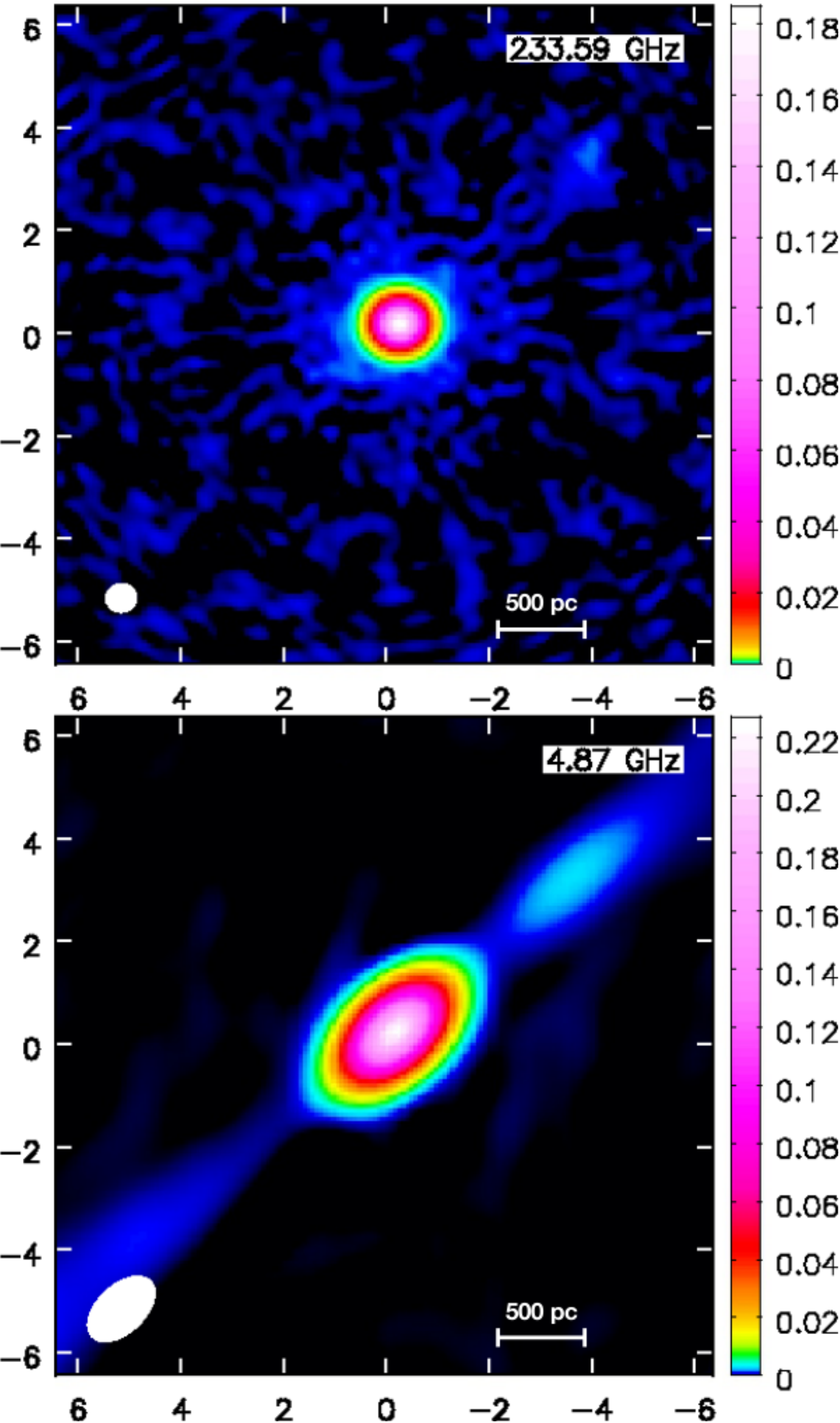}
\end{subfigure}
\hspace{3mm}
\begin{subfigure}[t]{.32\textwidth}
\centering
\caption{\textbf{NGC\,7075}}\label{fig:ngc7075_cont}
\includegraphics[width=\linewidth]{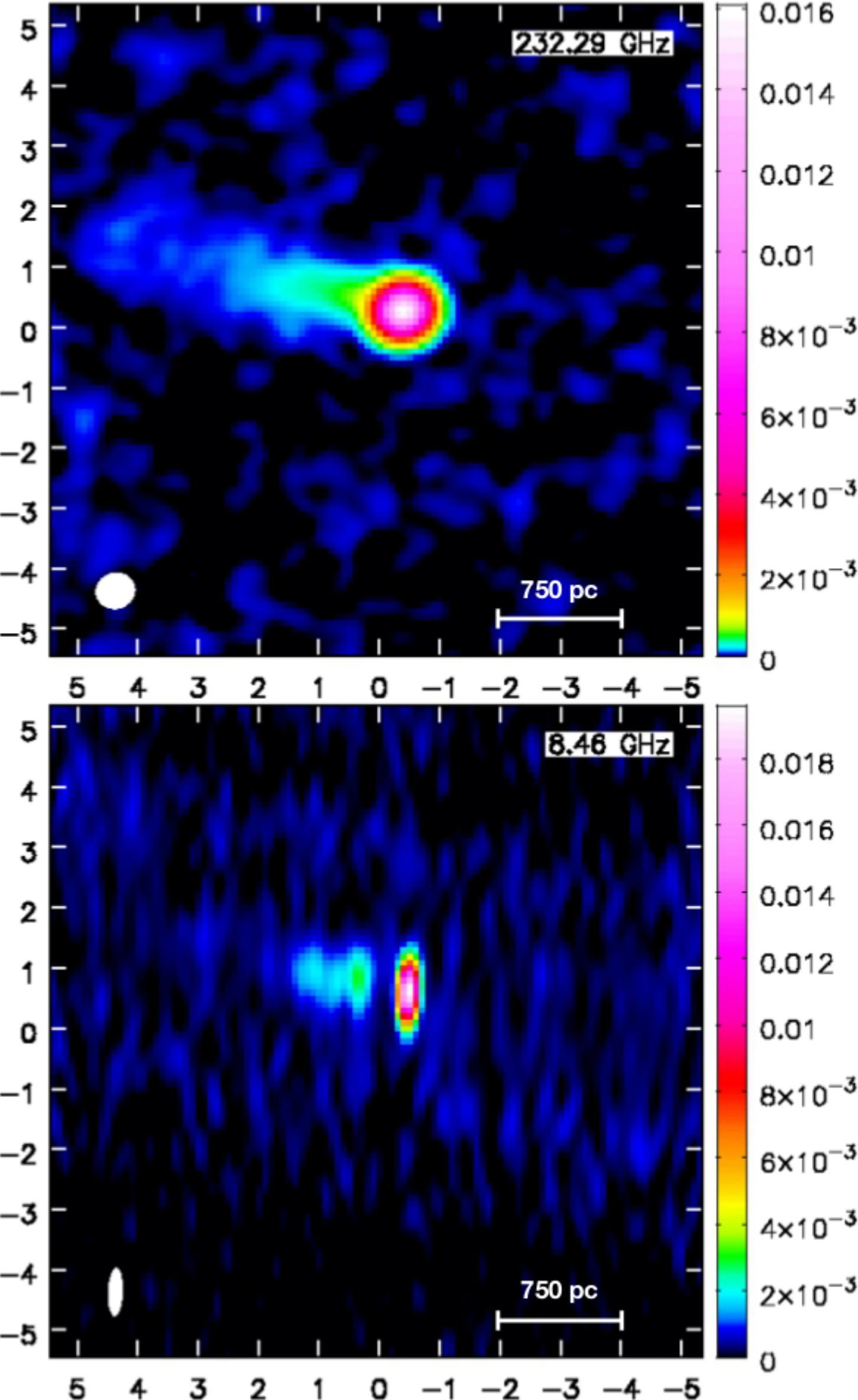}
\end{subfigure}
\hspace{3mm}
\begin{subfigure}[t]{.31\textwidth}
\centering
\caption{\textbf{IC\,1459}}\label{fig:ic1459_cont}
\includegraphics[width=\linewidth]{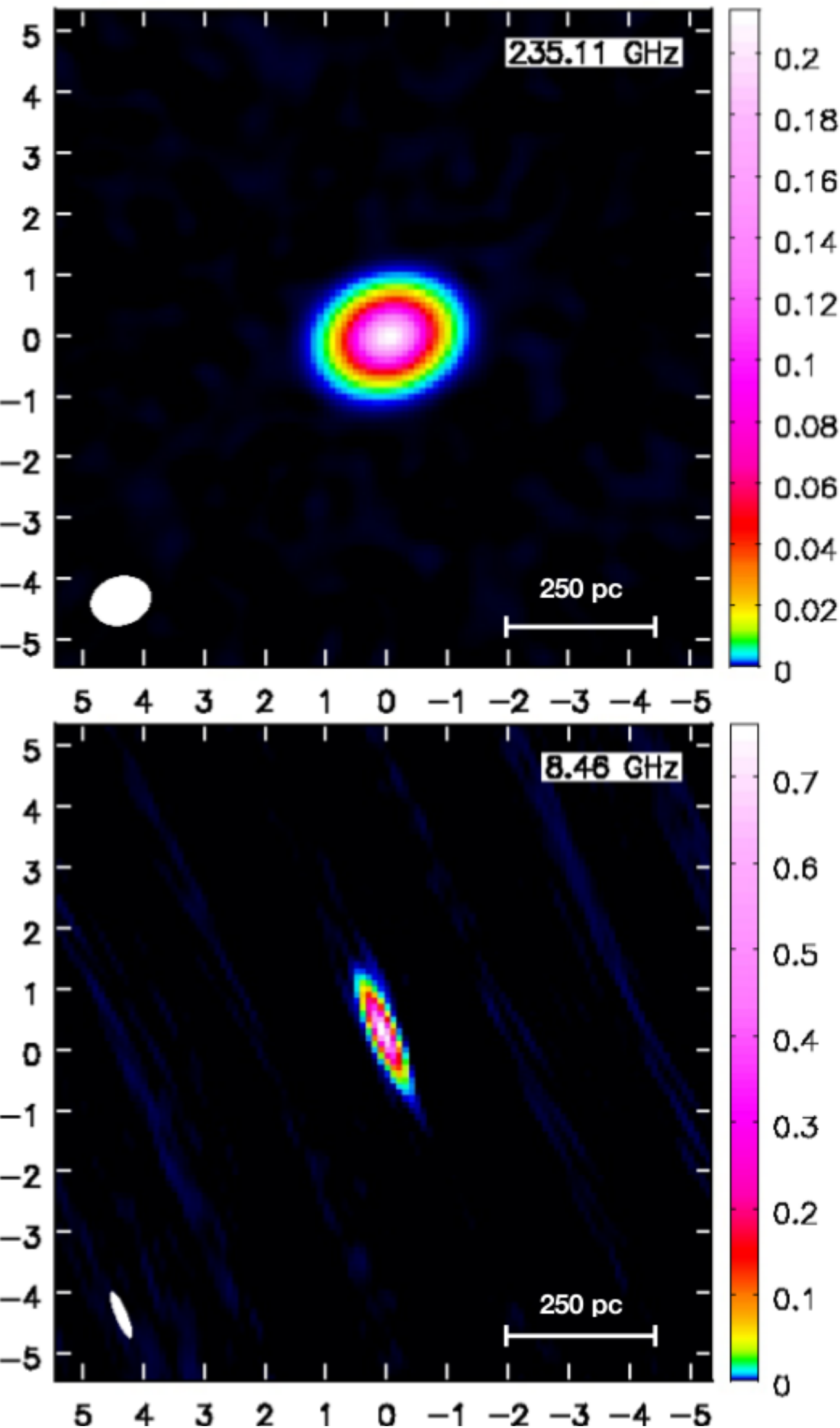}
\end{subfigure}
\caption{\textit{- continued.}\label{fig:continuum}}
\end{figure*}

\begin{figure*}
\centering
\begin{subfigure}[t]{0.3\textheight}
\centering
 \caption{}\label{fig:ic1531_mom0}
\includegraphics[scale=0.3]{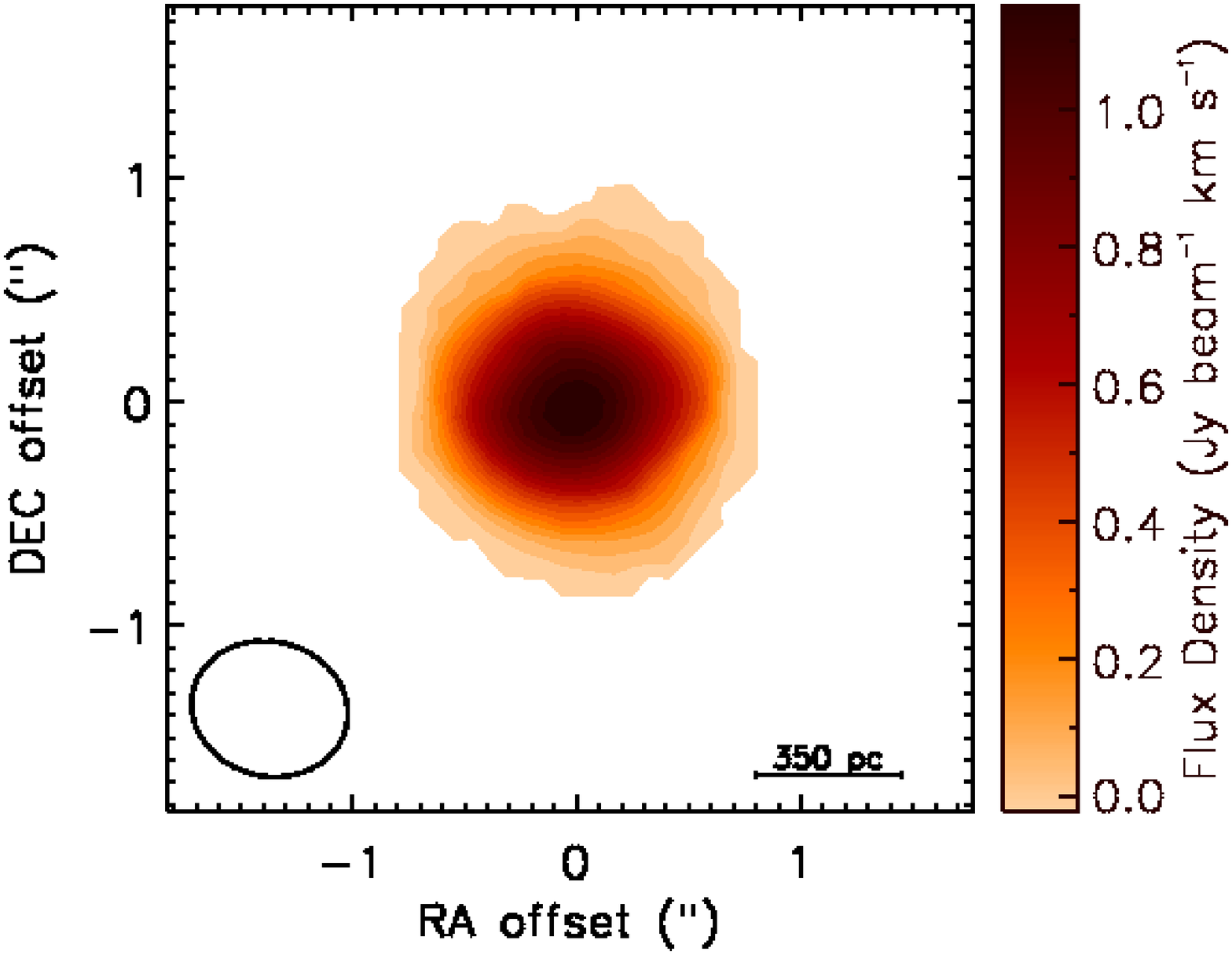}
\end{subfigure}
\hspace{6.5mm}
\begin{subfigure}[t]{0.3\textheight}
\centering
\caption{}\label{fig:ic1531_mom1}
\includegraphics[scale=0.3]{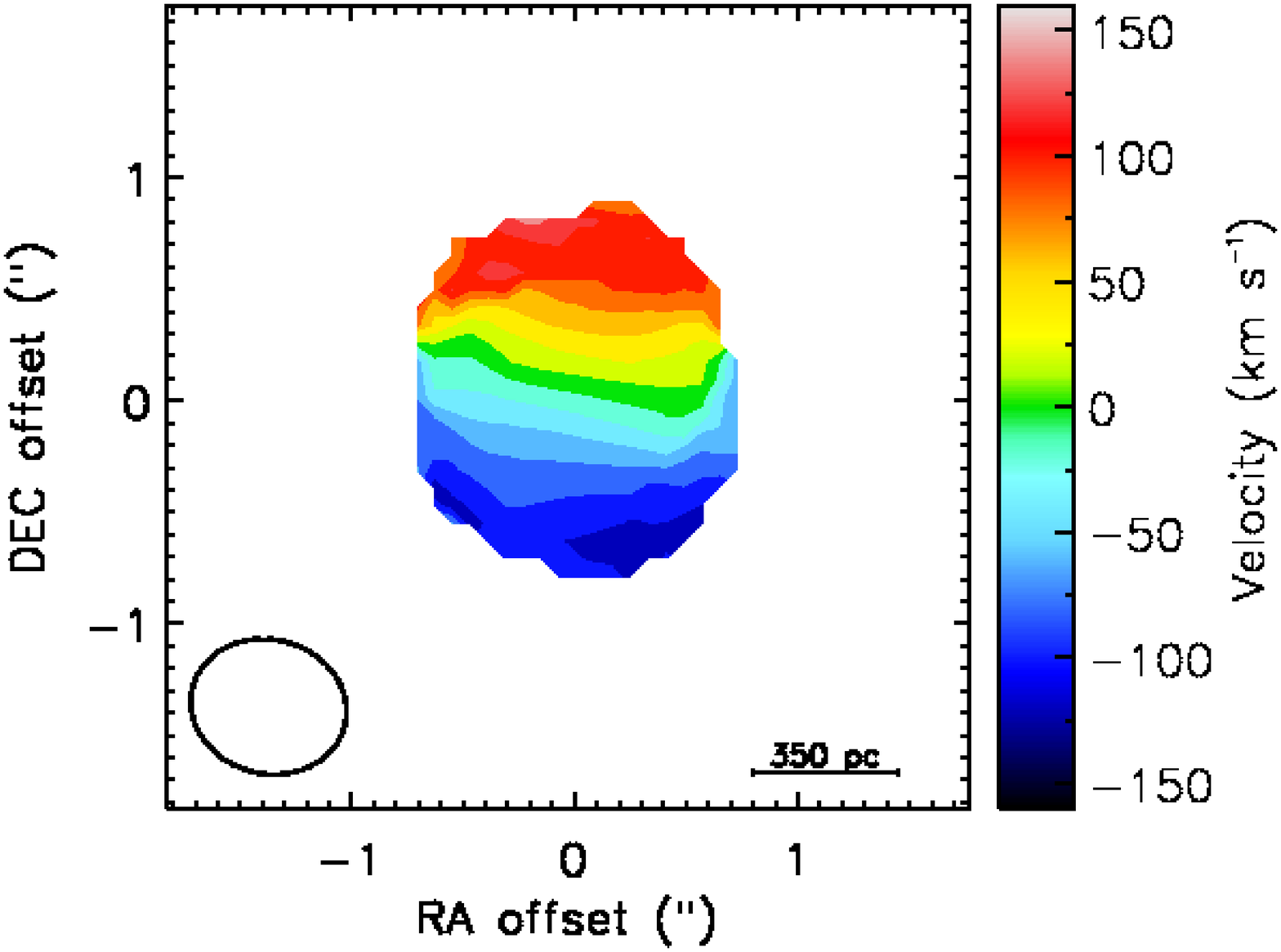}
\end{subfigure}

\medskip

\begin{subfigure}[t]{0.3\textheight}
\centering
\vspace{0pt}
\caption{}\label{fig:ic1531_mom2}
\includegraphics[scale=0.3]{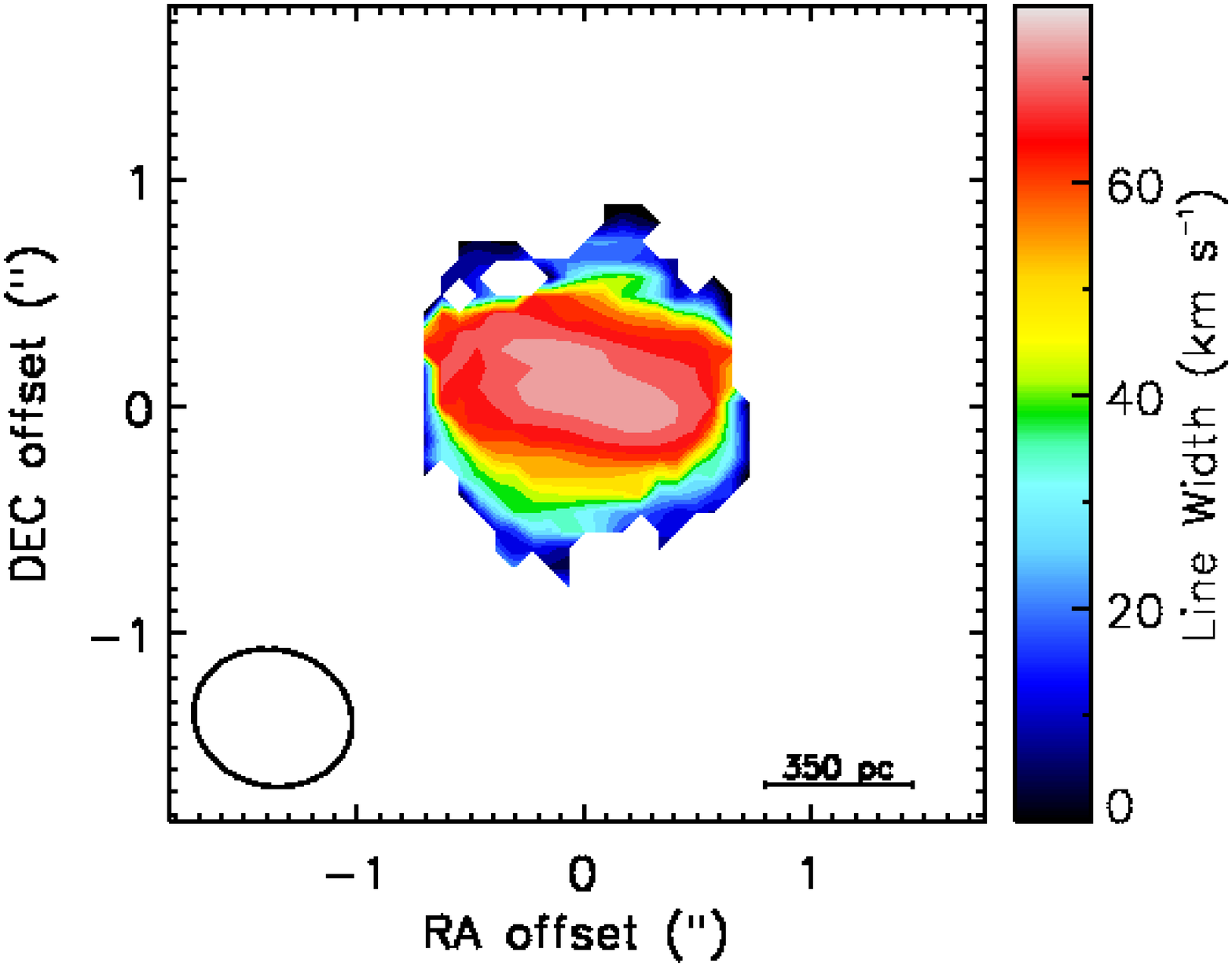}
\end{subfigure}
\hspace{6.5mm}
\begin{subfigure}[t]{0.3\textheight}
\centering
\vspace{0pt}
\caption{}\label{fig:ic1531_spectrum}
\includegraphics[scale=0.3]{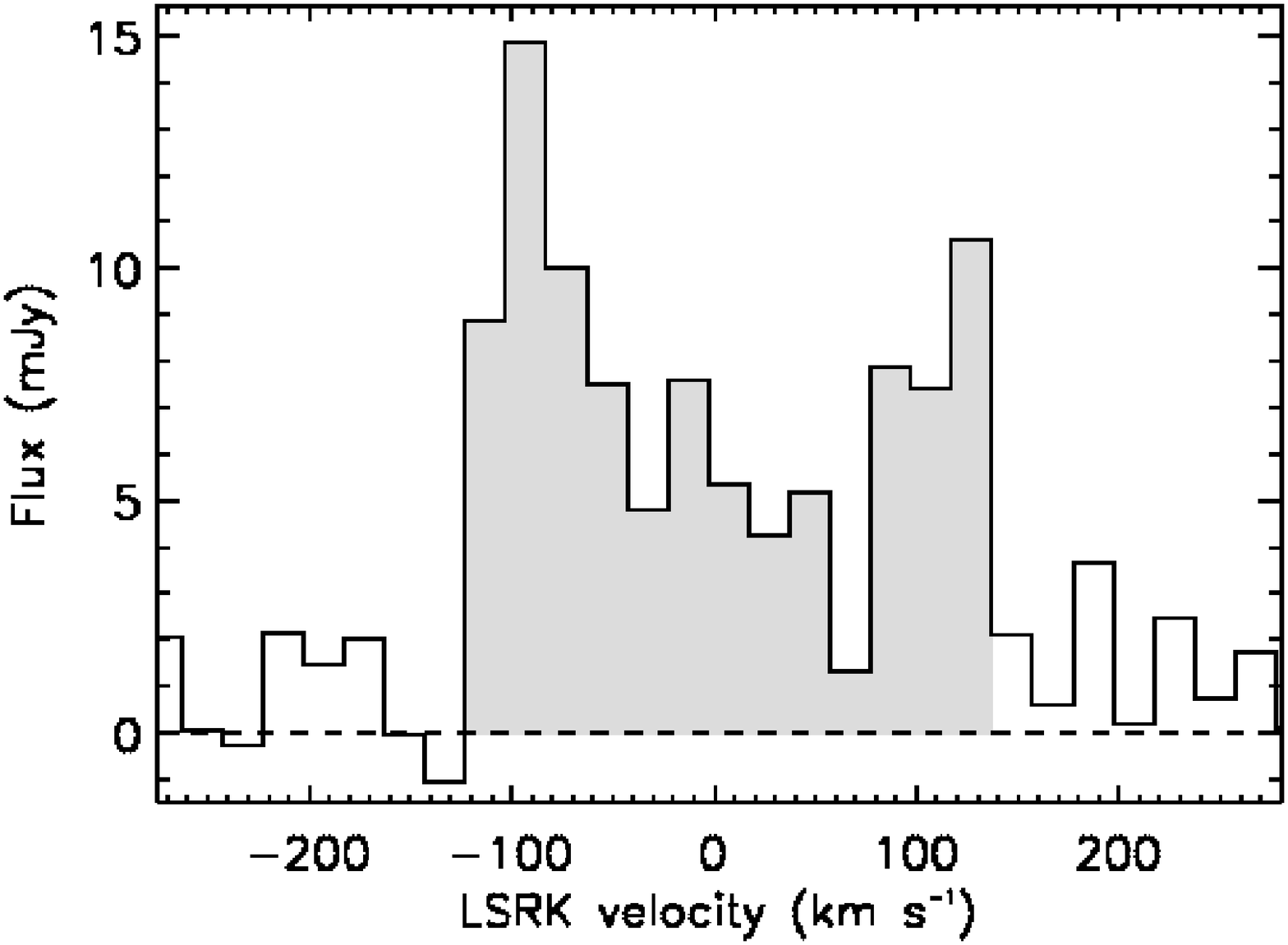}
\end{subfigure}
\caption{IC\,1531 moment 0 (\ref{fig:ic1531_mom0}), moment 1 (\ref{fig:ic1531_mom1}) and moment 2 (\ref{fig:ic1531_mom2}) maps created with the masked moment technique described in Section~\ref{sec:analysis} using a data cube with a channel width of 20~km~s$^{-1}$. The synthesised beam is shown in the bottom-left corner of each panel. The wedges to the right show the colour scale. In each panel, East is to the left and North to the top. The integrated spectral profile in panel~\ref{fig:ic1531_spectrum} was extracted within a box of $1.6 \times 1.6$\,arcsec$^2$, including all the CO emission. The spectral channels that are used to estimate the FWZI are highlighted in grey. The black dashed horizontal line indicates the zero flux level. In panels b -- d, velocities are measured in the source frame and the zero-point corresponds to the intensity-weighted centroid of the CO emission, equivalent to v\textsubscript{CO} in the LSRK frame (Table~\ref{tab:line parameters}).}\label{IC1531}
\end{figure*}

\begin{figure*}
\centering
\begin{subfigure}[t]{0.3\textheight}
\centering
 \caption{}\label{fig:ngc612_mom0}
\includegraphics[scale=0.3]{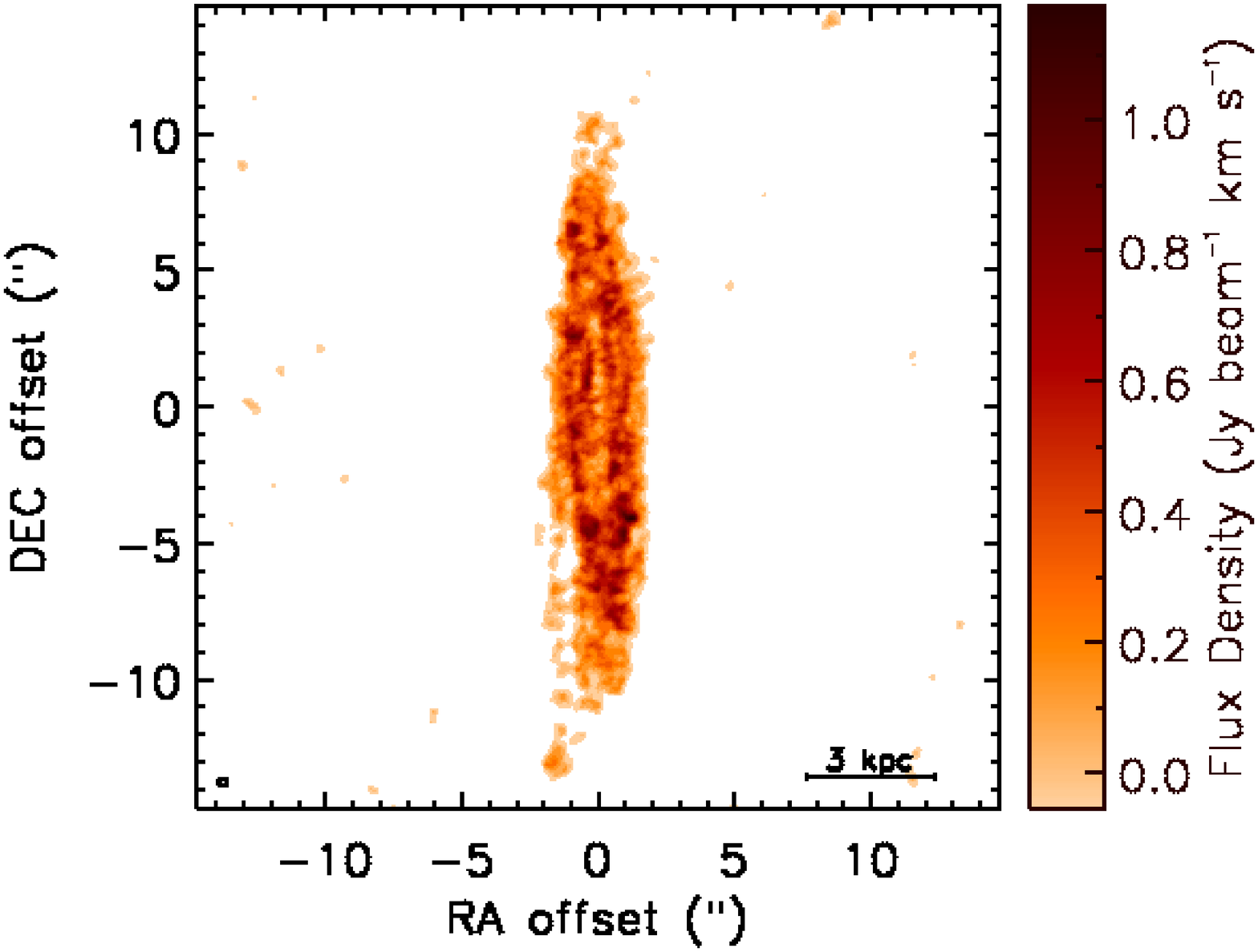}    
\end{subfigure}
\hspace{6.8mm}
\begin{subfigure}[t]{0.3\textheight}
\centering
\caption{}\label{fig:ngc612_mom1}
\includegraphics[scale=0.3]{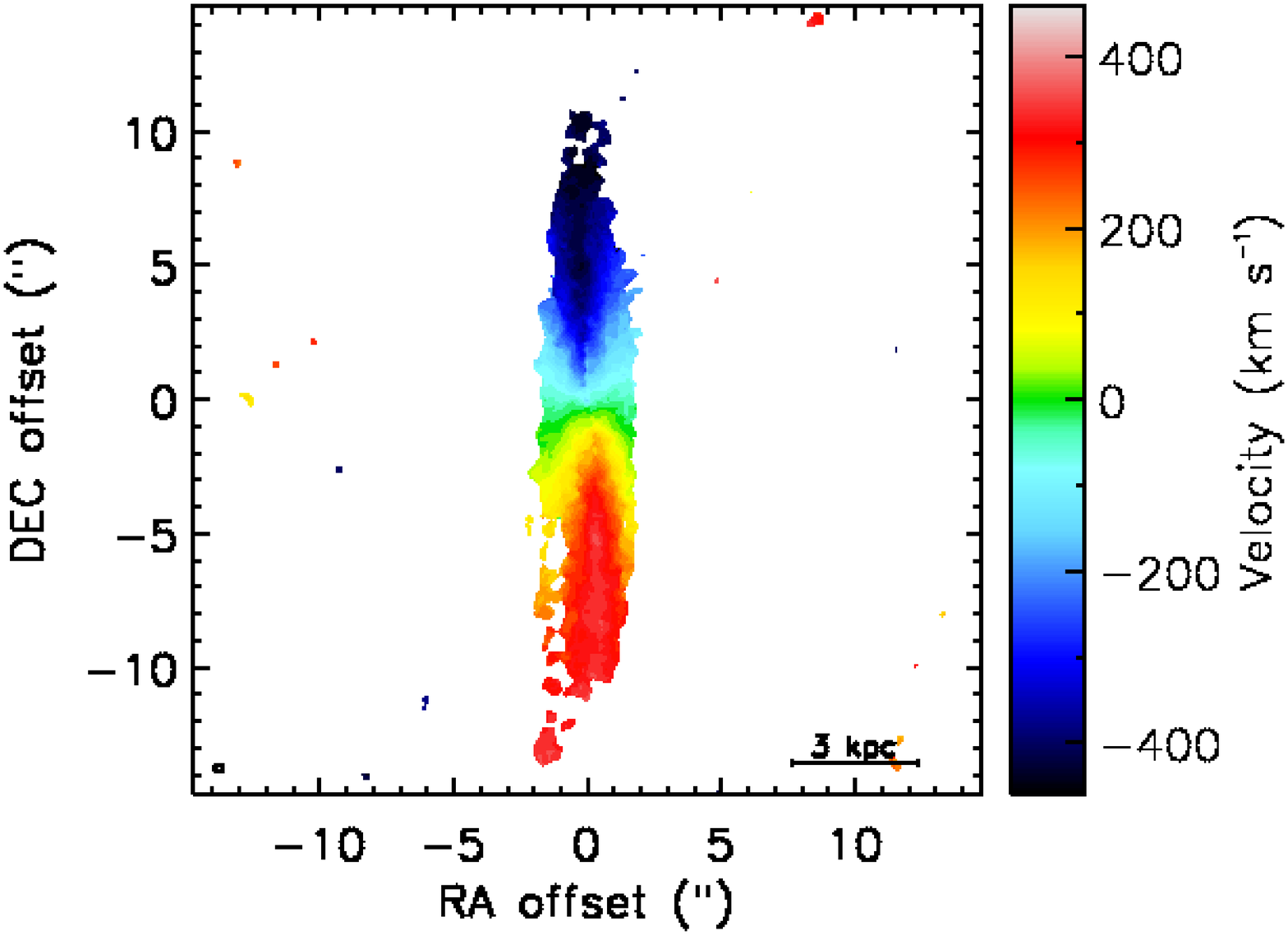}
\end{subfigure}

\medskip

\begin{subfigure}[t]{0.3\textheight}
\centering
\vspace{0pt}
\caption{}\label{fig:ngc612_mom2}
\includegraphics[scale=0.3]{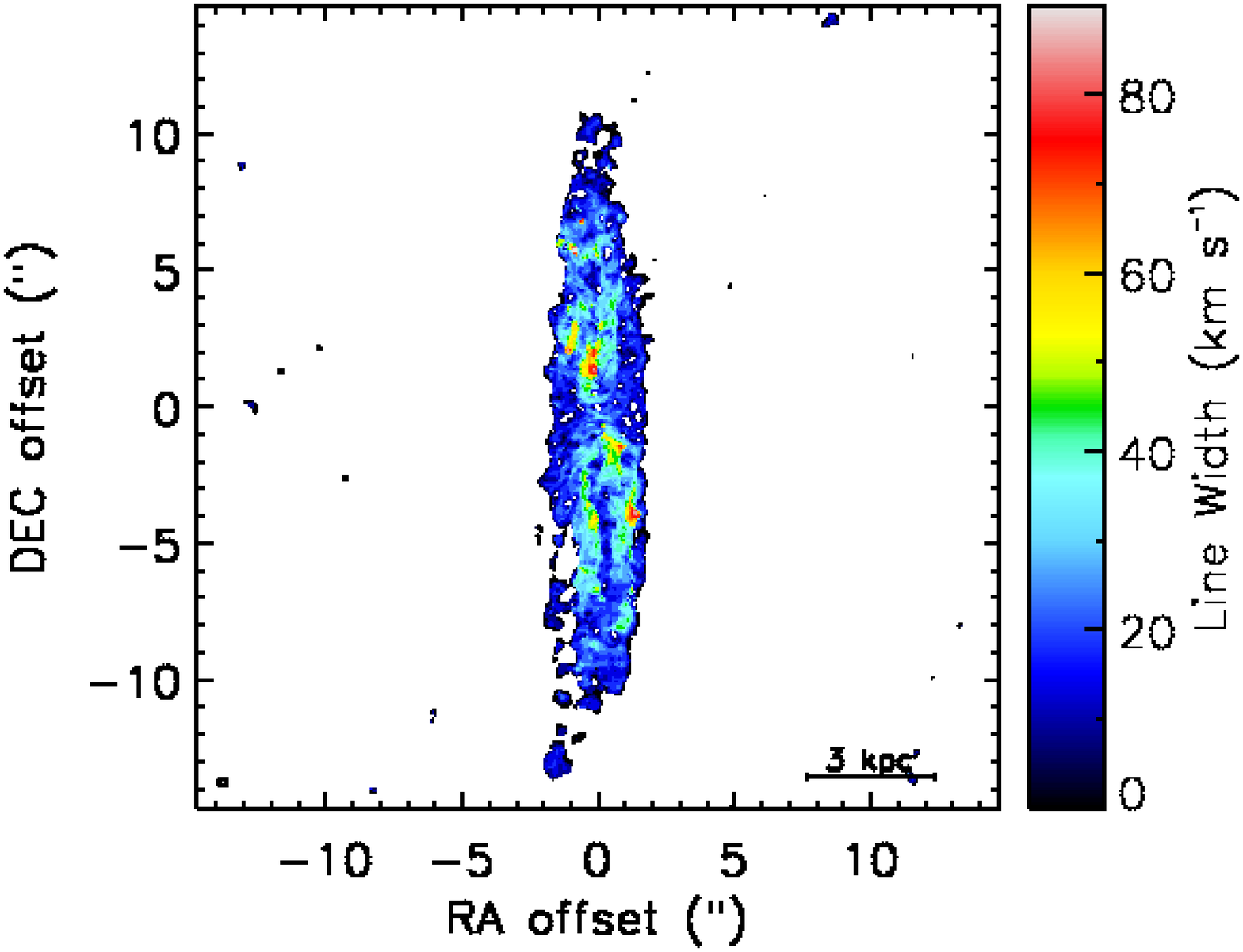}
\end{subfigure}
\hspace{6.5mm}
\begin{subfigure}[t]{0.3\textheight}
\centering
\vspace{0pt}
\caption{}\label{fig:ngc612_spectrum}
\includegraphics[scale=0.3]{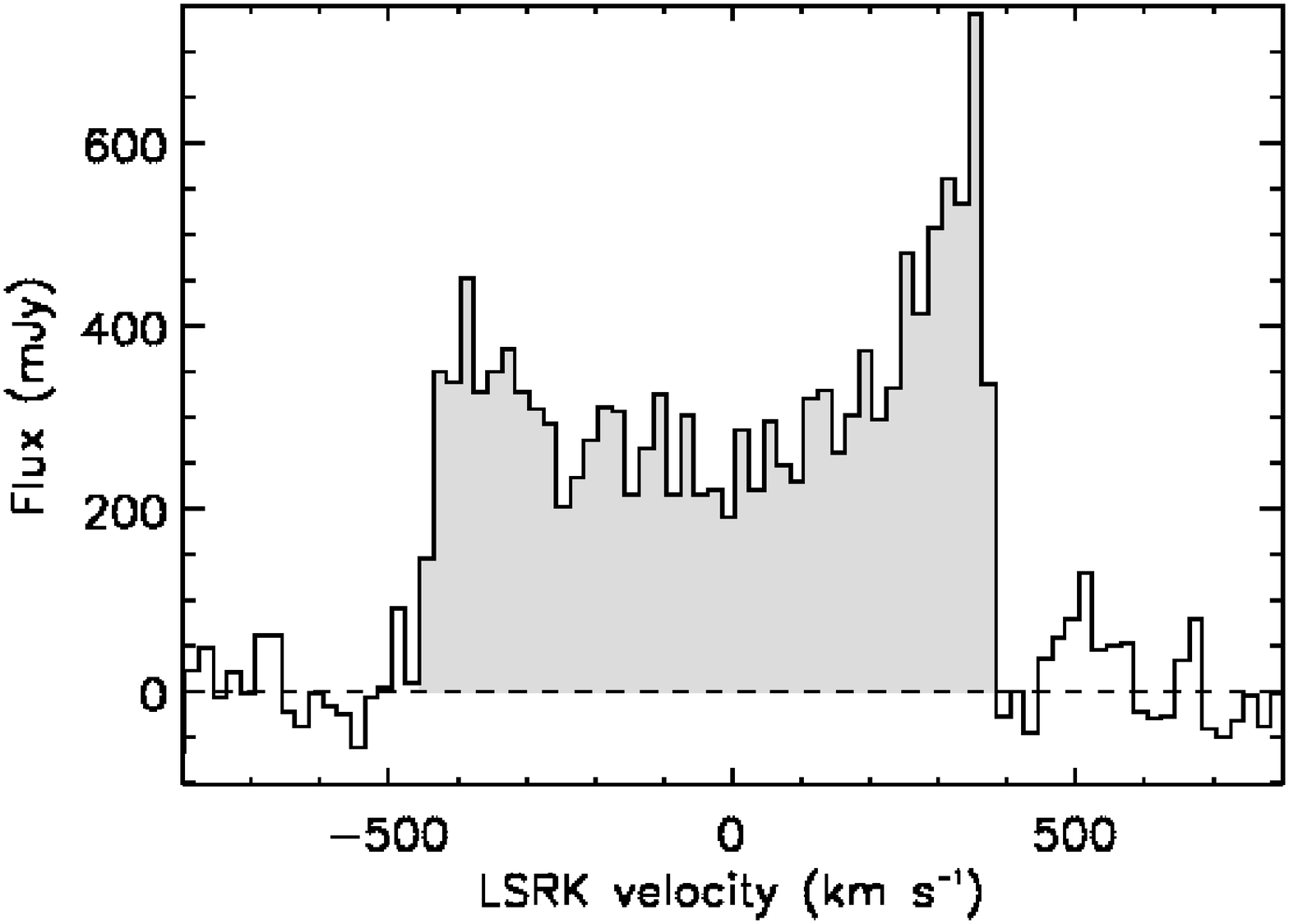}
\end{subfigure}
\caption{NGC\,612 moment maps and spectral profile as in Fig.~\ref{IC1531}, created using a data cube with a channel width of 20~km~s$^{-1}$. The integrated CO spectral profile was extracted within a box of $4 \times 20$\,arcsec$^2$.}\label{NGC612}
\end{figure*}

\begin{figure*}
\centering
\begin{subfigure}[t]{0.3\textheight}
\centering
 \caption{}\label{fig:ngc3100_mom0}
\includegraphics[scale=0.3]{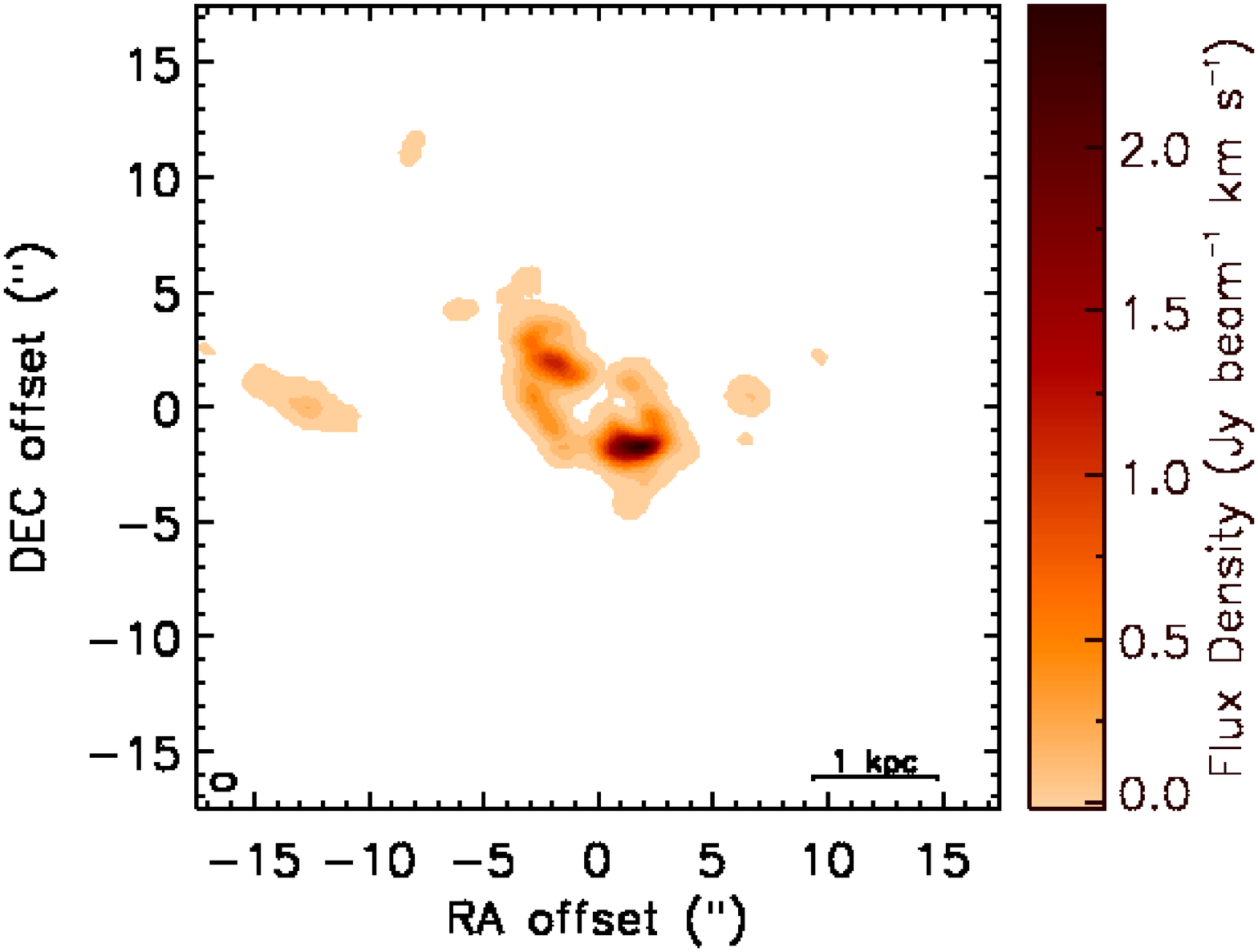}
\end{subfigure}
\hspace{7mm}
\begin{subfigure}[t]{0.3\textheight}
\centering
\caption{}\label{fig:ngc3100_mom1}
\includegraphics[scale=0.3]{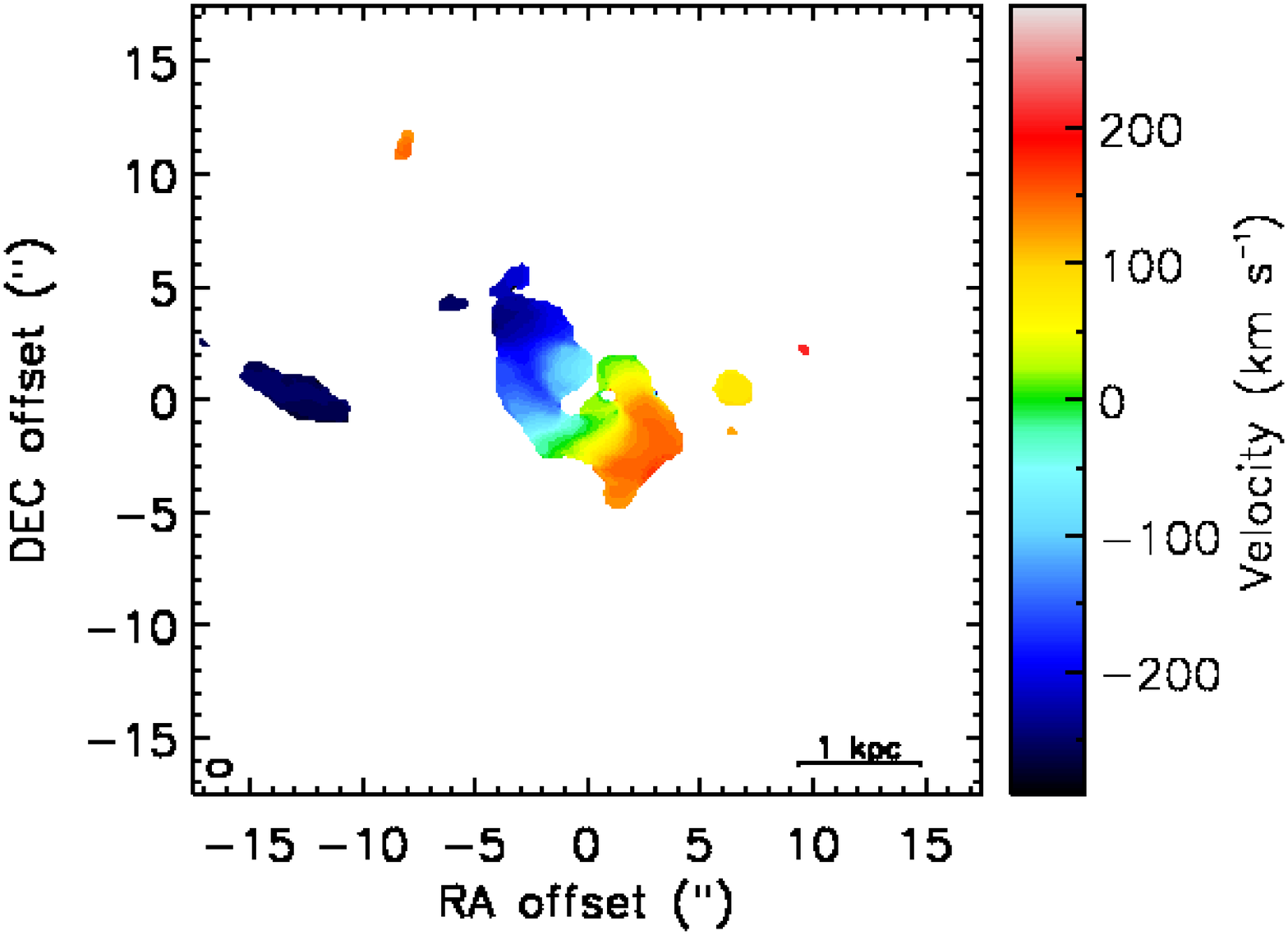}
\end{subfigure}
\medskip
\begin{subfigure}[t]{0.3\textheight}
\centering
\caption{}\label{fig:ngc3100_mom2}
\includegraphics[scale=0.3]{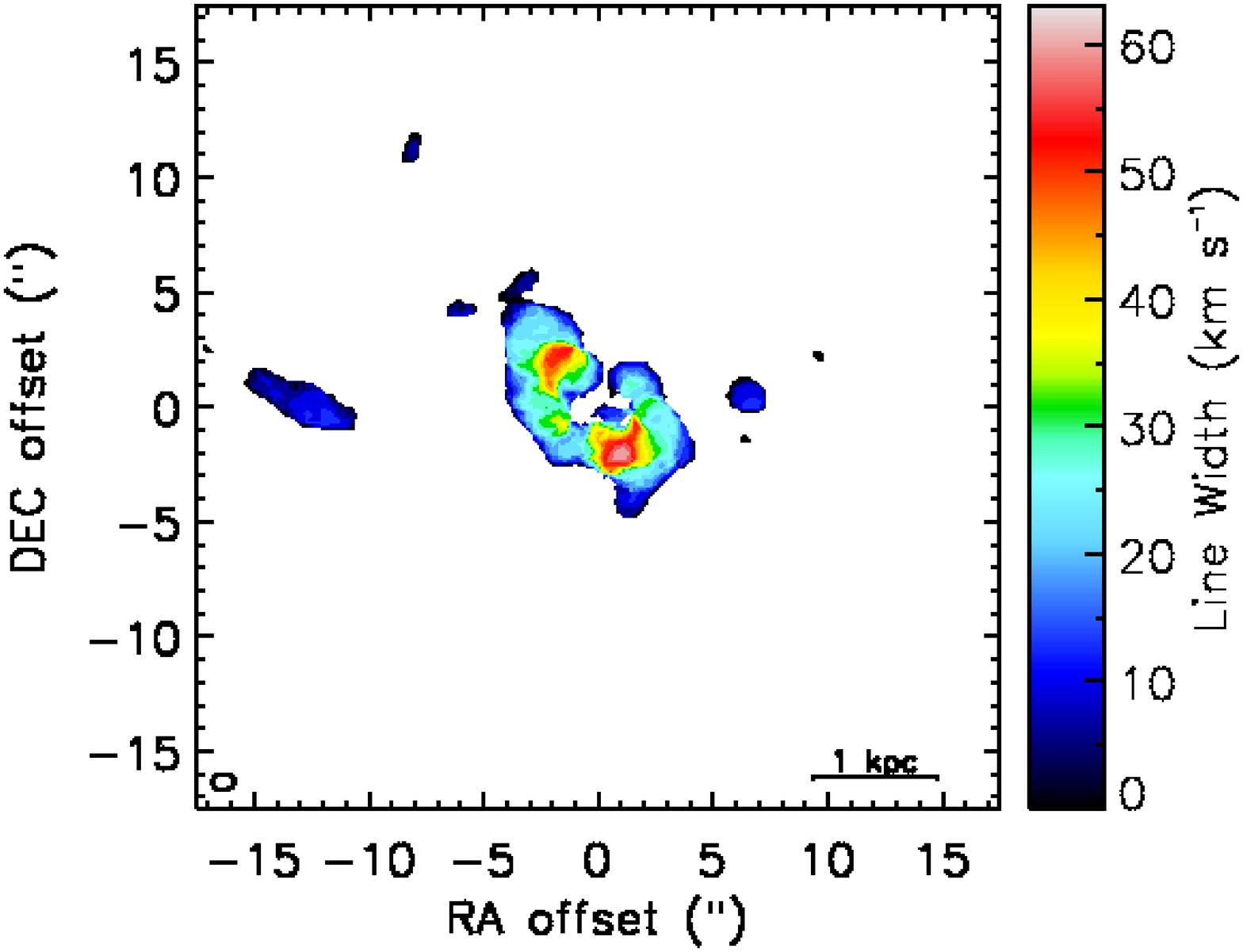}
\end{subfigure}
\hspace{7mm}
\begin{subfigure}[t]{0.3\textheight}
\centering
\caption{}\label{fig:ngc3100_spectrum}
\includegraphics[scale=0.3]{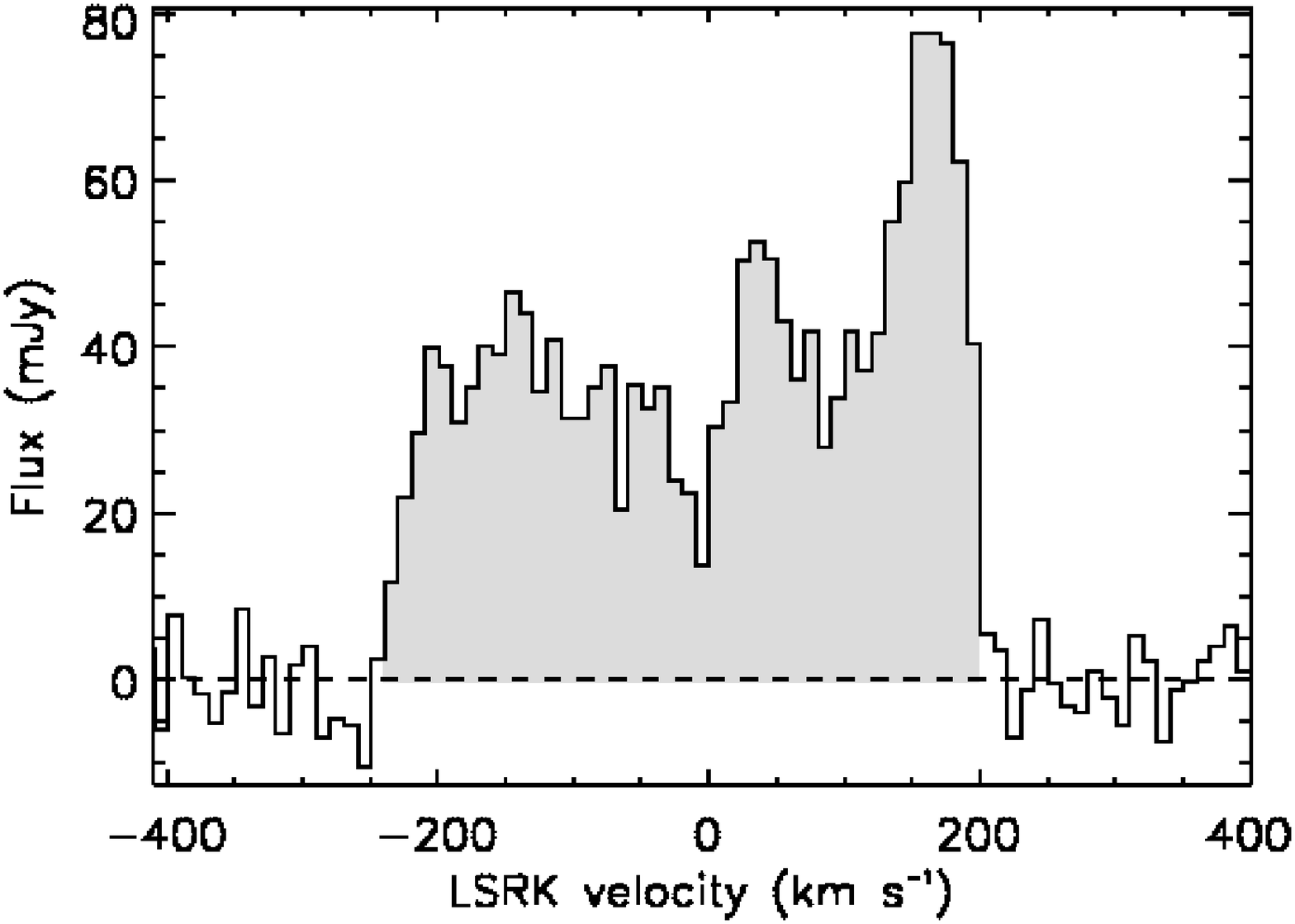}
\end{subfigure}
\caption{NGC\,3100 moment maps and spectral profile as in Fig.~\ref{IC1531}, created using a data cube with a channel width of 10~km~s$^{-1}$. The integrated CO spectral profile was extracted from the data cube within a 8.6 $\times 10$\,arcsec$^2$ box.}\label{fig:NGC3100}
\end{figure*}

\begin{figure*}
\centering
\begin{subfigure}[t]{0.3\textheight}
\centering
 \caption{}\label{fig:ngc3557_mom0}
\includegraphics[scale=0.3]{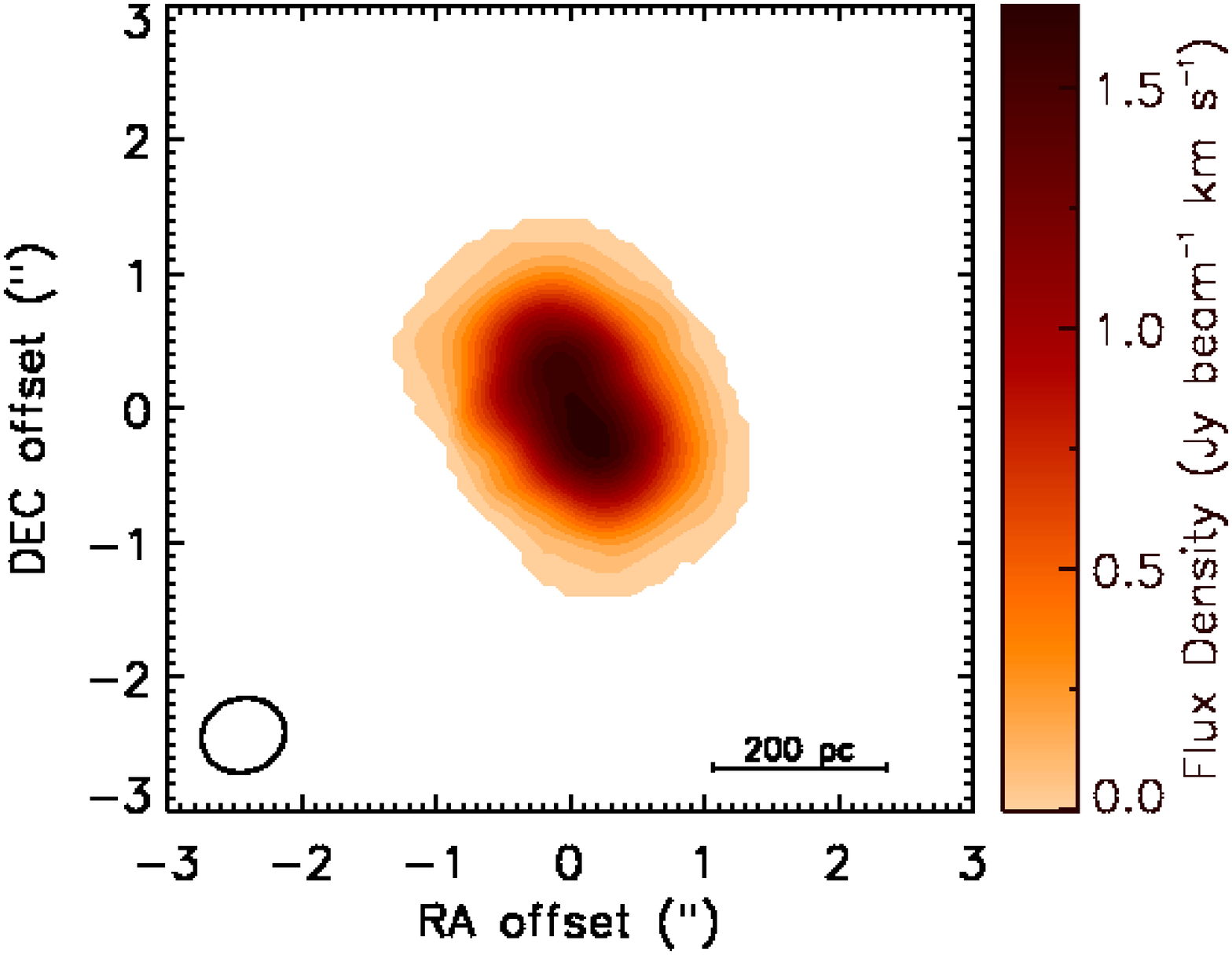}
\end{subfigure}
\hspace{6.5mm}
\begin{subfigure}[t]{0.3\textheight}
\centering
\caption{}\label{fig:ngc3557_mom1}
\includegraphics[scale=0.3]{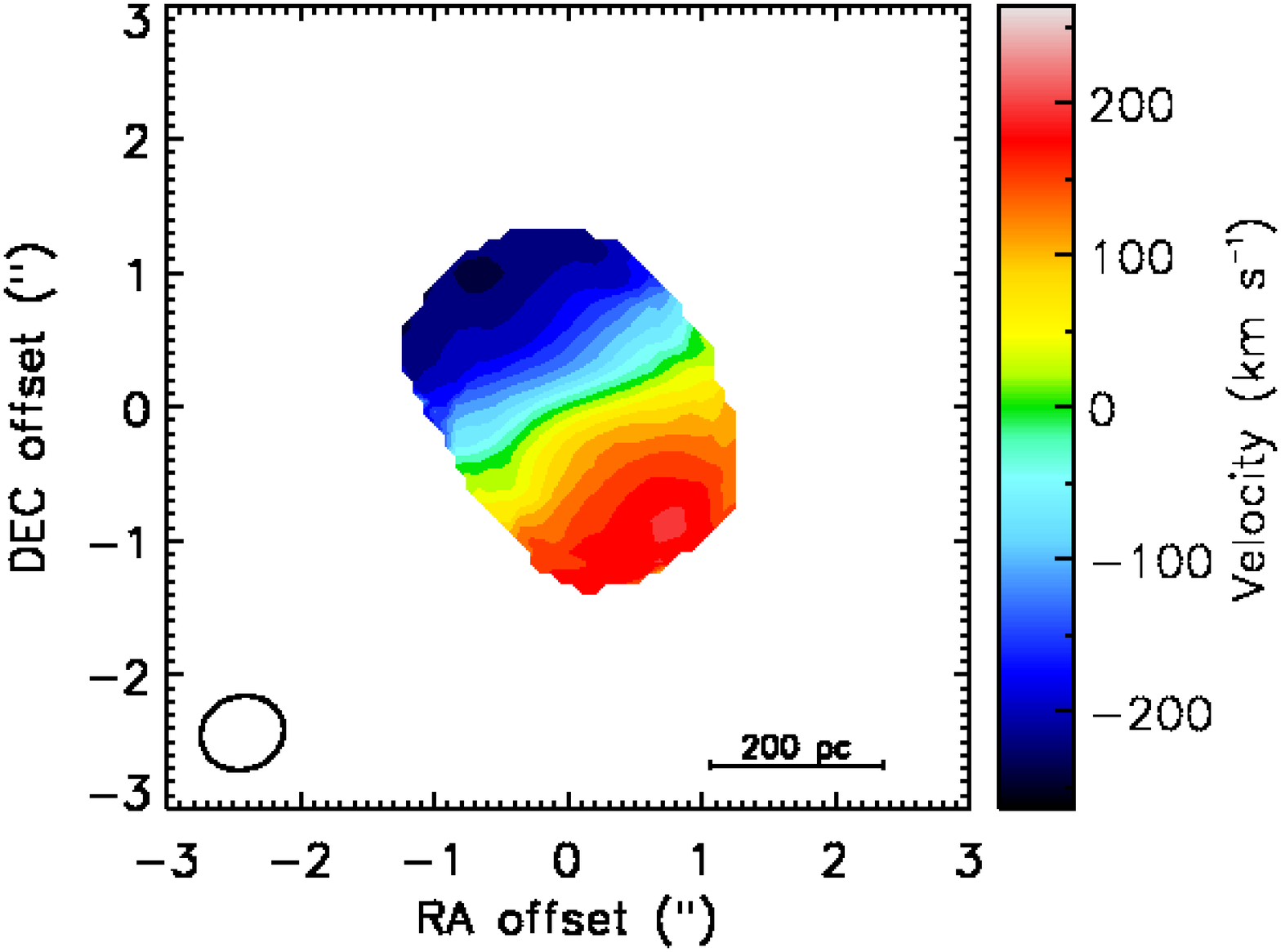}
\end{subfigure}

\medskip

\begin{subfigure}[t]{0.3\textheight}
\centering
\vspace{0pt}
\caption{}\label{fig:ngc3557_mom2}
\includegraphics[scale=0.3]{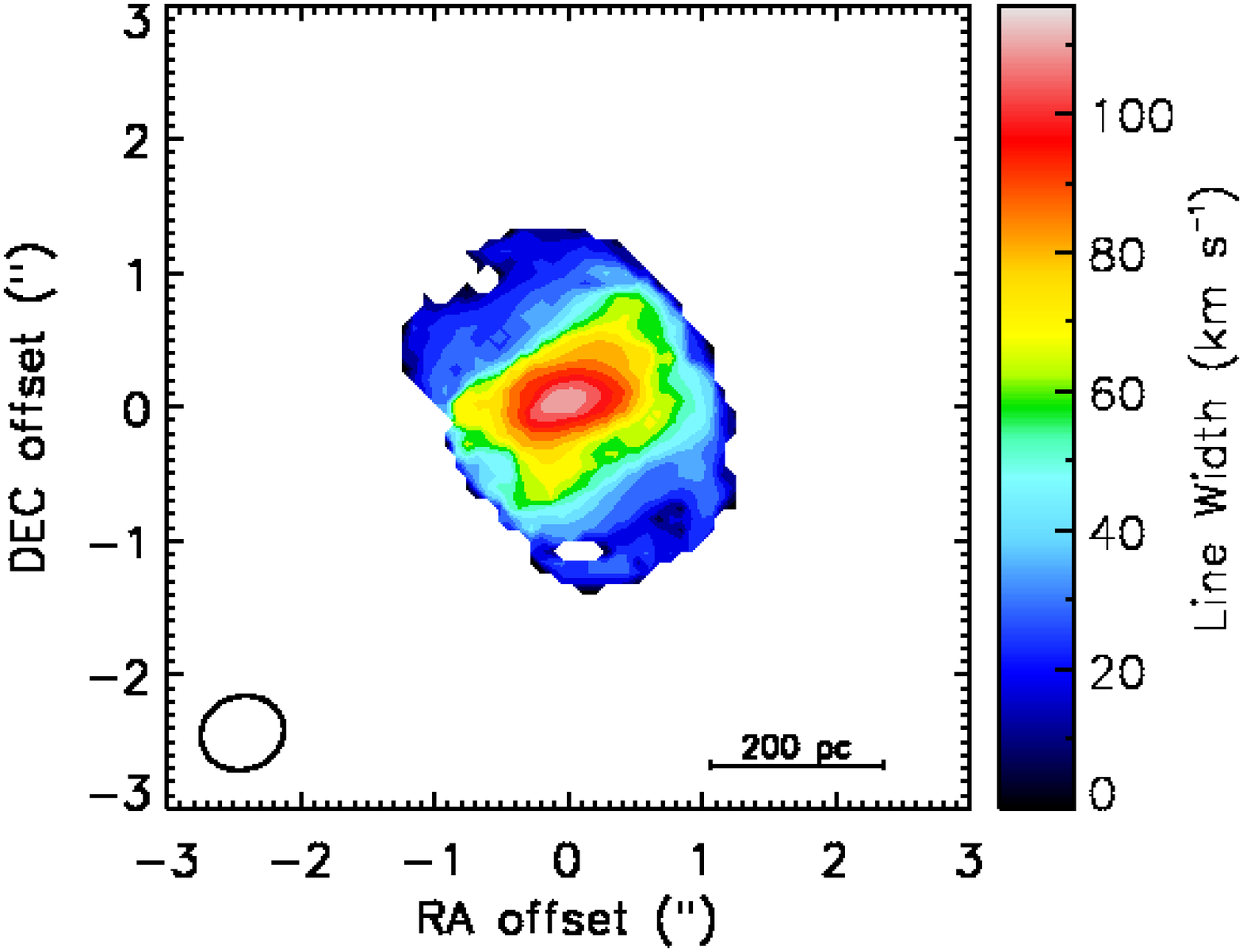}
\end{subfigure}
\hspace{6.5mm}
\begin{subfigure}[t]{0.3\textheight}
\centering
\vspace{0pt}
\caption{}\label{fig:ngc3557_spectrum}
\includegraphics[scale=0.3]{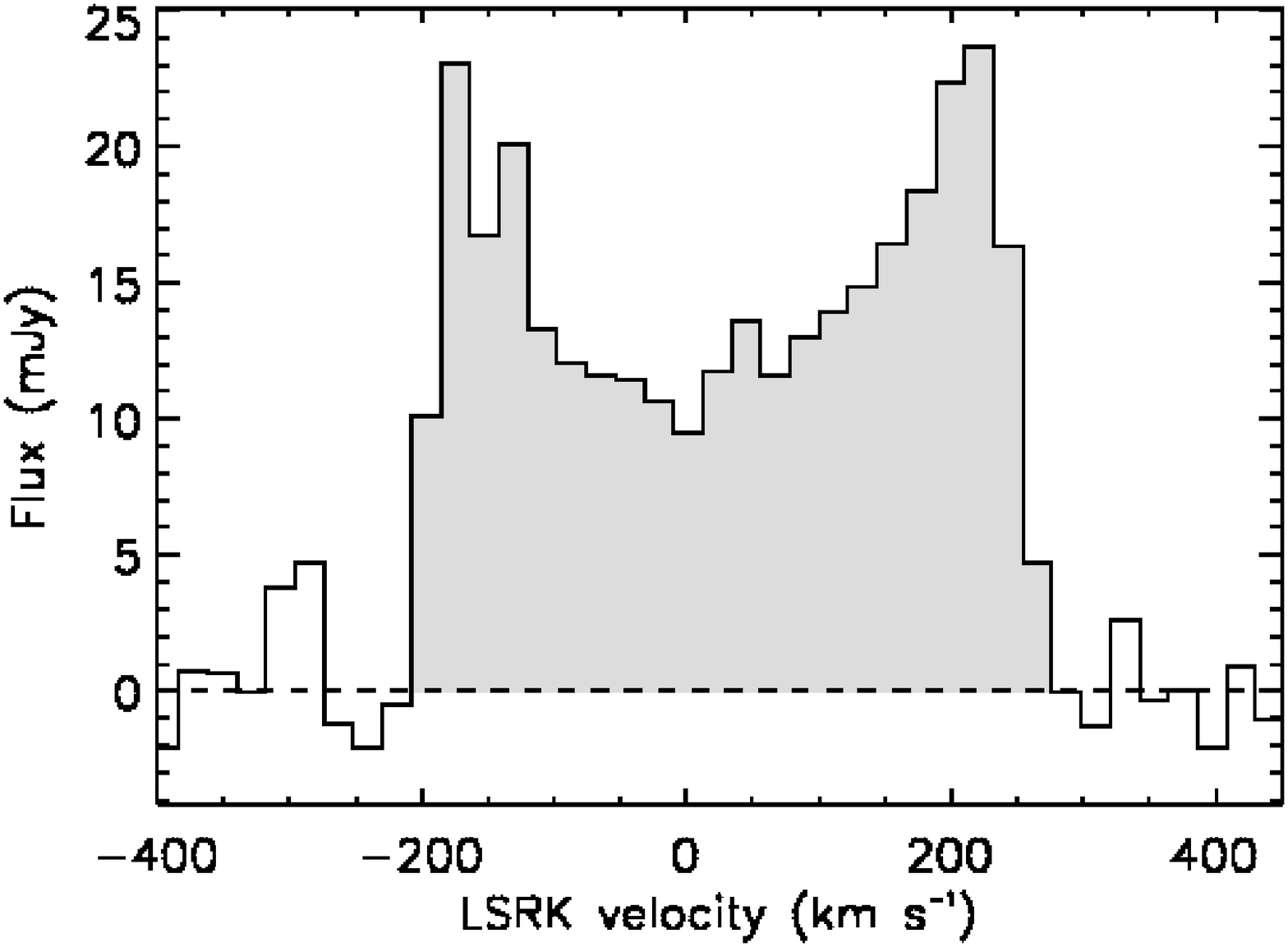}
\end{subfigure}
\caption{NGC\,3557 moment maps and spectral profile as in Fig.~\ref{IC1531}, created using a data cube with a channel width of 22~km~s$^{-1}$. The integrated CO spectral profile was extracted within a 3.0 $\times$ 3.0\,arcsec$^2$ box. 
}\label{fig:NGC3557}
\end{figure*}

\begin{figure*}
\centering
\begin{subfigure}[t]{0.3\textheight}
\centering
 \caption{}\label{fig:ic4296_mom0}
\includegraphics[scale=0.3]{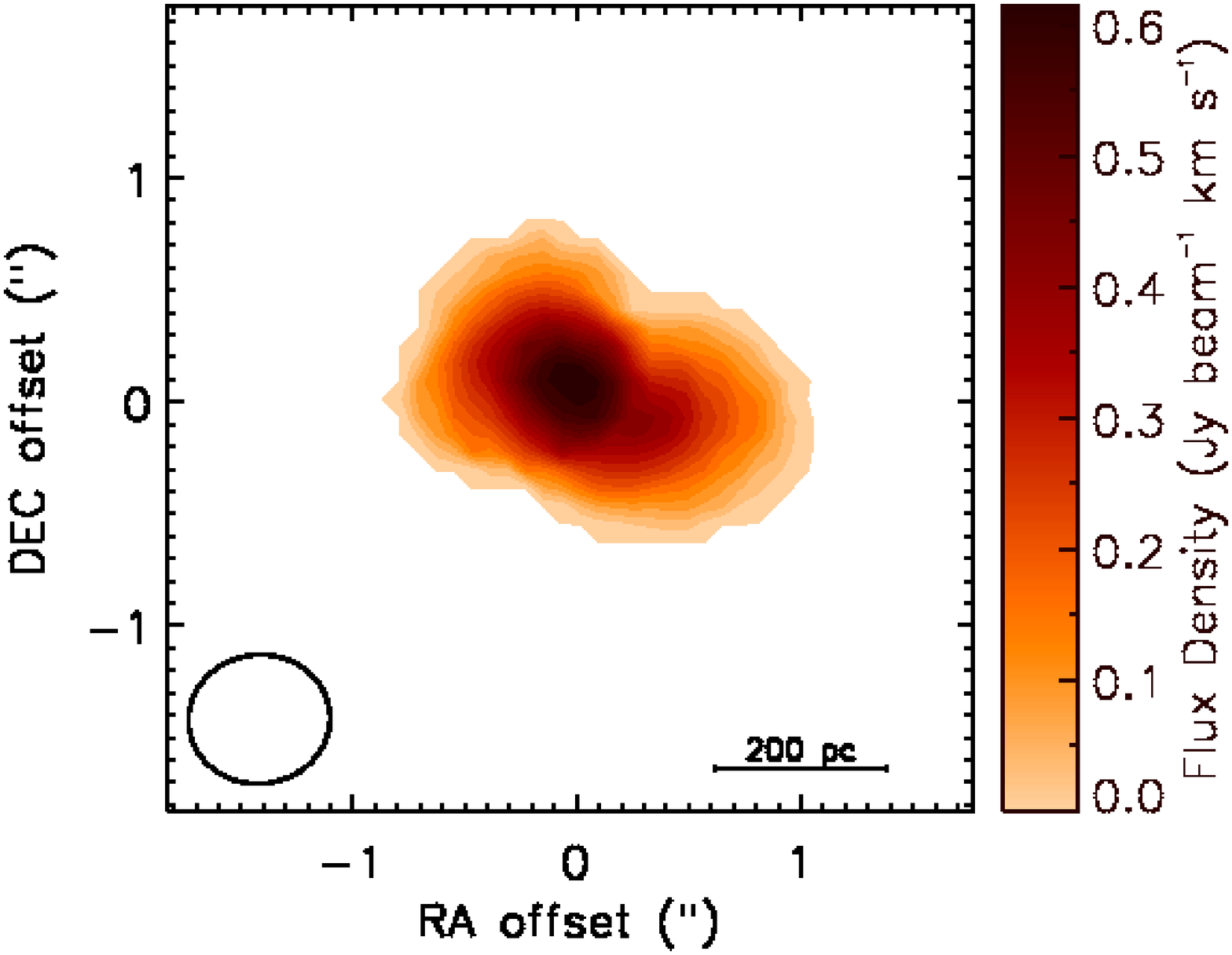}        
\end{subfigure}
\hspace{6.5mm}
\begin{subfigure}[t]{0.3\textheight}
\centering
\caption{}\label{fig:ic4296_mom1}
\includegraphics[scale=0.3]{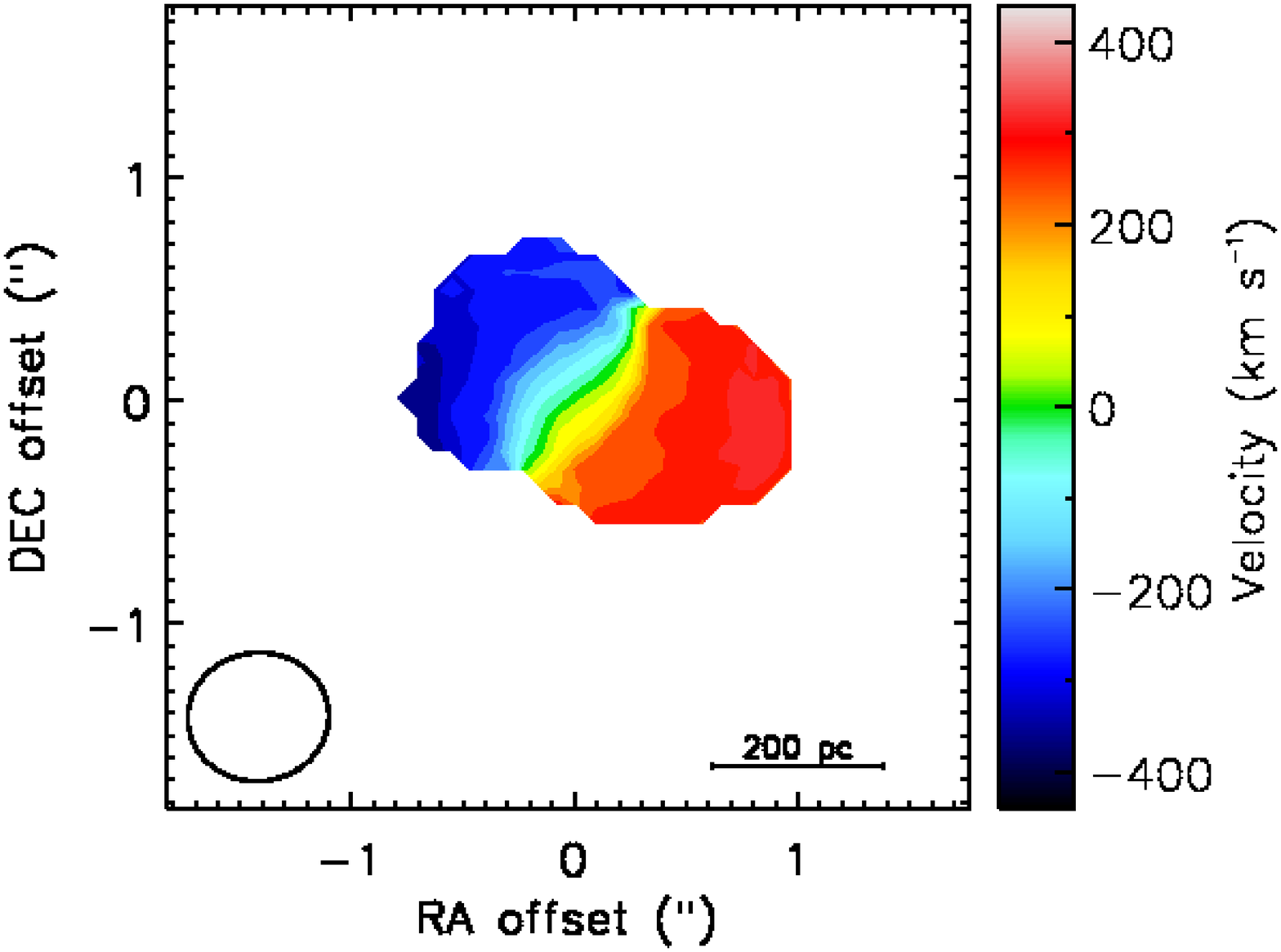}
\end{subfigure}

\medskip

\begin{subfigure}[t]{0.3\textheight}
\centering
\vspace{0pt}
\caption{}\label{fig:ic4296_mom2}
\includegraphics[scale=0.3]{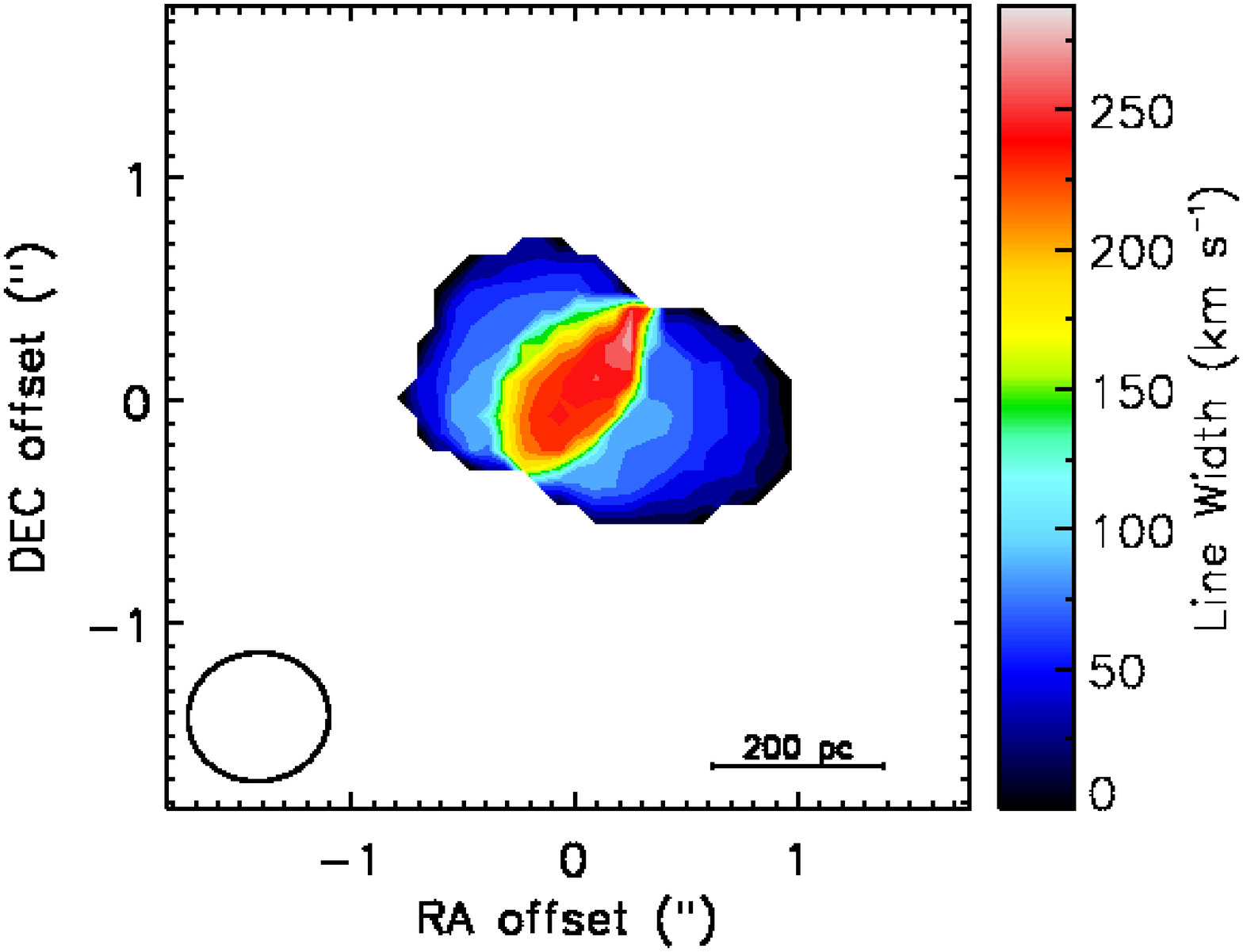}
\end{subfigure}
\hspace{6.5mm}
\begin{subfigure}[t]{0.3\textheight}
\centering
\vspace{0pt}
\caption{}\label{fig:ic4296_spectrum}
\includegraphics[scale=0.3]{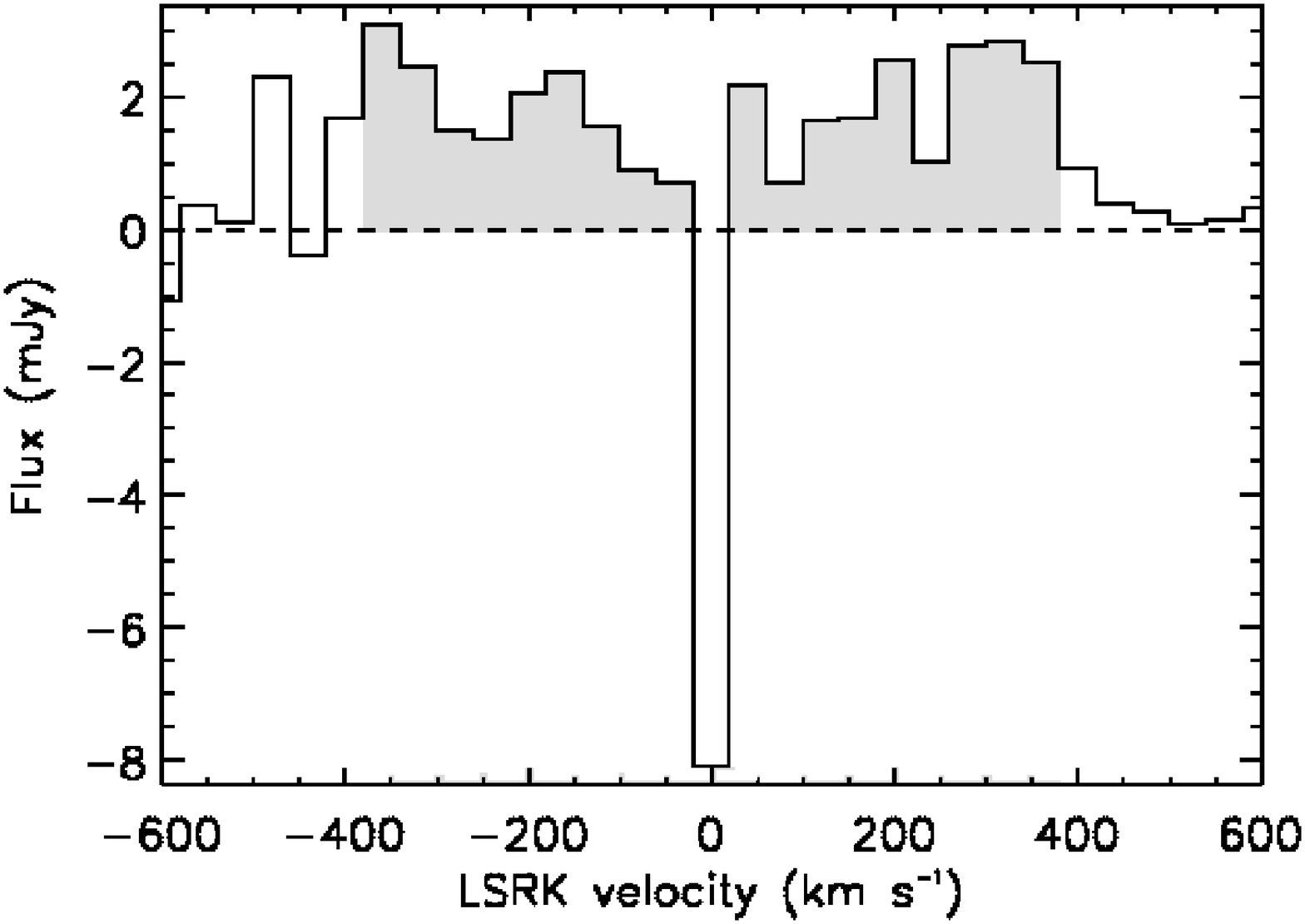}
\end{subfigure}
\caption{IC\,4296 moment maps and spectral profile as in Fig.~\ref{IC1531}, created using a data cube with a channel width of 40~km~s$^{-1}$. The integrated CO spectral profile was extracted within a $2 \times 1.5$\,arcsec$^2$ box.}\label{fig:IC4296}
\end{figure*}

\begin{figure*}
\centering
\begin{subfigure}[t]{0.3\textheight}
\centering
\caption{}\label{fig:ngc7075_mom0}
\includegraphics[scale=0.3]{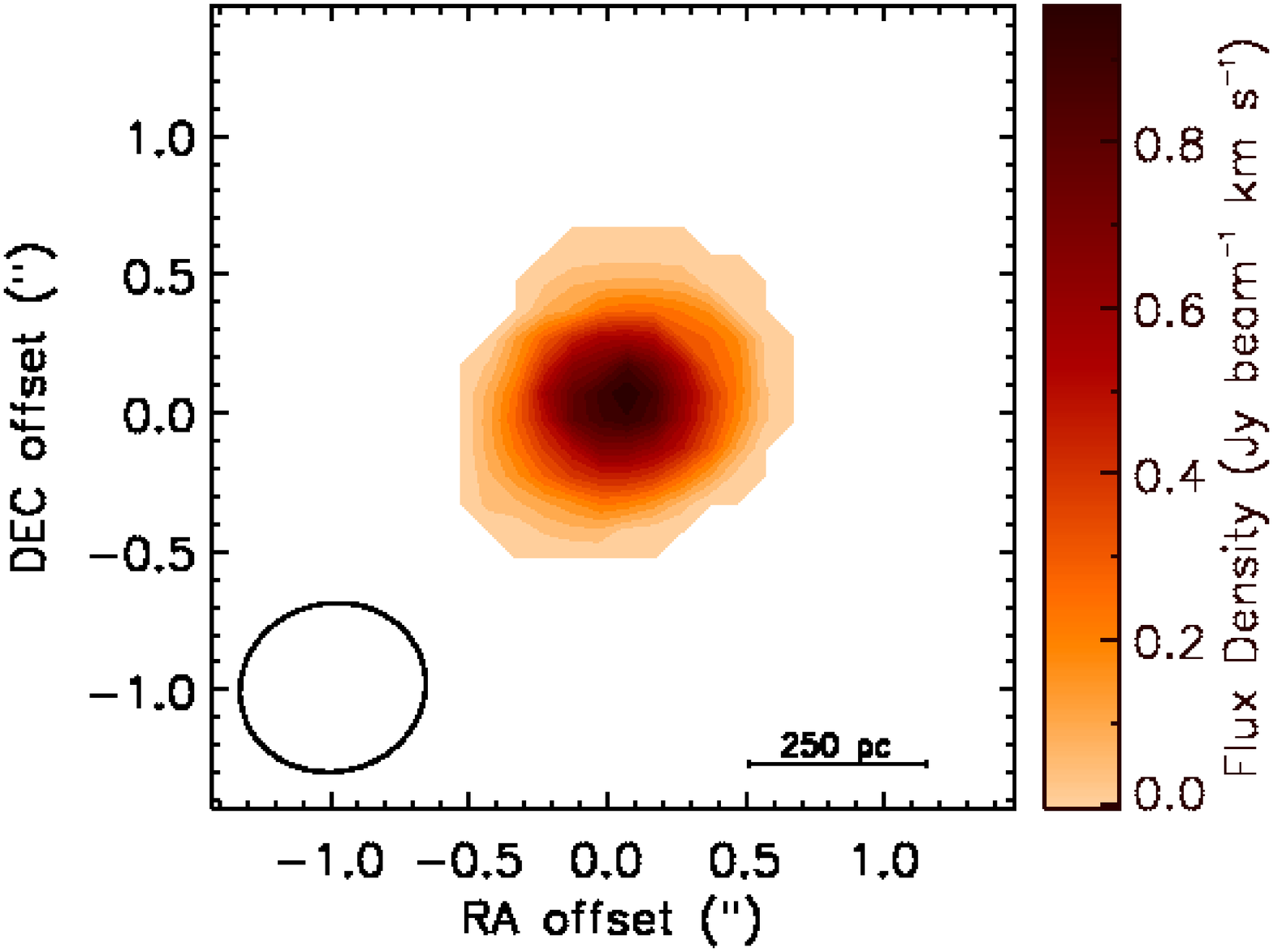}
\end{subfigure}
\hspace{8mm}
\begin{subfigure}[t]{0.3\textheight}
\centering
\caption{}\label{fig:ngc7075_mom1}
\includegraphics[scale=0.3]{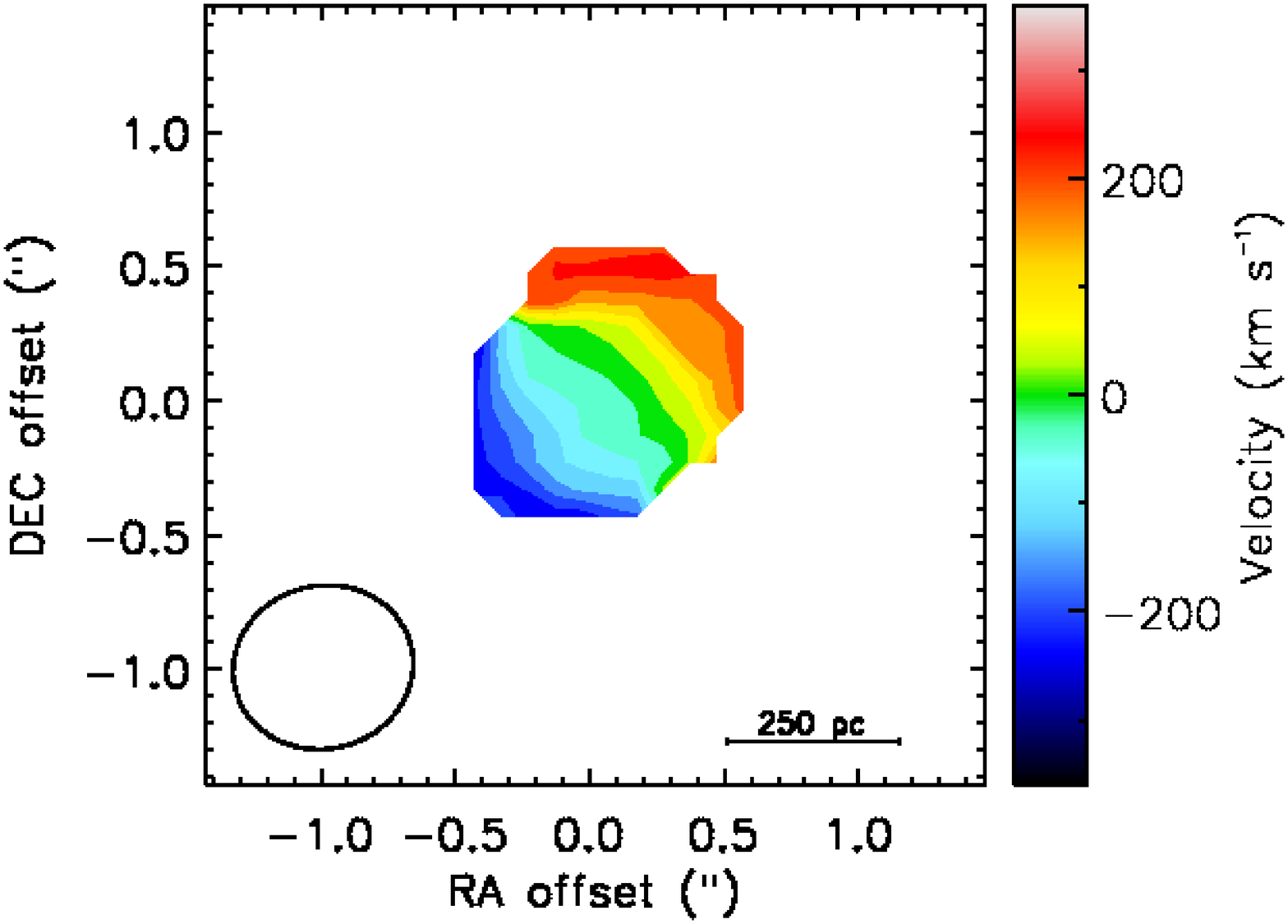}
\end{subfigure}

\medskip

\begin{subfigure}[t]{0.3\textheight}
\centering
\caption{}\label{fig:ngc7075_mom2}
\includegraphics[scale=0.3]{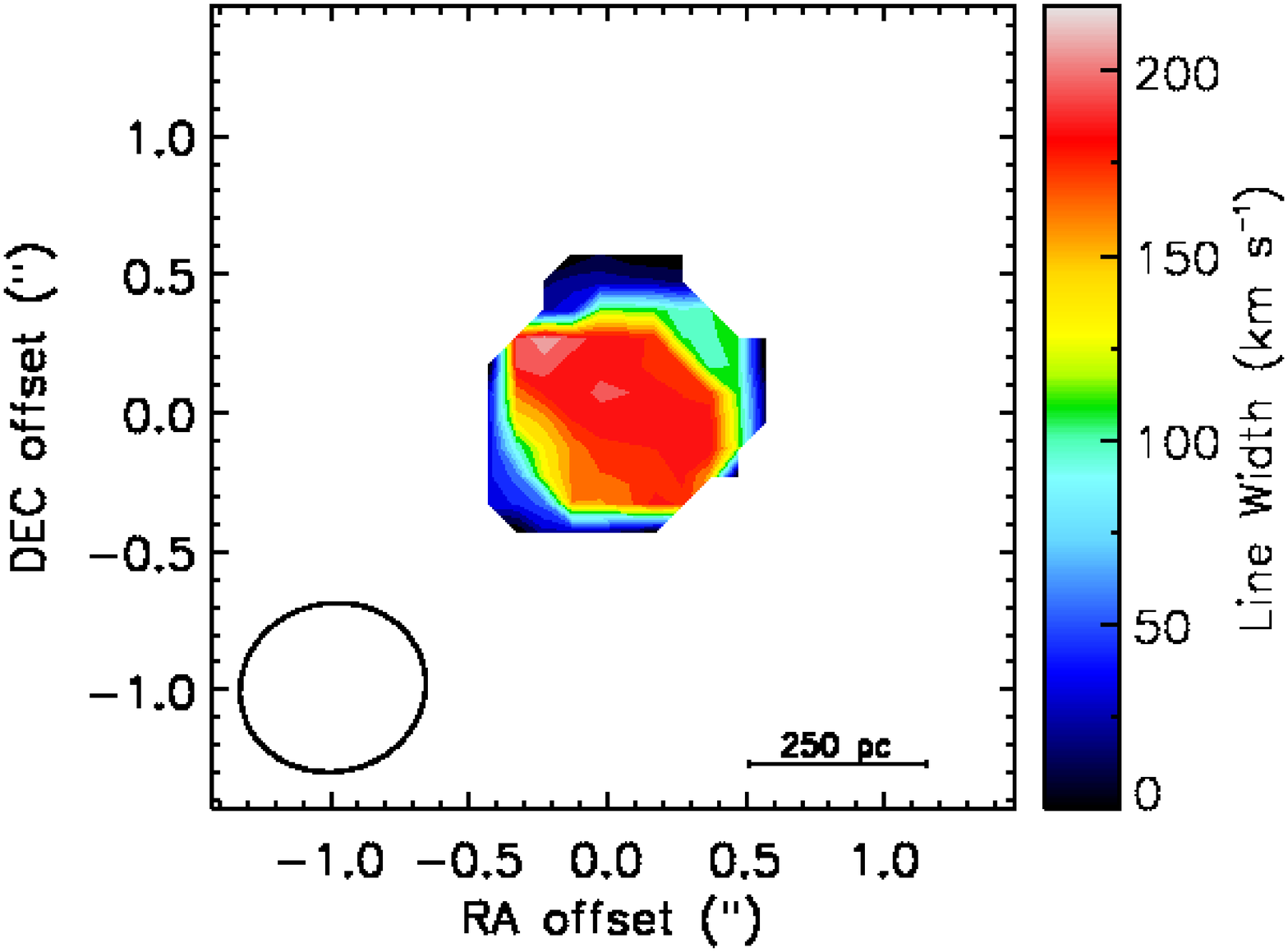}
\end{subfigure}
\hspace{8mm}
\begin{subfigure}[t]{0.3\textheight}
\centering
\caption{}\label{fig:ngc7075_spectrum}
\includegraphics[scale=0.3]{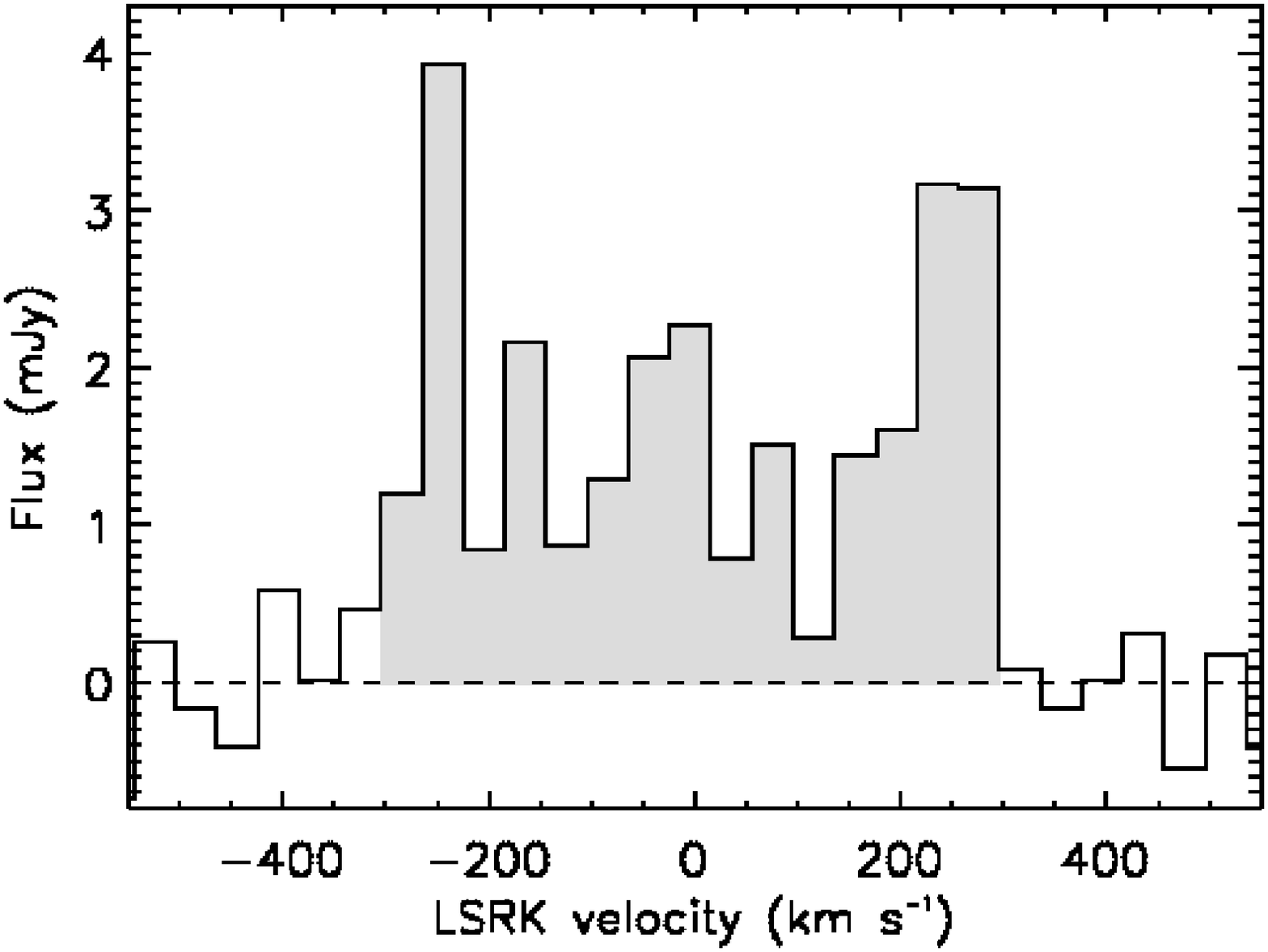}
\end{subfigure}
\caption{NGC\,7075 moment maps and spectral profile as in Fig.~\ref{IC1531}, created using a data cube with a channel width of 40~km~s$^{-1}$. The integrated CO spectral profile was extracted within a box of $1 \times 1$\,arcsec$^2$.}\label{NGC7075}
\end{figure*}

\section{Results: individual sources}\label{sec:results}

\subsection*{IC\,1531 (PKS 0007-325)}
IC\,1531 is a barred lenticular galaxy (SB0) in a low-density environment \citep{Osulli07}, with a FR\,I radio structure \citep{vanVelzen12}.

We detect a bright nuclear source in the continuum (Fig.~\ref{fig:ic1531_cont}, top panel). Emission from the South-East jet is also detected. The archival VLA image of IC\,1531 at 8.4~GHz (Fig.~\ref{fig:ic1531_cont}, bottom panel) shows similar core-jet structure. No counter-jet to the North-West is detected at either frequency.

A disc of molecular gas is detected at 18$\sigma$ significance (Fig.~\ref{fig:ic1531_mom0}). The disc is barely resolved in our observation, with a deconvolved major axis FWHM of 250~pc (Table~\ref{tab:line parameters}). The estimated molecular gas mass is 1.1$\times10^{8}$~M$_{\odot}$. The mean velocity map in Figure~\ref{fig:ic1531_mom1} shows a rotation pattern with an (s-shaped) distortion in the zero-velocity contour (i.e. the kinematic centre), possibly suggesting the presence of a warp in the molecular gas disc. This needs to be confirmed with higher resolution observations. The integrated spectral profile in Figure~\ref{fig:ic1531_spectrum} exhibits  the double-horned shape of a rotating disc. The line width (Fig.~\ref{fig:ic1531_mom2}) is likely to be dominated by beam smearing.

\subsection*{NGC\,612 (PKS 0131-36)}\label{sec:ngc612}
NGC\,612 is a peculiar lenticular galaxy viewed close to edge-on, characterised by an extensive and strongly warped dust distribution in the equatorial plane and by a massive stellar disc \citep{West66,Fasano96,Asabere16}. The presence of a stellar disc is exceptional for extended radio galaxies, at least in the luminosity range $10^{22} - 10^{25}$ \citep[e.g.][and references therein]{Ekers78,Veron01,Emonts08,Morganti11,Mao15}. NGC 612 is one of the few examples known to date. A 140~kpc-wide disc of atomic hydrogen (HI) with M\textsubscript{HI}$=1.8\times10^{9}$~M$_{\odot}$ was detected by \citet{Emonts08}. A faint outer bridge of HI connects NGC\,612 with the barred galaxy NGC\,619, roughly 400~kpc away. This suggests that a past interaction between the two galaxies (or a minor merger event) may have channelled large amounts of gas and dust into NGC\,612, also triggering the associated radio source. The large-scale radio map at 4.9~GHz published by \citet[][a re-imaged version is presented in Fig.~A1]{Morganti93} shows a weak core, a bright hot-spot at the outer edge of the more prominent eastern lobe, and a more diffuse western lobe with a total radio power $P$\textsubscript{1.4~GHz}$=1.5\times10^{25}$~W~Hz$^{-1}$.

We detect a marginally-resolved nuclear source in the continuum map of NGC\,612 (Fig.~\ref{fig:ngc612_cont}, upper panel). Continuum emission from the central region of the extended radio source is also visible in an archival VLA  4.9~GHz map (Fig.~\ref{fig:ngc612_cont}, bottom panel) but at much lower resolution.
We detect a large-scale disc of molecular gas with an extent of 9.6~kpc along the major axis, by far the largest in our sample. The molecular gas distribution appears clumpy (Fig.~\ref{fig:ngc612_mom0}). The estimated molecular gas mass is 2.0$\times10^{10}$~M$_{\odot}$, larger by about two orders of magnitude than that of any other sample member (see Table~\ref{tab:line parameters}). The mean velocity map (Fig.~\ref{fig:ngc612_mom1}) shows a regularly rotating disc with some asymmetries at its extreme edges, where the major axis of the velocity field changes orientation, suggesting the presence of a warp on large scales. A well-defined double-horned shape is visible in the integrated spectrum  (Figure~\ref{fig:ngc612_spectrum}), with some asymmetries reflecting those in the gas distribution, the higher peak in the integrated spectrum at positive velocities being associated with the larger extent of the disc to the South. The moment 2 map (Fig.~\ref{fig:ngc612_mom2}) shows the ``x-shaped'' morphology characteristic of a rotating disc \citep[e.g.][]{Davis17}. The velocity dispersion varies from $\sim$10 to $\sim$100~km~s$^{-1}$, but the regions characterised by line widths $>$40~km~s$^{-1}$ are highly localised, while the bulk of the disc has a dispersion $<$30~km~s$^{-1}$. 


\subsection*{PKS 0718-34}
PKS~0718-34 is a radio source hosted by an elliptical galaxy in a poor environment \citep{Govoni00}. The 4.9~GHz VLA map published by \citet{Ekers89} shows a poorly resolved, double-sided source. 

A barely resolved nuclear source is detected in the 230~GHz continuum (Fig.~\ref{fig:pks0718_cont}, upper panel). Faint, double-sided emission from the jets is also detected, extending up to 3.2~kpc to the South-West and 1.8~kpc to the North-East of the nucleus. The 230~GHz continuum emission traces the radio structure observed at lower resolution in the archival 8.5~GHz VLA map (Fig.~\ref{fig:pks0718_cont}, lower panel).

This object is undetected in CO, with estimated M\textsubscript{mol}$<6.7\times10^{6}$~M$_{\odot}$ on the assumption of a point source, or  $\Sigma$\textsubscript{CO}$<2.7\times10^{2}$~M$_{\odot}$~pc$^{2}$ (Table~\ref{tab:line parameters}).

\subsection*{NGC\,3100 (PKS 0958-314)}
NGC\,3100 is classified as a late-type S0 galaxy: it is characterised by a patchy dust distribution, a bright nuclear component, a nuclear bulge, and weak asymmetric arm-like structures in the outer disc \citep{Sandage79,Lau06}. It is located in a poor group and forms a pair with NGC\,3095 \citep{Devac76}.

The 230~GHz continuum map of NGC\,3100 (Fig.~\ref{fig:ngc3100_cont}, upper panel) shows a bright nuclear source. Extended emission from a two-sided jet is also detected. The northern and southern jets extend more than 400~pc and $\approx300$~pc from the nucleus, respectively.  The continuum structures visible at 230~GHz and at 4.9~GHz match very well (Fig.~\ref{fig:ngc3100_cont}).

We detect well-resolved CO(2-1) emission with a FWHM of 1.6~kpc along the major axis. The moment 0 image (Fig.~\ref{fig:ngc3100_mom0}) shows an incomplete ring with a gap to the North-West of the nucleus and flux density enhancements to the South-West and North-East. The mean velocity map (Fig.~\ref{fig:ngc3100_mom1}) shows that the ring is rotating, but with some distortions in the rotation pattern. The iso-velocity contours are tilted and the major axis position angle clearly changes moving from the egde to the center of the ring, indicating the possible presence of a warp and/or non-circular motions. The velocity dispersion map (Fig.~\ref{fig:ngc3100_mom2}) shows CO line broadening on either side of the central hole, roughly consistent in position with the flux density enhancements. As discussed in Section~\ref{sec:jets_gas}, these features are suggestive of a physical interaction between the jets and the molecular gas disc. The integrated CO spectrum (Fig.~\ref{fig:ngc3100_spectrum}) exhibits the double-horned shape typical of a rotating disc, but with asymmetries reflecting those in the gas distribution.

Fig.~\ref{fig:ngc3100_mom0} also shows the presence of two structures at $\approx$1.3~kpc west and $\approx$2.3~kpc east from the outer edges of the central ring, detected at 7.5$\sigma$ and 14$\sigma$, respectively. Including these two structures, we estimate a total molecular gas mass of M\textsubscript{mol}$=1.2\times10^{8}$~M$_{\odot}$. The velocity field in Fig.~\ref{fig:ngc3100_mom1} shows that the western and eastern regions have redshifted and blueshifted velocities, respectively, consistent with the nearest edges of the central ring but with different position angles.
This leads us to speculate that they may trace the presence of a larger, warped molecular gas disc whose outer emission is below the detection threshold of our observations. Alternatively we may be seeing molecular clumps in a disc of atomic gas.

\subsection*{NGC\,3557 (PKS 1107-372)}
NGC\,3557 is a regular elliptical galaxy in a group \citep{Govoni00}. Its optical properties have been studied extensively \citep{Colbert01,Lauer05,Capetti05,Balmaverde06}. In particular, high-resolution HST observations clearly show the presence of a prominent dust ring in the central regions \citep[e.g.][see also the right panel of Fig.~\ref{fig:ngc3557_ic4296_optical}]{Lauer05}.

NGC\,3557 is the host galaxy of the double-sided radio source PKS~1107-372, which was first imaged by \citet{Birki85}.

The continuum map of NGC\,3557 (Fig.~\ref{fig:ngc3557_cont}, upper panel) shows emission from the core and a two-sided jet. The eastern and western jets extend $\approx$900~pc and $\approx$1.2~kpc, respectively, from the nucleus.
The continuum emission detected at 230~GHz traces the inner part of the radio structure visible on larger scales in the archival 4.9~GHz VLA map (Fig.~\ref{fig:ngc3557_cont}, lower panel).

We detect a CO disc (Fig.~\ref{fig:ngc3557_mom0}) with M\textsubscript{mol}$=6.2\times10^{7}$~M$_{\odot}$. The disc is barely resolved in our observations, which have a beamwidth of  0.6\,arcsec FWHM. The deconvolved major axis of the disc is $\approx300$~pc FWHM.
The mean velocity map (Fig.~\ref{fig:ngc3557_mom1}) shows that the gas is rotating regularly. This is also consistent with the symmetric double-horned shape of the integrated spectral profile (Fig.~\ref{fig:ngc3557_spectrum}). The velocity dispersion (Fig.~\ref{fig:ngc3557_mom2}) decreases from $\approx$100~km~s$^{-1}$ in the centre to $\approx$20~km~s$^{-1}$ at the edges of the disc and shows a boxy profile; both features are likely to be due to beam smearing. 
\subsection*{ESO 443-G 024 (PKS 1258-321)}
ESO~443-G~024 is an elliptical galaxy in the cluster Abell 3537 \citep{Govoni00}. It is the host galaxy of the FR\,I radio source PKS~1258-321, characterised by a double-sided radio morphology (Fig.~A2, upper panel).

The source is detected in the continuum, showing a bright nuclear component and also faint extended emission from a two-sided jet (Fig.~\ref{fig:eso443_cont}, top panel). The South-East and North-West jets extend $\approx$900~pc and $\approx$1.6~kpc, respectively, from the nucleus.
The extended emission detected at 230~GHz matches that observed at lower sensitivity in the archival 15~GHz VLA map (Fig.~\ref{fig:eso443_cont}, bottom panel).

ESO~443-G~024 is undetected in CO with estimated M\textsubscript{mol}$<3.5\times10^{6}$~M$_{\odot}$ assuming a point source, or $\Sigma$\textsubscript{CO}$<3.9\times10^{2}$~M$_{\odot}$~pc$^{2}$ (Table~\ref{tab:line parameters}).
\subsection*{IC\,4296 (PKS 1333-33)}
IC\,4296 is an elliptical galaxy and is the brightest member of the group HCG 22 \citep{Huchra82}. HST observations reveal a prominent nuclear dust disc \citep{Lauer05}. 

IC\,4296 hosts the FR\,I radio source PKS~1333-33. The core is marginally resolved on scales of few parsecs \citep{Venturi00}. A large-scale, symmetric, double-sided jet extends up to 5~arcmin (77~kpc) from the nucleus, connecting with the outer lobes at 30~arcmin \citep{Killeen86,Burke09}. The inner jet structure of IC\,4296 is discussed further in Appendix~A (Fig.~A3).

A bright continuum nuclear source is detected at 230~GHz (Fig.~\ref{fig:ic4296_cont}, upper panel), with a major axis FWHM of 30~pc. The inner jets are undetected at 230~GHz, but faint (7$\sigma$) emission is detected at a distance of about 950~pc North-West of the nucleus, coincident with the brighter knot of the North-West jet visible in the archival 4.9~GHz VLA map (Fig.~\ref{fig:ic4296_cont}, bottom panel).

We detect CO(2-1) emission with an estimated molecular gas mass of 2.0$\times10^{7}$~M$_{\odot}$.
The CO integrated intensity map (Fig.~\ref{fig:ic4296_mom0}) shows a disc with a somewhat asymmetric morphology. The velocity field (Fig.~\ref{fig:ic4296_mom1}) shows an s-shaped 
zero-velocity contour, suggesting the presence of a warp in the disc, although better resolution observations are necessary to confirm this hypothesis. The moment 2 map (Fig.~\ref{fig:ic4296_mom2}) is likely to be dominated by beam smearing. The integrated spectral profile (Fig.~\ref{fig:ic4296_spectrum}) reveals the presence of a strong absorption feature that is discussed in more detail in Section~\ref{sec:absorption_analysis}.

\citet{Boizelle17} presented Cycle 2 ALMA observations of IC\,4296 in the same CO transition. They reported a 5$\sigma$ CO(2-1) detection with an integrated flux of 0.76 Jy km s$^{-1}$, about a factor of two lower than that measured in this work. However, their  CO(2-1) observation is a factor of two noisier than that presented in this paper (for the same channel width) and this may have caused them to miss some of the emission. We also note that they used different R$_{21}$ and X\textsubscript{CO} values. Their mass estimate is therefore a factor of three lower than ours. The integrated spectral profile of IC\,4296 presented in \citet{Boizelle17} is qualitatively similar to ours, with a line width of $\approx\pm480$~km~s$^{-1}$.

\subsection*{NGC\,7075 (PKS 2128-388)}
NGC\,7075 is an elliptical galaxy with an optically unresolved component in the core \citep{Govoni00b}. It is the host galaxy of the FR\,I radio source PKS~2128-388: a low-resolution 4.9~GHz VLA map of NGC\,7075 showing its large-scale radio structure is presented in Appendix A (Fig.~A2, lower panel).

A nuclear source is detected in our 230~GHz continuum map (Fig.~\ref{fig:ngc7075_cont}, upper panel), with a major axis FWHM of 40~pc. Faint emission from the eastern jet is also detected, extending to $\approx$1.9~kpc from the nucleus; the western (counter-) jet is undetected. The eastern jet is also visible in the archival 8.5~GHz VLA map (Fig.~\ref{fig:ngc7075_cont}, lower panel).

We detect a barely resolved CO(2-1) disc (Fig.~\ref{fig:ngc7075_mom0}), with an estimated molecular gas mass of 2.9$\times10^{7}$~M$_{\odot}$. The mean velocity map (Fig.~\ref{fig:ngc7075_mom1}) shows regular gas rotation. The line width (Fig.~\ref{fig:ngc7075_mom2}) is likely to be dominated by beam-smeared rotation.
\subsection*{IC\,1459 (PKS 2254-367)}
IC\,1459 is an elliptical galaxy in a poor group of spiral galaxies. 
A faint dust lane is detected in the HST image of IC\,1459, mostly visible towards the galaxy outskirts; its morphology is interpreted as a transitional stage between a patchy chaotic structure and a nuclear ring \citep{Sparks85,Lauer05}. IC\,1459 also contains one of the most prominent counter-rotating cores observed in an elliptical galaxy \citep{Verdoes00,Cappellari02,Ricci15b}, probably the result of a major merger between gas-rich spiral galaxies \citep[e.g.][]{Hernquist91}.

IC\,1459 is the host galaxy of the radio source PKS~2254$-$367, which was classified by \citet{Tingay03} as a compact (arcsec-scale), GHz-peaked radio source (GPS) in a high-density environment.

We detect continuum emission from a bright nuclear source (Fig.~\ref{fig:ic1459_cont}, upper panel) with an estimated major axis FWHM of 10~pc. This marginally resolved continuum emission matches that visible in the archival 8.5~GHz VLA map (Fig.~\ref{fig:ic1459_cont}, lower panel). The angular resolution (0.9$''$) of our observations does not allow us to resolve the pc-scale double-sided radio source detected with Very-Long Baseline Interferometry by \citet{Tingay15}.

Surprisingly, IC\,1459 is undetected in CO with ALMA, with estimated M$_{\rm mol}<1.0\times10^{6}$~M$_{\odot}$ assuming a point source, or $\Sigma$\textsubscript{CO}$<7.6\times10^{2}$~M$_{\odot}$~pc$^{2}$ (Table~\ref{tab:line parameters}). 


\section{Discussion}
\label{sec:discussion}

\begin{figure*}
\centering
\includegraphics[scale=0.35]{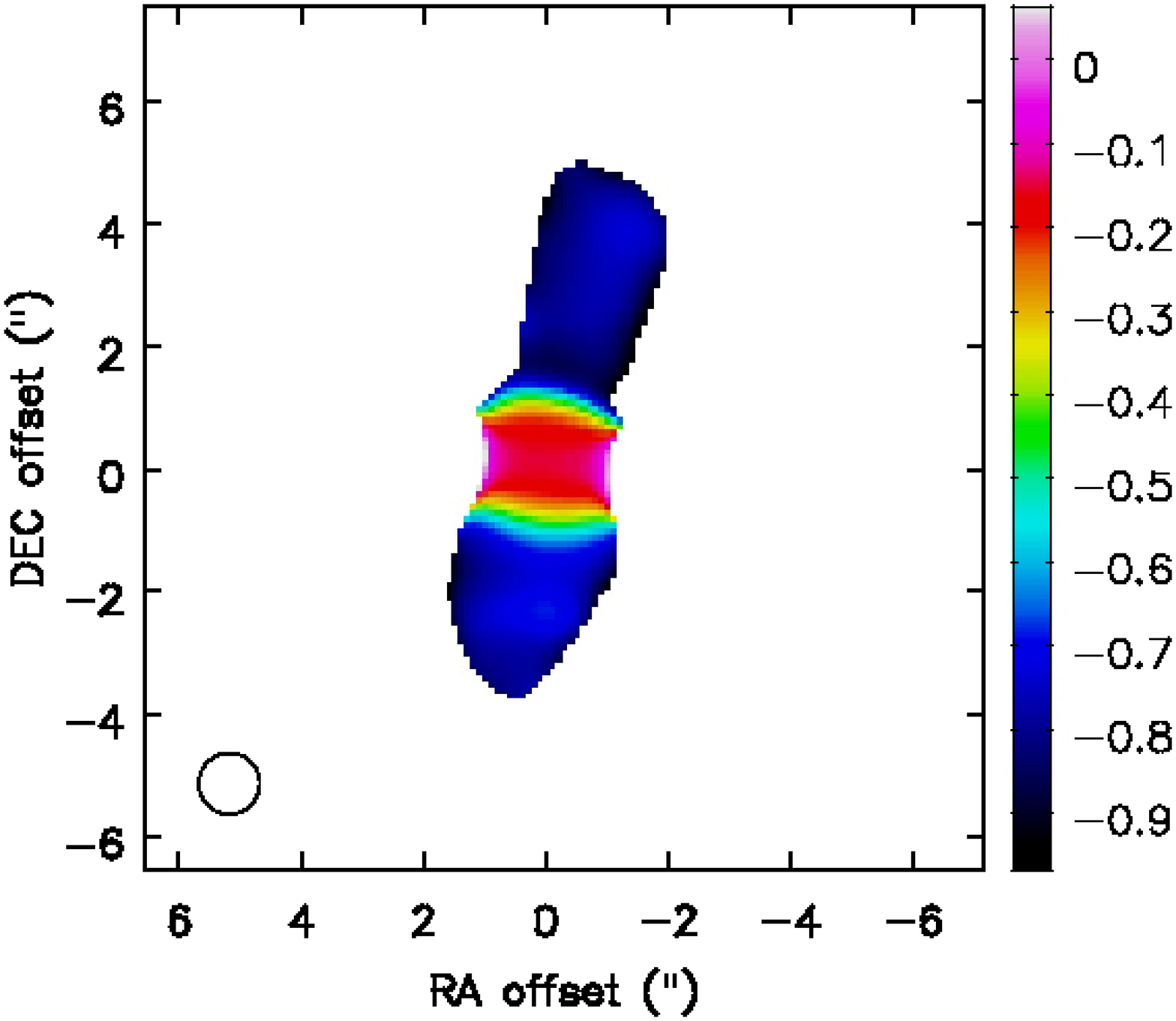}
\quad
\includegraphics[scale=0.35]{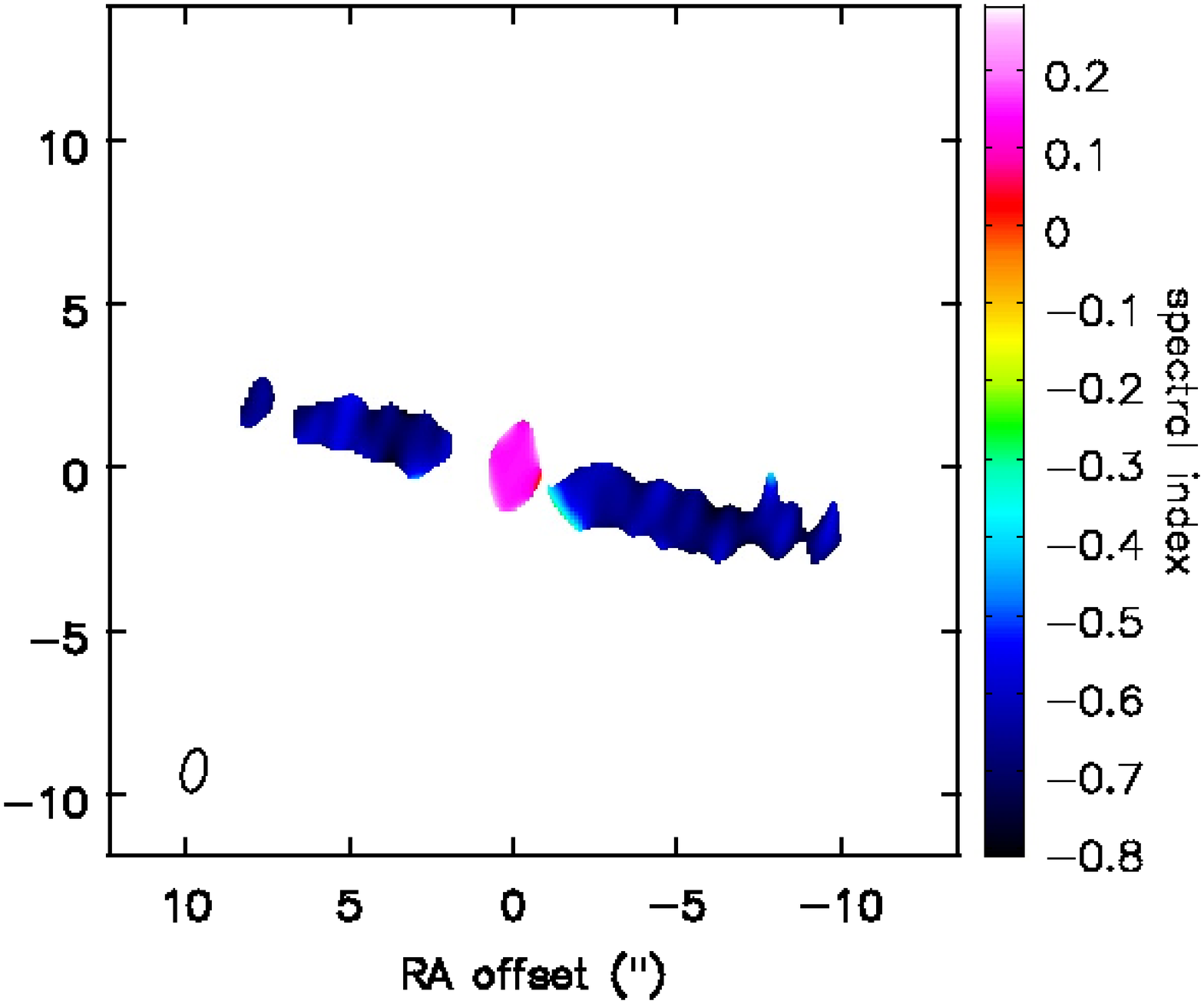}
\caption{\small{Spectral index maps of NGC\,3100 (left panel) and NGC\,3557 (right panel), obtained from our ALMA 230~GHz and archival VLA 4.9~GHz maps. The wedge on the right of each map indicates the colour scale. Coordinates are reported as relative position with respect to the image phase centre, in arcsec; East is to the left and North to the top.}}\label{fig:spix}
\end{figure*}

\subsection{Origin of the 230 GHz continuum emission}
\label{sec:continuum_analysis}

The continuum emission at 230 GHz is morphologically very similar to the emission observed at frequencies between 4.9 and 15\,GHz in the archival VLA images (Fig.~\ref{fig:continuum}). It is likely to be dominated by radio synchrotron emission from the core and jet structures. There is no evidence of thermal emission associated with extended dust or CO, although we cannot rule out the possibility of some contribution to the unresolved core emission from this mechanism.

For two sources, NGC\,3100 and NGC\,3557, the available VLA archive data at 4.9-GHz (Fig.~\ref{fig:ngc3100_cont} and \ref{fig:ngc3557_cont}, bottom panels) are sufficiently well matched to the ALMA data in resolution (see Tables~\ref{tab:ALMA observations summary} and A1) and uv coverage to enable us to derive spectral-index maps. The ALMA and VLA input maps were first re-imaged using the {\sc casa} \textsc{clean} task in multi-frequency synthesis (MFS) mode (nterms$=1$) with the same uv ranges and natural weighting. The maps were restored with the same synthesized beam and spectral index maps were then produced using the {\sc casa} task \textsc{immath}. The results are shown in Fig.~\ref{fig:spix}. As expected, the radio cores show flat spectra (-0.2$<\alpha<$0.2, for $S \propto \nu^{\alpha}$), while the jet spectra are steeper ($\alpha \approx -0.7$). This is consistent with synchrotron emission,  partially optically thick (self-absorbed) in the core and optically thin in the jets. The jet spectral indices are slightly steeper than those typically seen between 1.4 and 4.9\,GHz in FR\,I jets ($\alpha \approx -0.6$; \citealt{LB13}), but there is no sign of abrupt high-frequency steepening due to synchrotron or inverse Compton losses. 

We also estimated the spectral indices of the radio cores of all the sources using the core flux densities determined from the ALMA 230~GHz and archival VLA continuum maps shown in Fig.~\ref{fig:continuum}. The VLA maps were chosen to match the ALMA images as closely as possible in resolution (Table~A1), but we note that for NGC\,612 the only available VLA map has much lower resolution. The estimated core spectral indices are listed in Table~\ref{tab:spixes}. All of the spectra are flat, as expected for partially optically thick synchrotron emission from the inner jets, except for NGC\,7075 and IC\,1459. The latter has a sub-arcsecond FR\,I radio structure which is unresolved in our observations \citep{Tingay15}, so we expect the emission to be dominated by an optically-thin component.

\begin{table}
\centering
\caption{Core spectral index.}
\label{tab:spixes}
\begin{tabular}{lrr}
\hline
\multicolumn{1}{c}{ Target} &
\multicolumn{1}{c}{ $\alpha$\textsubscript{core} }&
\multicolumn{1}{c}{$\nu_{\rm VLA}$}\\
\multicolumn{1}{c}{  } &
\multicolumn{1}{c}{  } &
\multicolumn{1}{c}{ (GHz) } \\
\multicolumn{1}{c}{ (1) } &
\multicolumn{1}{c}{ (2) } &
\multicolumn{1}{c}{ (3) } \\
\hline
 IC\,1531  & 0.17 & 8.5 \\
 NGC\,612   & 0.10 & 4.9   \\
  PKS 0718-34  & 0.12 & 8.5   \\
  NGC 3100 & -0.2 & 4.9 \\
  NGC 3557 & 0.2 & 4.9  \\
 ESO 443-G 024  &  0.25 & 14.9    \\
  IC\,4296  & 0.05 & 4.9 \\
 NGC\,7075 & $-$0.43 & 8.5 \\
 IC\,1459  & $-$0.40 & 8.5 \\ 
\hline
\end{tabular}
\parbox[t]{8.5cm}{ \textit{Notes.} $-$ Columns: (1) Target name. (2) Core spectral index between 230~GHz and $\nu_{\rm VLA}$. (3) Frequency of the VLA radio map.}
\end{table}

\subsection{CO absorption}\label{sec:absorption_analysis}

\subsubsection{Search for absorption}

We searched for absorption features in all the galaxies detected in CO by extracting integrated spectra in small boxes around the bright nuclear continuum sources. The integrated spectrum of IC\,4296 shows a deep and narrow absorption feature (already clear in Fig.~\ref{fig:ic4296_spectrum}). This is discussed in detail below (Section~\ref{sec:absorption_IC4296}). No significant absorption features were found in any of the other sources. In particular, we do not detect CO absorption against the nucleus of NGC\,612, where \citet{Morganti01} found absorption in HI.

\subsubsection{Absorption in IC\,4296}
\label{sec:absorption_IC4296}

\begin{figure}
\centering
\includegraphics[scale=0.3]{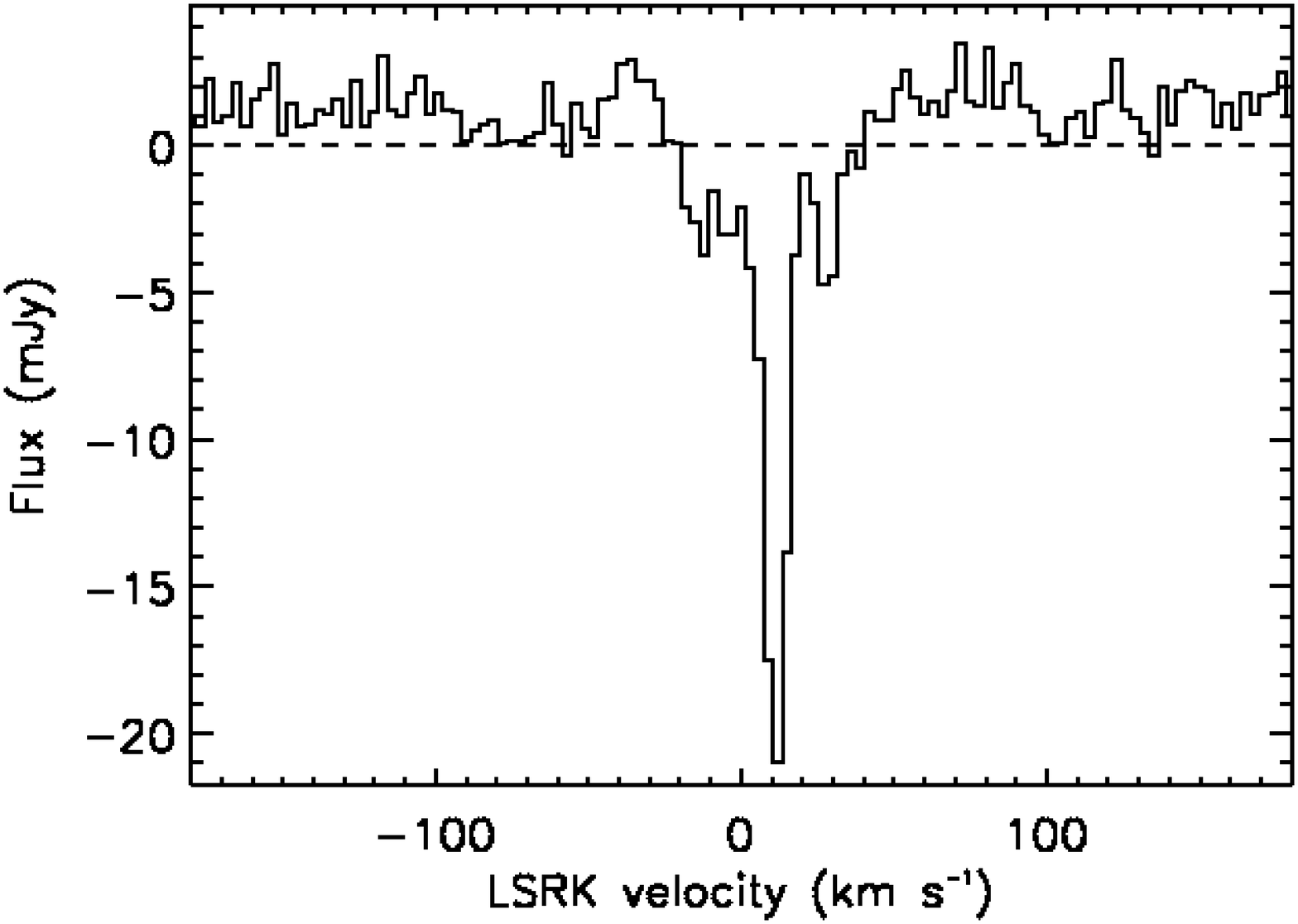}
\caption{\small{CO spectral profile of IC\,4296 extracted within a $0.4 \times 0.4$\,arcsec$^2$ box around the core. The spectrum has a channel width of 3\,\kms. The black dashed horizontal line indicates the zero flux level. Some unresolved CO emission is visible above the zero flux level. The line absorption feature has a FWHM of 9~km~s$^{-1}$, with an absorption peak of  $\approx-$20.2~mJy. }}\label{fig:absorption}
\end{figure}

In order to investigate  the absorption feature in IC\,4296 in more detail, we re-imaged the visibility data into a cube with a channel width of 3~km~s$^{-1}$ (i.e. approximately twice the raw channel width). We then extracted the CO spectrum from a $0.4 \times 0.4$\,arcsec$^2$ ($\approx 100 \times 100$\,pc) box, centred on the 230~GHz core continuum emission. The resulting spectrum is shown in Figure~\ref{fig:absorption}. The absorption peak is at $3720\pm 3$\,\kms, consistent within the errors with the most accurate determination of the optical systemic velocity ($3737\pm 10$\,km\,s$^{-1}$; see Table~\ref{tab:Southern Sample}). The maximum absorption depth measured from this spectrum is $-$20.2~mJy and the FWHM is $\approx$9~km~s$^{-1}$, resulting in an integrated absorption flux of $\approx$0.14~Jy~km~s$^{-1}$. \citet{Morganti01} presented a 3$\sigma$ upper limit of $\tau < 0.041$ for the optical depth of the HI absorption in IC\,4296. Following \citet{Morganti01}, we measured the peak optical depth, $\tau$, using the core flux density derived from the continuum image (Table~\ref{tab:Continuum images}): the resulting optical depth is $\tau \approx 0.12$. If we assume an HI spin temperature of 100\,K \citep[e.g.][]{Morganti01}, a CO excitation temperature of 10\,K \citep[e.g.][]{Heyer09}, and an HI to H\textsubscript{2} column density ratio of $\sim10$\textsuperscript{-2}, as estimated from HI and CO absorption in the radio emitting LINER PKS B1718$-$649  \citep{Maccagni14,Maccagni18}, we estimate an HI optical depth of $\tau \approx 1.2\times10$\textsuperscript{-4}, well below the  upper limit calculated by \citet{Morganti01}.

The spectral profile in Figure~\ref{fig:absorption} also exhibits fainter absorption features on either side of the peak (broader at blueshifted velocities). Similar structures are visible in the integrated spectrum of IC\,4296 presented by \citet{Boizelle17} with comparable channel width and extraction region. The observed asymmetric features may be a signature of the disc
warping inferred on larger scales from the velocity field of CO in emission (see Section~\ref{sec:results}). Alternatively, they may be interpreted as an indication of the presence of non-circular motions in the central 100~pc of the CO disc. Higher-resolution observations are needed to differentiate between these two scenarios.

\subsection{Jets and molecular gas}
\label{sec:jets_gas}
\begin{figure*}
\centering
\begin{subfigure}[t]{0.3\textheight}
\centering
\caption{\textbf{IC\,1531}}\label{fig:ic1531_mom1_cont}
\includegraphics[scale=0.38]{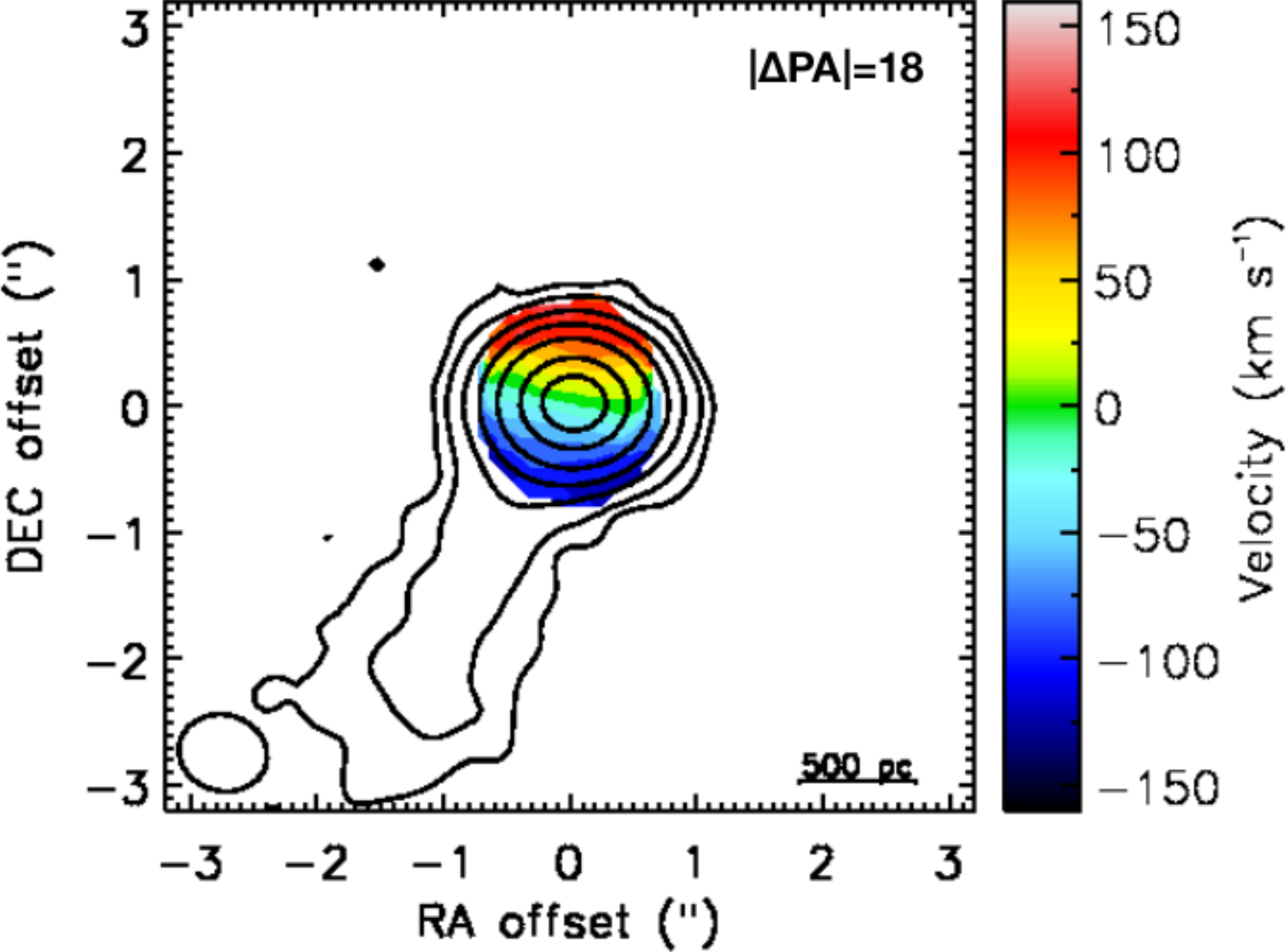}
\end{subfigure}
\begin{subfigure}[t]{0.3\textheight}
\centering
\caption{\textbf{NGC\,612}}\label{fig:ngc612_mom1_cont}
\includegraphics[scale=0.31]{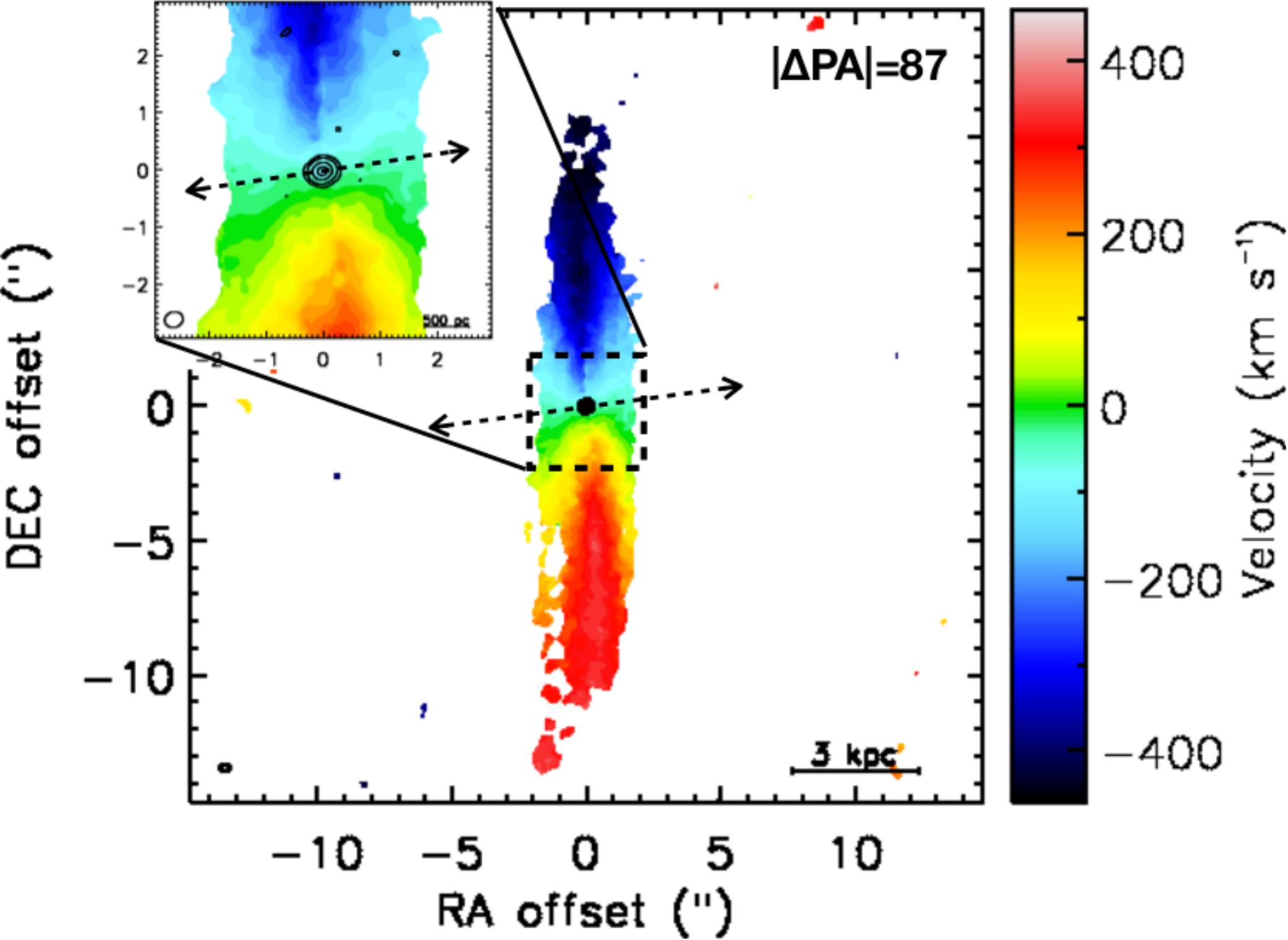}
\end{subfigure}
\medskip
\begin{subfigure}[t]{0.3\textheight}
\centering
\vspace{2mm}
\caption{\textbf{NGC\,3100}}\label{fig:ngc3100_mom1_cont}
\includegraphics[scale=0.38]{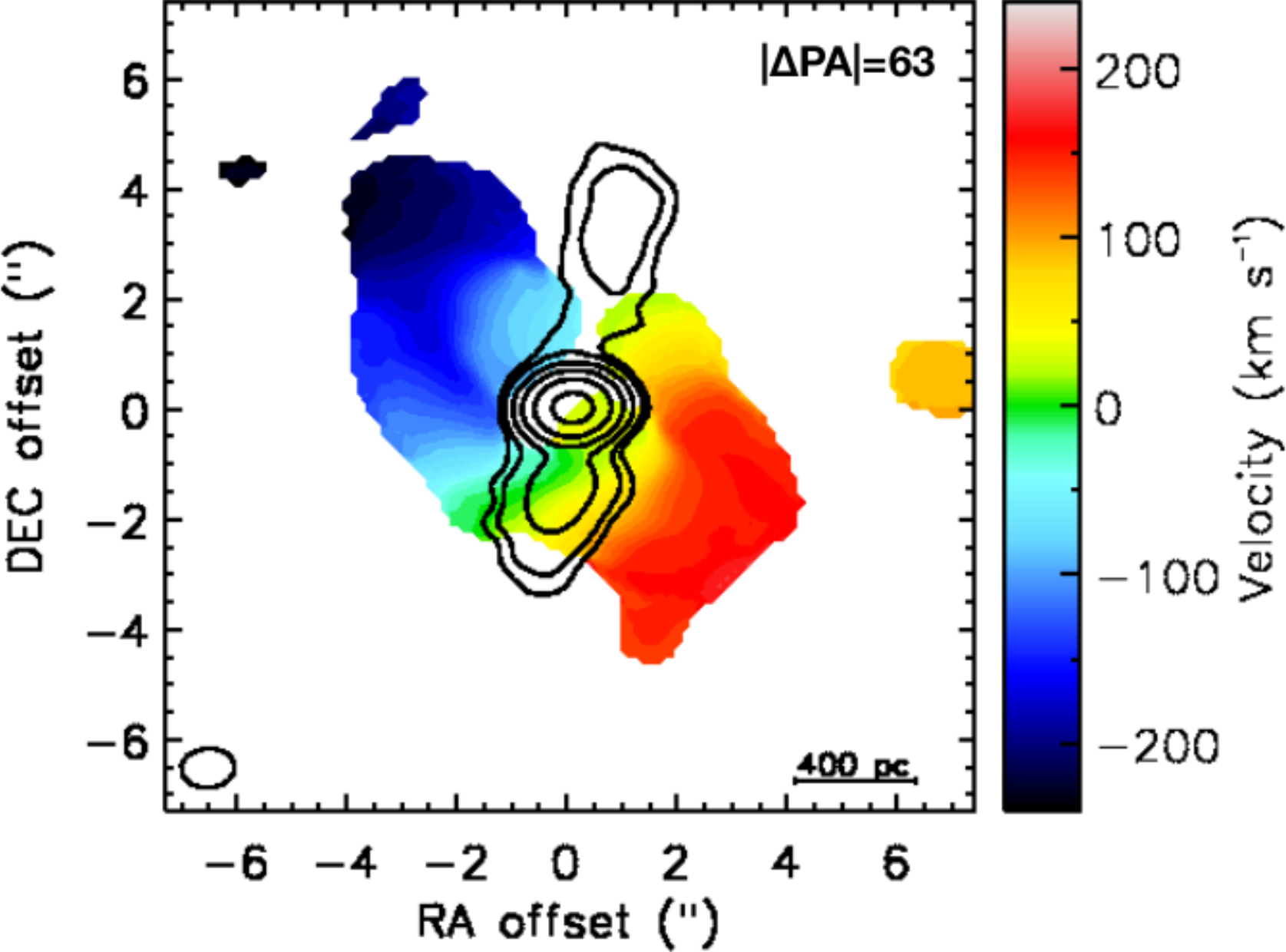}
\end{subfigure}
\begin{subfigure}[t]{0.3\textheight}
\centering
\vspace{2mm}
\caption{\textbf{NGC\,3557}}\label{fig:ngc3557_mom1_cont}
\includegraphics[scale=0.38]{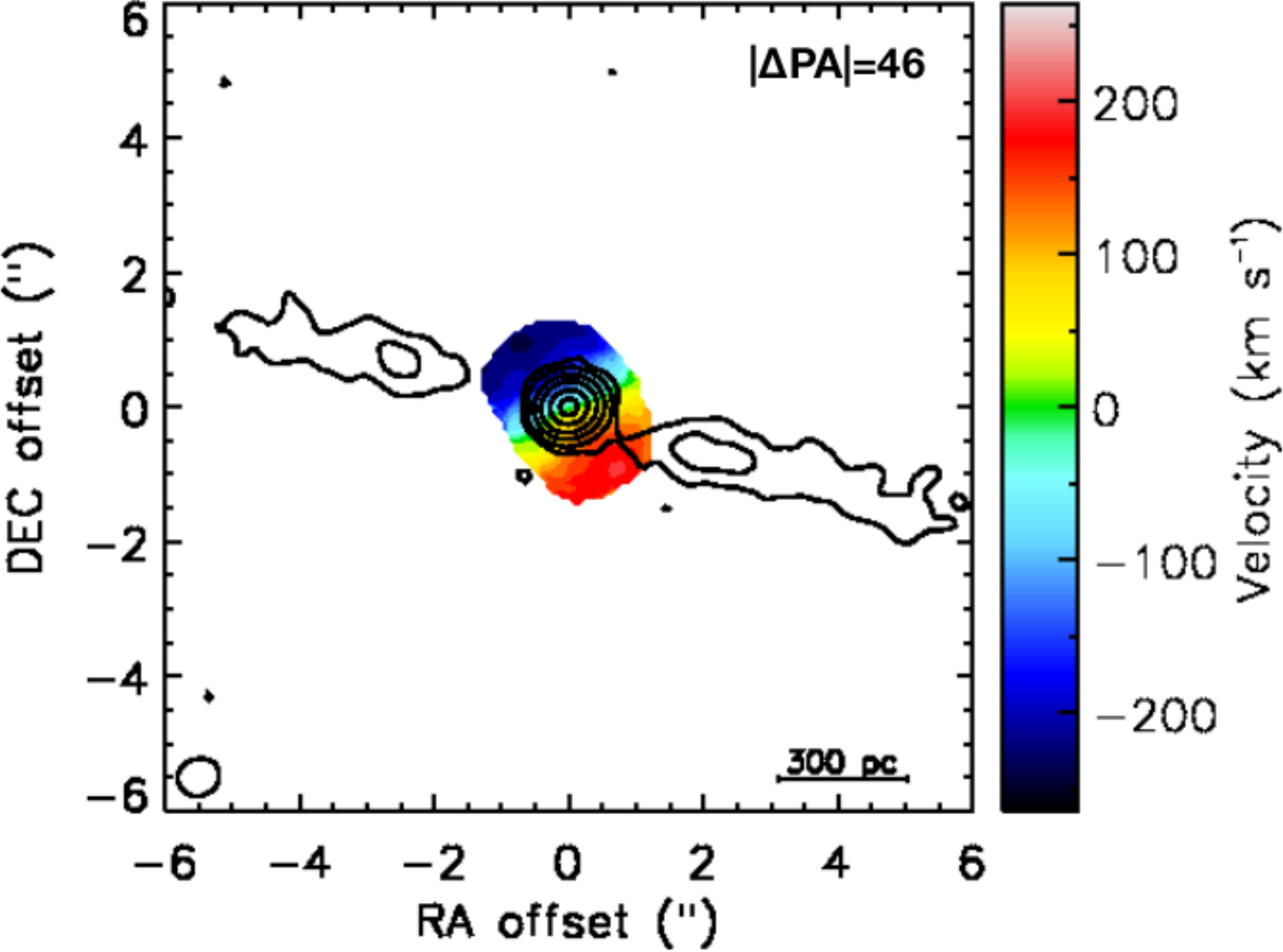}
\end{subfigure}
\medskip 
\begin{subfigure}[t]{0.3\textheight}
\centering
\caption{\textbf{IC\,4296}}\label{fig:ic4296_mom1_cont}
\includegraphics[scale=0.31]{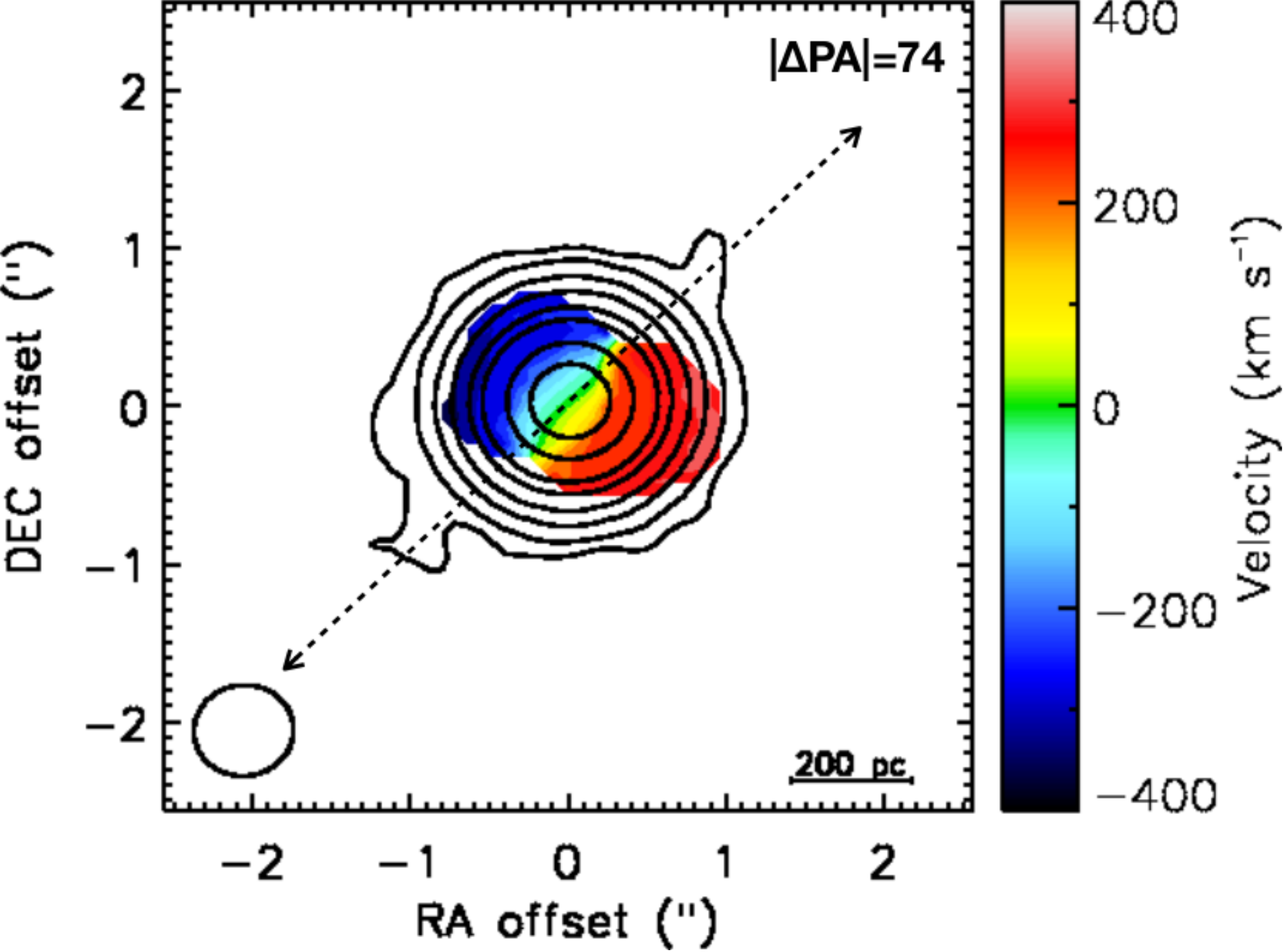}
\end{subfigure}
\begin{subfigure}[t]{0.3\textheight}
\centering
\caption{\textbf{NGC\,7075}}\label{fig:ngc7075_mom1_cont}
\includegraphics[scale=0.38]{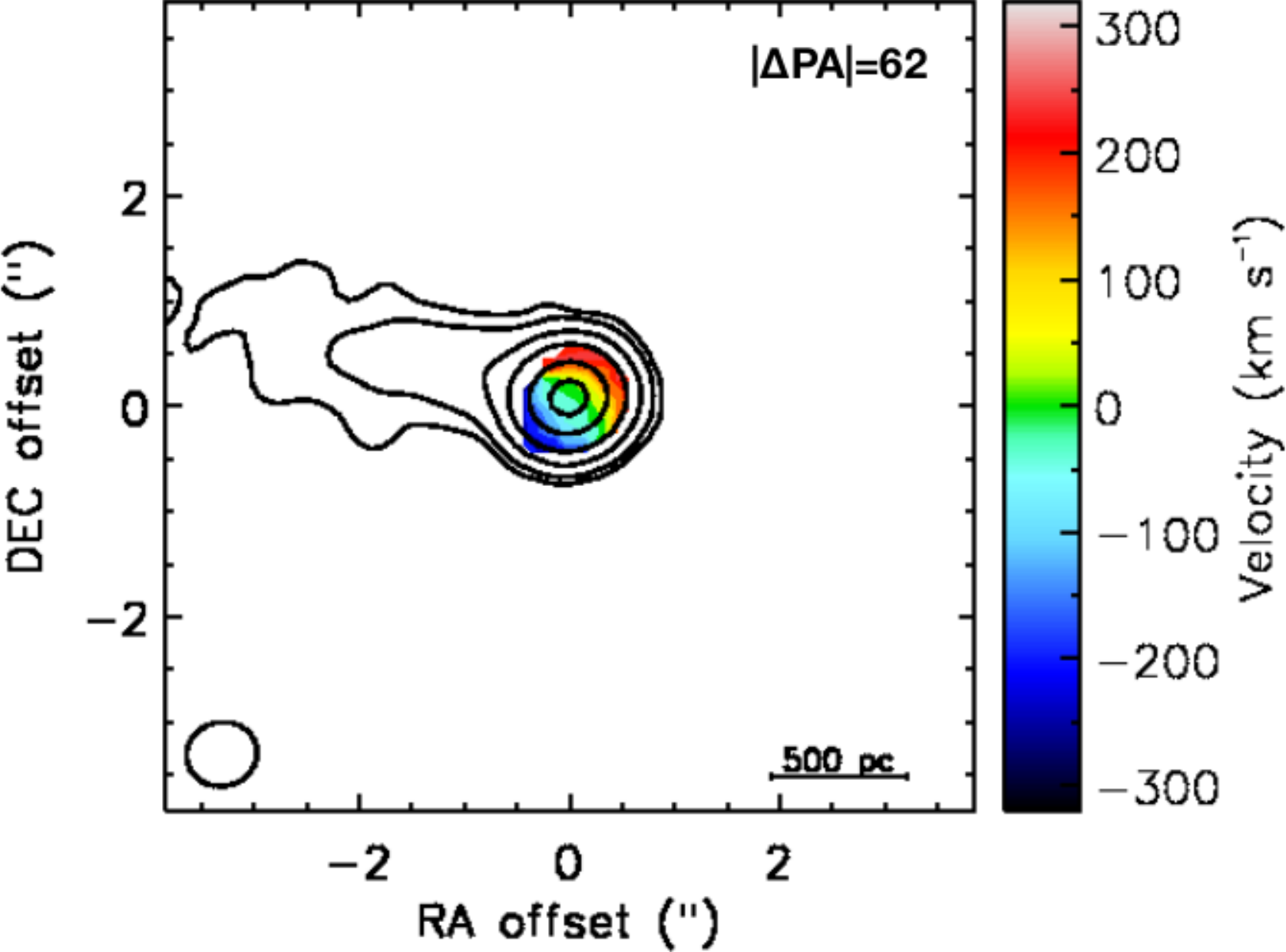}
\end{subfigure}
\caption{Mean velocity (moment 1) maps of the six \co\ detections, with 230~GHz continuum contours superimposed. Contours are drawn at 1,3,9..times the 3$\sigma$ rms noise level. The wedge on the right shows the colour scale of the CO velocity maps. The beam and the physical scale bars are drawn in the bottom-left and bottom-right corner of each panel, respectively. Black dashed arrows indicating the jet axes are also included in panels b and e. The alignment angle between the jet axis and the CO disc, $|\Delta PA|$, is given in the top-right corner of each panel (see the text for details).}\label{fig:CO_cont}
\end{figure*}

Figure~\ref{fig:CO_cont} shows the mean CO(2-1) velocity maps for the detected galaxies with the corresponding 230~GHz continuum contours superimposed. For those sources in which emission from the resolved jets was not detected at 230\,GHz, we added dashed arrows to indicate the jet axes as determined from archival VLA data. The position angles  of the jet axis and the CO disc major axis are listed in Table~\ref{tab:jet_discs}. Following \citet{deKoff00} and \citet{deRuiter02}, we estimated the alignment angles as $|\Delta PA|=|PA\textsubscript{CO}-PA\textsubscript{jet}|$ ranged in $0 - 90^\circ$. This is a measure of the relative orientation of CO discs and jets, at least in projection. A more comprehensive analysis would require 3D modelling of both the radio jets and the molecular gas disc, in order to estimate their inclinations with respect to the plane of the sky and hence their true (mis-)alignments.

It is clear from our analysis that the jet/disc relative orientations span a wide range. In four of the six detected galaxies (67$\pm33$\%) we measure large misalignements ($|\Delta PA|\geq60^{\circ}$), meaning that the gas discs are roughly orthogonal to the jets in projection (i.e.\ the rotation axes of the gas and the jets appear almost parallel); two sources (33$\pm24$\%), however, show smaller alignment angles ($|\Delta PA|<60^{\circ}$), suggesting a significant misalignment between the jet and the disc rotation axes, at least in projection. These results are statistically consistent with earlier findings based on the analysis of jets and dust lanes, although a larger sample is obviously needed to draw significant conclusions about the form of the misalignment distribution, for example bimodality (see Section~\ref{sec:dust_general} for a more detailed discussion). 
 
 NGC 3100 is a special case. While the jet and disc rotation axes appear almost aligned in projection, there are indications of a possible interaction between the gas and the jets. The ring-like CO distribution shows a clear discontinuity to the North of the nucleus, in the direction of the northern jet (Fig.~\ref{fig:ngc3100_mom1_cont}). Enhancements of both line brightness and width are visible adjacent to the jet (Figs~\ref{fig:ngc3100_mom0} and \ref{fig:ngc3100_mom2}), as well as distortions in the rotation pattern (Fig.~\ref{fig:ngc3100_mom1}), that may hint at warps and/or non circular motions.  We also note that the VIMOS IFU [OIII]$\lambda$5007 map of NGC\,3100 shows clear equivalent width broadening (by a factor of 2) at the positions of the peaks of both radio jets (Warren et al., in preparation), reinforcing the case for a jet/gas interaction.

\begin{table}
\centering
\caption{Alignment between jets and CO discs}
\label{tab:jet_discs}
\begin{tabular}{lrrr}
\hline
\multicolumn{1}{c}{ Target} &
\multicolumn{1}{c}{ PA\textsubscript{CO} }&
\multicolumn{1}{c}{ PA\textsubscript{jet}}&
\multicolumn{1}{c}{ $|\Delta PA|$}\\
\multicolumn{1}{c}{  } &
\multicolumn{1}{c}{ (deg) } &
\multicolumn{1}{c}{ (deg) } &
\multicolumn{1}{c}{ (deg) } \\
\multicolumn{1}{c}{ (1) } &
\multicolumn{1}{c}{ (2) } &
\multicolumn{1}{c}{ (3) } &
\multicolumn{1}{c}{ (4) } \\
\hline
 IC\,1531  & 176 & 158 & 18 \\
 NGC\,612   & 4 & 97 & 87 \\
  NGC 3100 & 48 & 165 & 63 \\
  NGC 3557 & 31 & 77 & 46 \\
  IC\,4296  & 57 & 131 & 74 \\
 NGC\,7075 & 130 & 68 & 62 \\
\hline
\end{tabular}
\parbox[t]{8.5cm}{ \textit{Notes.} $-$ Columns: (1) Target name. (2) Kinematic position angle of the CO disc measured counterclockwise from North to the approaching side of the velocity field. (3) Position angle of the jet axis (from North through East) derived from the 230~GHz continuum images (see also Table~\ref{tab:Continuum images}) or the archival radio images presented in this work. (4) Alignment angles between the jet axes and CO discs, $|\Delta PA|=|PA\textsubscript{CO}-PA\textsubscript{jet}|$ ranged in $0 - 90^\circ$.}
\end{table}

\subsection{Comparison of CO and dust distributions}\label{sec:dust_comparison}

It is expected that dust obscuration and line emission from cold molecular gas will trace the same component of the interstellar medium of ETGs. The association between the presence of nuclear dust discs and double-horned CO line profiles is well established \citep{Prandoni07,Prandoni10,Young11} and the imaging observations of \citet{Alatalo13} demonstrated a clear morphological correspondence. This co-spatiality seems also to be confirmed for the four of our CO detected sources for which optical imaging is available (see Sections~A2 and A3 for details), although higher resolution observations would be needed in some cases to draw strong conclusions.

\begin{figure*}
\begin{subfigure}[c]{0.3\textheight}
\caption{\textbf{IC\,4296}}\label{ic4296_optical}
\includegraphics[scale=0.36]{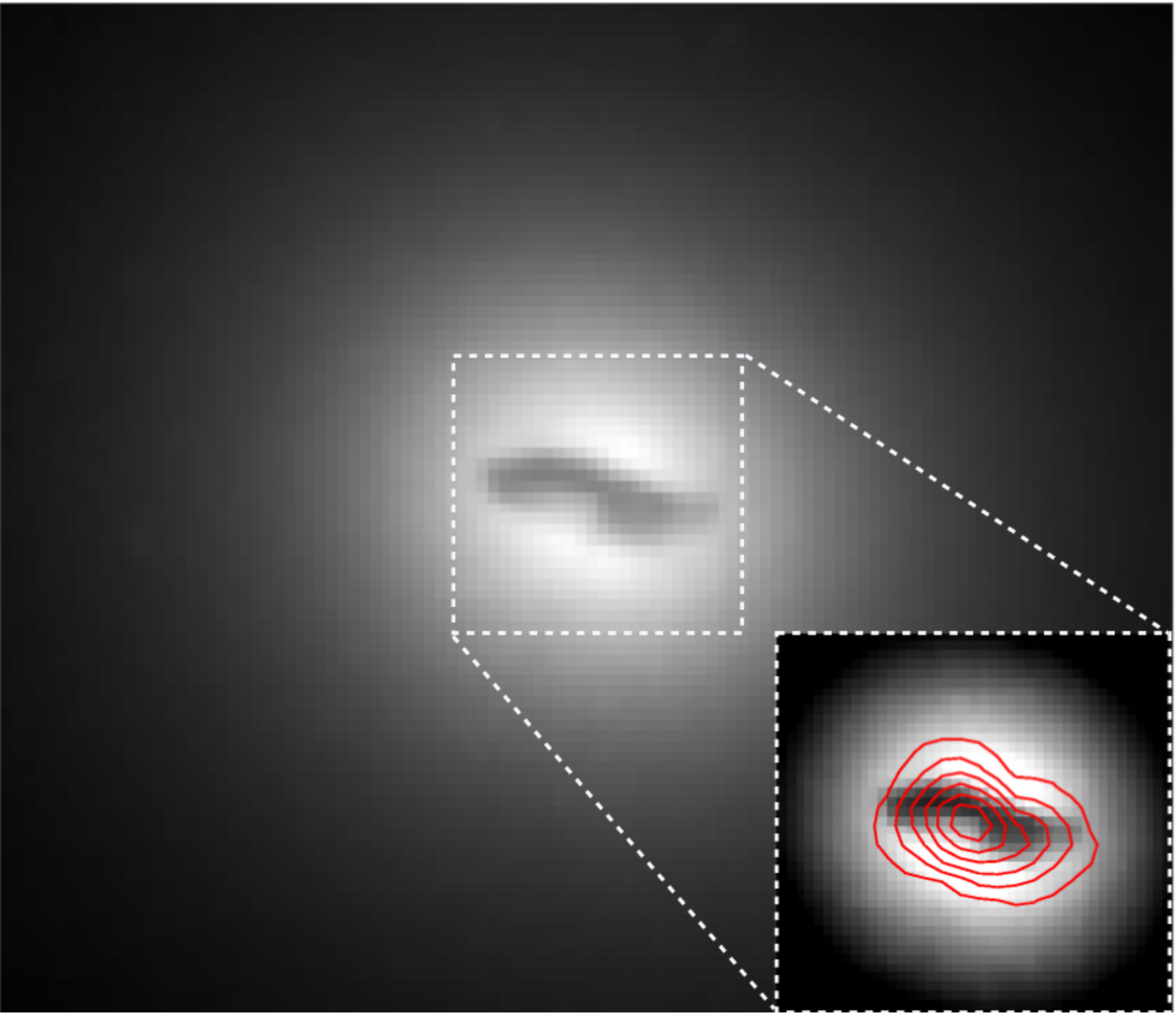}
\end{subfigure}
\hspace{5mm}
\begin{subfigure}[c]{0.3\textheight}
\caption{\textbf{NGC\,3557}}\label{ngc3557_optical}
\includegraphics[scale=0.37]{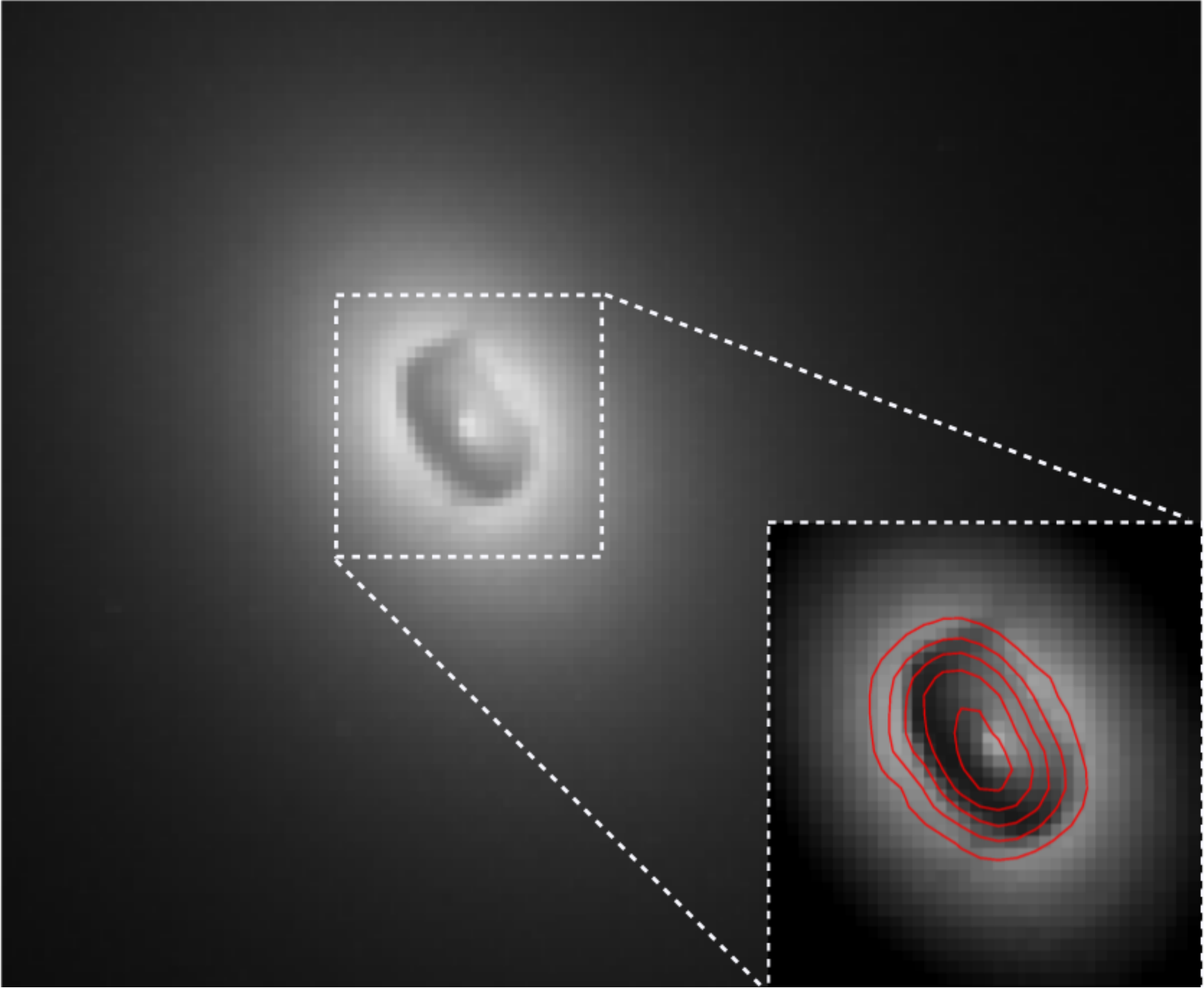}
\end{subfigure}
\caption{\small{Archival HST images of \textbf{(a)} IC\,4296 ($7\times7$~arcsec$^2$) and \textbf{(b)} NGC\,3557 ($11\times7$~arcsec$^2$) taken in the F555W filter. In both images, the pixel scale and the image FWHM are 0.1~arcsec~pixel$^{-1}$ and 0.08\,arcsec, respectively. The insets show superposed red contours of CO integrated intensity drawn at 1,3,9 times the 3$\sigma$ rms noise level}. Additional information on the optical images is provided in Appendix A.}\label{fig:ngc3557_ic4296_optical}
\end{figure*}

\begin{figure*}
\begin{subfigure}[c]{0.3\textheight}
\caption{\textbf{NGC\,3100}}\label{ngc3100_optical}
\includegraphics[scale=0.35]{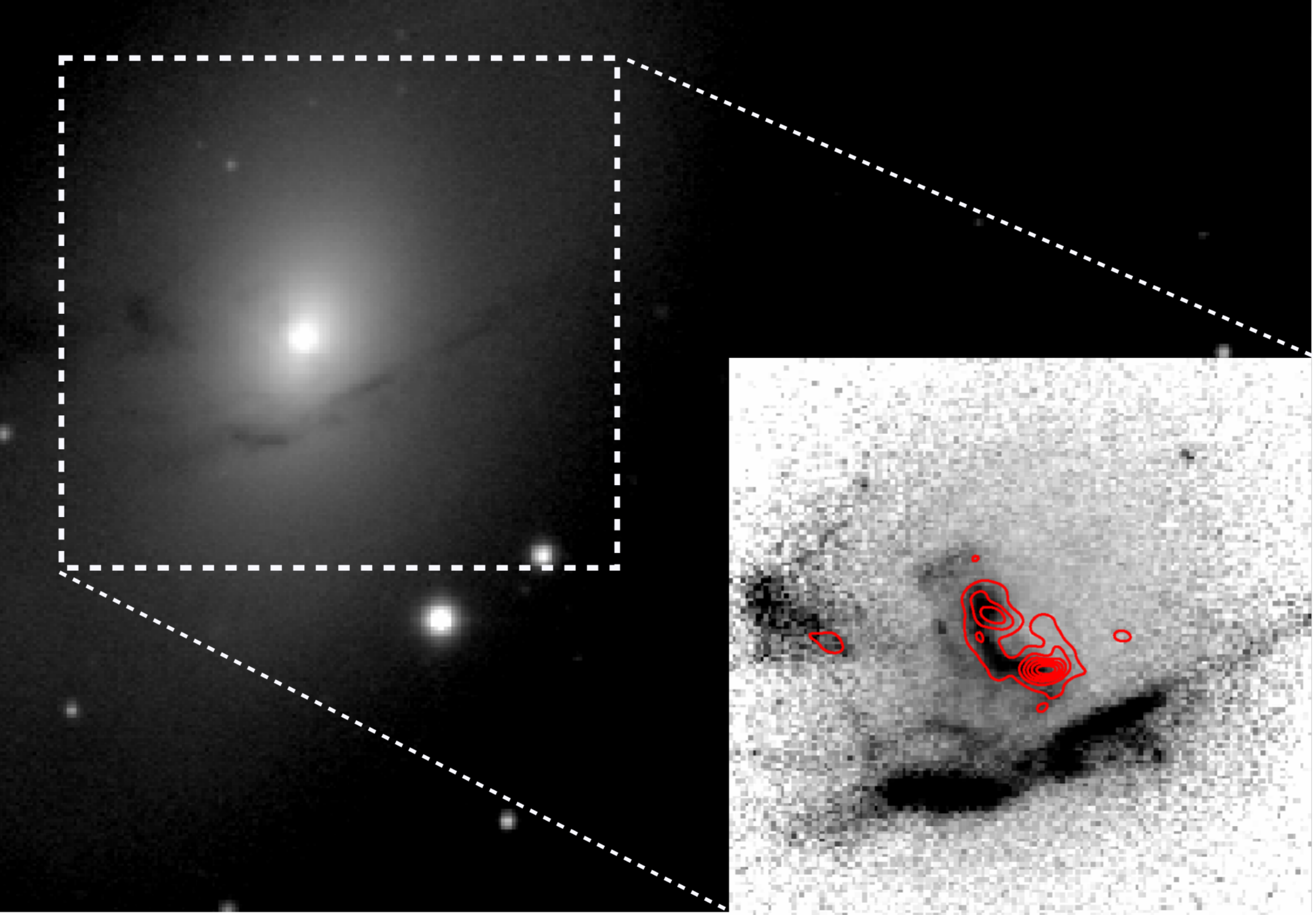}
\end{subfigure}
\hspace{20mm}
\begin{subfigure}[c]{0.3\textheight}
\caption{\textbf{NGC\,612}}\label{ngc612_optical}
\includegraphics[scale=0.355]{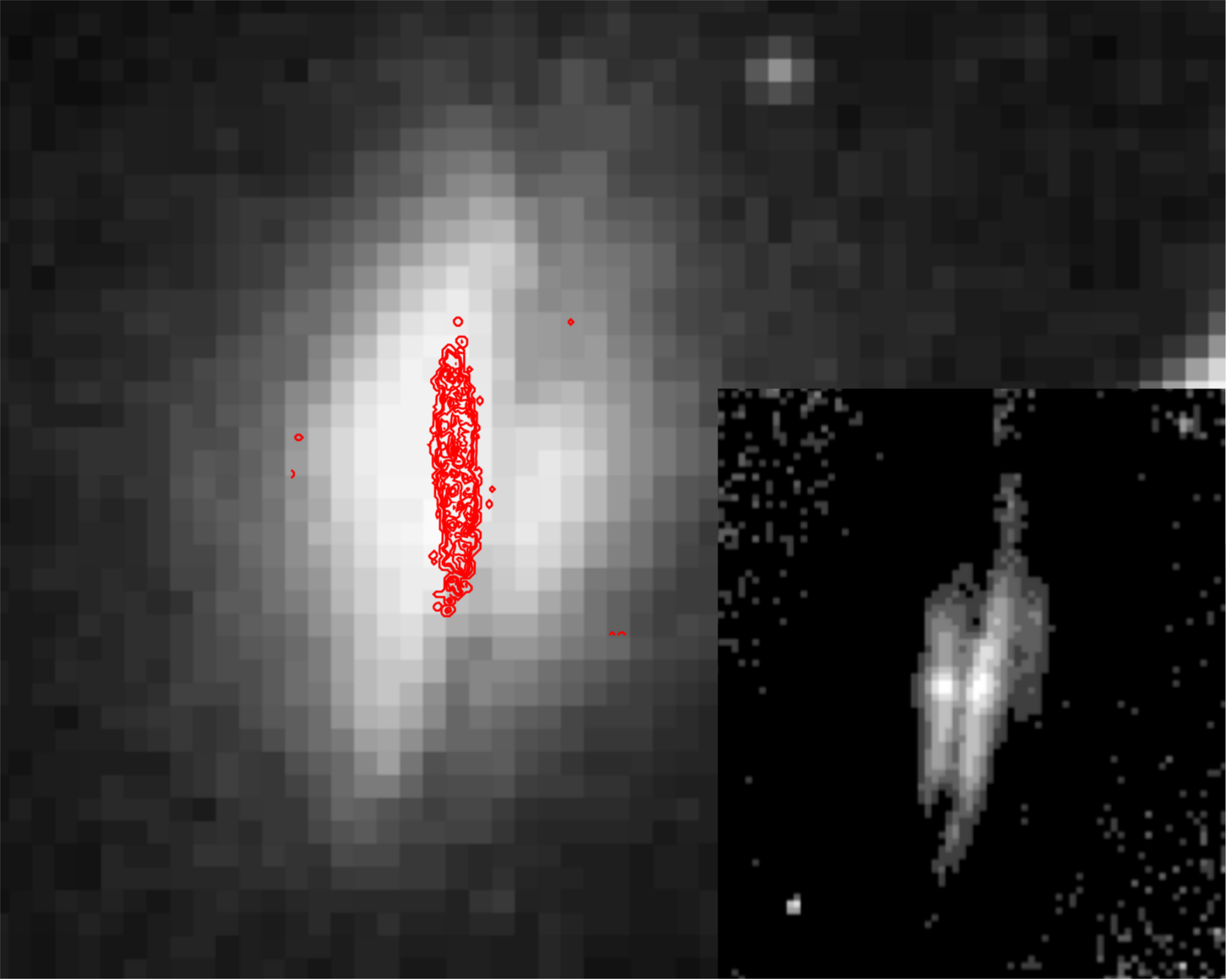}
\end{subfigure}
\caption{\small{\textbf{(a):} Archival optical image of NGC\,3100 taken with the Dupont 2.5\,m Telescope using a blue filter (300-400~nm). The image scale and the resolution are 0.26~~arcsec~pixel$^{-1}$ and 0.77 arcsec, respectively. The size of the panel is $112\times80$~arcsec$^2$. The inset in the bottom right corner shows the B$-$I colour (dust absorption) map in a box of $35\times35$~arcsec$^2$. The CO integrated intensity contours superimposed in red are drawn at 1,3,9....times the 3$\sigma$ rms noise level. 
\textbf{(b):}  Archival optical image of NGC\,612 taken with the UK Schmidt telescope at 468~nm. The image scale is 1.7~arcsec~pixel$^{-1}$ and the image size is $80\times65$~arcsec$^2$. The contours of the CO integrated intensity map are superimposed in red and are drawn at 1,3,9....times the 3$\sigma$ rms noise level. The figure in the bottom right corner shows the B$-$I colour (dust absorption) map adapted from \citet{Veron01}. Additional information on the optical images is provided in Appendix~A.}}\label{fig:ngc3100_ngc612_optical}
\end{figure*}

Figure~\ref{fig:ngc3557_ic4296_optical} shows archival HST images of IC\,4296 and NGC\,3557, with  the CO moment 0 contours overlaid. In both cases, the dust and molecular gas are clearly co-spatial. The distorted CO morphology of IC\,4296 accurately follows that of the dust (Fig.~\ref{ic4296_optical}), supporting the hypothesis of a warp as suggested by the CO velocity field (Fig.~\ref{fig:ic4296_mom1}). The CO distribution in NGC\,3557 appears to be settled in a regular disc, whereas the dust shows a well-defined ring-like structure (Fig.~\ref{ngc3557_optical}). In this case, it is possible that the CO distribution would be resolved into a ring by observations with FWHM \la 0.3\,arcsec,

Figure~\ref{ngc3100_optical} shows the CO moment 0 contours of NGC\,3100 overlaid on a B$-$I dust absorption map (see Section~A3, for details). The CO ring of NGC\,3100 appears to be co-spatial with a nuclear dust ring, while the outer structure detected to the East seems to trace a diffuse dust patch on larger scales. Additional dust is visible beyond the detected CO emission, possibly indicating the presence of a larger CO distribution which could be below the detection threshold or beyond the FOV of our observations ($\approx26"$).

Figure~\ref{ngc612_optical} shows the CO moment 0 contours of NGC\,612 overlaid on an archival photographic B band optical image; a B$-$I dust absorption map is shown in the bottom right corner (see Section~A3 for details). The CO disc seems to be located slightly eastward of and skewed with respect to the dominant dust lane visible in the B band image. The B$-$I colour map shows the presence of an inner dust lane, which may be co-spatial with the CO disc, although there is still a slight apparent difference in position angle on the sky.  Given the  low spatial resolution of the B-band image and the absence of absolute astrometry for the colour map, the connection between CO and dust in this object is not clear: higher-resolution optical observations are needed. 

\subsection{Dust and molecular gas in radio galaxies}
\label{sec:dust_general}

There is evidence that radio galaxies contain significantly more cold gas and dust than radio-quiet ETG's. A correlation between dust mass and radio power was found by \citet{deRuiter02}. LERGs with 1.4~GHz radio luminosities $P_{1.4} \ga 10^{22}$ W Hz$^{-1}$ also contain significantly larger masses of cold molecular gas than radio-weak or radio-silent ETGs (Paper~II), but the dependence of dust and gas mass on $P_{1.4}$ does not appear to extend to lower radio luminosities (Paper~II; \citealt{Baldi15}).

\citet{deKoff00} found the dust properties of radio galaxies to be strongly correlated with FR classification, in the sense that FR\,II sources have larger masses of dust in chaotic distributions, whereas FR\,I's tend to have smaller masses of dust in kpc-scale discs. Similar results were subsequently found by \citet{deRuiter02}. Given the almost one-to-one correspondence between FR class and emission-line classification for those particular samples (FR\,I = LERG; FR\,II = HERG), the results are equally consistent with a fundamental relation between optical spectral type (or accretion rate) and dust mass, in the sense that HERGs have more dust than LERGs. This has a natural explanation in the framework of recent ideas on the fuelling of the two classes of radio galaxy \citep{Heckman14} and therefore seems physically more plausible than a direct relation between dust mass and FR class.

Our results are consistent with the observation that dust in LERGs is most usually in disc-like distributions \citep{deKoff00,deRuiter02}. To a good first approximation, our CO detections show disc or ring-like structures on kpc or sub-kpc scales (NGC\,612 is significantly larger). It therefore seems likely that the gas and dust are mostly in regular orbits, with NGC\,612, NGC\,3100 and IC\,4296 perhaps still being in the process of settling, as in the episodic model described by \citet{Lauer05}.

\citet{KE79} first suggested that dust discs and jets tend to be orthogonal and \citet{deKoff00} supported this result, albeit with clear outliers (see also \citealt{vDF95}). \citet{deRuiter02} found that 79\% of the radio galaxies in their sample (mostly LERGs) have alignment angles $\geq60^{\circ}$, further supporting this scenario. \citet{Schmitt02} found less tendency to orthogonality and \citet{vdKdZ} suggested that orthogonality is restricted to galaxies with irregular dust lanes (as opposed to more regular, disc-like distributions like those found in our sample).  Although the small sample size does not allow us to draw strong conclusions, our results on CO/jet (mis-)alignments are statistically consistent with previous studies: in four cases the gas discs are roughly orthogonal to the jets in projection; in two objects there are gross misalignments.

Simulations of jet formation by black holes accreting at $\ll 0.01$ \.{M}$_{\rm Edd}$, as inferred for LERGs, confirm that jets are launched along the spin axes of the holes and their inner accretion discs, primarily powered by electromagnetic energy extraction \citep{BZ77,McKinney12}.  For a simple axisymmetric system we might expect a common rotation axis for the black hole, inner accretion disc and kpc-scale molecular disc. In this case, the jets and molecular discs should be accurately orthogonal.  This is clearly not always true, and models in which the jets can be misaligned with respect to the rotation axis of either the accretion disc or the larger-scale dust/molecular gas disc have therefore been discussed extensively in the literature (e.g.\ \citealt{Kinney00,Schmitt01,Schmitt02,vdKdZ,Gallimore06,King18}).

One possibility is that the molecular gas results from a minor merger or interaction and has not yet settled into a principal plane of the host galaxy potential (e.g.\ \citealt{Lauer05,Shabala12,Voort15,Voort18}). In this case, there is no reason for the angular momentum vector of the gas to be aligned with that of the central black hole. \citet{Schmitt02} and \citet{vdKdZ} argue against this idea on the grounds that regular dust discs appear to rotate around the short axes of oblate-triaxial gravitational potentials and have therefore settled. Both of the galaxies in our sample with extreme disc-jet misalignments (IC\,1531 and NGC\,3557) show regular disc rotation (at the resolution of our ALMA observations) and the gas and stars rotate together (see below). At least in these two cases, the gas is likely to have settled. Gas in younger (and therefore smaller) radio galaxies might be more likely to be in the settling phase, but the majority of our sample (including the two very misaligned cases) are mature radio galaxies with large-scale jets. Furthermore, the spectral ages and alignment angles for the B2 radio-galaxy sample (mostly LERGs; \citealt{Colla75}) are not correlated \citep{Parma99,deRuiter02}.   A misaligned inflow of molecular gas therefore seems not to explain all of the observations, although it may be relevant for some objects.

The misalignment might instead occur between the inner edge of the molecular disc and the jet formation scale. One  obvious possibility is that the jet is launched along the spin axis of the black hole, which in turn is determined by earlier merger events and is not aligned with an axis of the stellar gravitational potential. Alternatively, if the jet direction is defined by the inner accretion disc, then  warping of the disc may cause misalignment \citep{Schmitt02}. The inner part of a tilted thin accretion disc is expected to become aligned  with the black hole mid-plane via the Bardeen-Petterson effect \citep{BP75}, although the hole spin eventually becomes parallel to the angular momentum vector of the accreted matter \citep{Rees78,SF96}. Simulations  by \citet{Liska18} show that jets are launched along the angular momentum vector of the outer tilted disc in this case.  It is not clear whether this mechanism can work for the thick accretion discs thought to occur in LERGs, however \citep{Zhuravlev14}.

As emphasised by \citet{Schmitt02}, projection effects significantly affect the observed distribution of misalignment angles, so we plan to investigate this issue further only after performing 3D modelling of both discs and jets.

The emerging picture from observations of CO emission and dust is that LERGs contain substantial masses of molecular gas, always associated with dust, but often in stable orbits.  The low accretion rate in LERGs need not be determined by a complete absence of cold molecular fuel, but rather by a low infall rate from a substantial gas reservoir. A possible implication is that a perturbation of the gas could rapidly increase the accretion rate and convert the galaxy to a HERG, even without further gas supply from (e.g.) major mergers.

The origin of the cold gas is an open question: it can either be internal (stellar mass loss, cooling from the hot gas phase) or external (minor/major merger, interaction, accretion). Observationally, galaxies in poor environments or in the field appear to accrete their ISM reservoirs from external sources \citep[e.g.][]{Young05,Davis11,Davis15}. \citet{Ocana10} found no correlation between the optical luminosity and the molecular gas mass of their sample of radio galaxies, so they favoured an external origin for the molecular gas, probably from minor mergers. \citet{Davis11} used the criterion of misalignment (by at least 30$^{\circ}$) between the gaseous and stellar kinematic major axes to establish the origin of the gas. Using this method, they favoured an external origin for at least 15/40 (38\%) of the radio-quiet ETGs imaged in CO with CARMA. This criterion implies an external origin for at least two of our radio galaxies,  NGC\,3100 and NGC\,7075, where a direct comparison with VIMOS IFU spectroscopy (Warren et al., in preparation) reveals a strong kinematic misalignment between the CO and the stellar components ($\sim$120~deg) in both cases. In the remaining four galaxies the CO and stellar rotation axes are aligned: this is consistent either with an internal origin or with the gas having settled into stable orbits in the gravitational potential.  Patchy or distorted dust morphologies are usually indicative of recent disturbances, such as interactions or merger events \citep{Lauer05,Alatalo13}, but the situation is complicated by potential interactions between gas and radio jets: indeed, both processes may be at work in NGC\,3100.

\section{Summary and conclusions}\label{sec:conclusion}
This is the first paper of a project studying a complete, volume- and flux-limited ($z<0.03$, S\textsubscript{2.7 GHz}$\leq0.25$~Jy) sample of eleven LERGs, selected from the sample of \citet{Ekers89}.

In this paper, we presented Cycle 3 ALMA\co\ and 230~GHz continuum observations of nine of the eleven sample members, together with a first comparison with archival observations at other wavelengths (radio and optical). The results can be summarised as follows.
\begin{itemize}
\item CO was detected in six out of nine sources with a S/N ranging from 8 to 45 and typical molecular gas masses of  $10^{7} - 10^{8}$~M$_{\odot}$. Upper limits (obtained assuming the gas is concentrated in the inner few hundred parsecs of the galaxy) are of the order of $10^{6}$~M$_{\odot}$.
\item  To a first approximation, the CO is distributed in rotating disc-like structures, on typical scales from a few hundred parsecs to a few kpc. 
\item NGC\,612 is exceptional: it  shows a massive ($2 \times 10^{10}$\,M$_{\odot}$) molecular gas disc extending $\approx$10~kpc along the major axis, and co-spatial with previously-known HI and stellar discs. 
\item NGC\,3100 is characterized by a central ring-like CO morphology, with distortions and patchy structures on larger scales. The CO disc in IC\,4296 is also slightly distorted. 
\item A comparison with available optical images shows that dust absorption and CO emission trace the same ISM component, as expected.
\item Double-horned integrated CO spectral profiles are observed in all of the sources, consistent with the resolved morphology and kinematics. IC\,4296 also shows a deep and narrow absorption feature against the bright continuum nuclear radio source. The CO absorption optical depth is $\tau \approx 0.12$. From this value we inferred an HI absorption optical depth $\tau \approx 1.2\times 10^{-4}$, consistent with the non-detection of HI absorption by \citet{Morganti01}.
\item The nuclei of all of the sources were detected in the 230~GHz continuum images. Six objects also show emission from jets, four of which are double-sided. Both core and extended emission components are morphologically very similar to those observed at GHz frequencies in archival VLA maps. They are likely to be dominated by synchrotron emission.  No evidence of thermal emission correlated with extended dust or CO was found. 
\item Spectral index maps were produced for two sources (NGC\,3100 and NGC\,3557) having matched-resolution ALMA and VLA continuum maps (see Tables~\ref{tab:ALMA observations summary} and A1). The radio cores show flat spectra (-0.2$<\alpha<$0.2, for S$\sim \nu^{\alpha}$), while the jet spectra are steeper ($\alpha \approx$ -0.7). This is consistent with synchrotron emission (self-absorbed in the core and optically thin in the jets. The cores of the remaining objects typically show flat spectra, consistent with partially self-absorbed synchrotron emission from the inner jets.
\item A comparison between the CO mean velocity maps and the mm or cm-wavelength continuum emission gives the relative orientation of the gas rotation and jet axes, in projection. In four cases (67\%) the gas discs are roughly orthogonal to the jets in projection. In two sources (33\%) the disc angular momentum axis and the jet appear significantly misaligned. Despite the poor statistics, our results are consistent with previous studies, in particular with those reported by \citet[][]{deRuiter02} on a similar sample of objects.
\item In NGC\,3100 the ring-like CO distribution shows a clear disruption to the North of the nucleus, in the direction of the northern jet, as well as signs of deviation from regular rotation (see Section~\ref{sec:results}): an interaction between the CO disc and the jets is very likely in this case.
  \end{itemize} 

Detailed 3D modelling of both the molecular gas discs detected with ALMA and the radio jets (using newly acquired high-resolution JVLA observations) will be presented in forthcoming papers. This will allow us to investigate further the issue of the relative orientation of the molecular gas and the radio jets, taking into proper account their inclination with respect to the plane of the sky. Comparison with resolved stellar kinematics will also help to constrain the relationship between 
  the jet and/or disc axes and the principal axes of the galactic potential well. In addition we are making a detailed study of NGC~3100, our best candidate for a jet/gas disc interaction. For this source we have obtained follow-up ALMA observations in Cycle 6 (project ID: 2018.1.01095.S, PI: I. Ruffa), with the primary aim of assessing the impact of the radio jets on the surrounding environment by probing the physical conditions of the molecular gas. 

\section*{Acknowledgements}
We thank the referee for useful comments.
This work was partially supported by the Italian Ministero dell'Istruzione, Universit\`{a} e Ricerca, through the grant Progetti Premiali 2012 -- iALMA (CUP C52I13000140001). IP acknowledges support from INAF under PRIN SKA/CTA `FORECaST'. This paper makes use of the following ALMA data: ADS/JAO.ALMA\#[2015.1.01572.S]. ALMA is a partnership of ESO (representing its member states), NSF (USA) and NINS (Japan), together with NRC (Canada), NSC and ASIAA (Taiwan), and KASI (Republic of Korea), in cooperation with the Republic of Chile. The Joint ALMA Observatory is operated by ESO, AUI/NRAO and NAOJ. The National Radio Astronomy Observatory is a facility of the National Science Foundation operated under cooperative agreement by Associated Universities, Inc. The scientific results reported in this article are also based on photographic data obtained using The UK Schmidt Telescope. The UK Schmidt Telescope was operated by the Royal Observatory Edinburgh, with funding from the UK Science and Engineering Research Council, until 1988 June, and thereafter by the Anglo-Australian Observatory.  Original plate material is copyright (c) the Royal Observatory Edinburgh and the Anglo-Australian Observatory.  The plates were processed into the present compressed digital form with their permission. This paper has also made use of the NASA/IPAC Extragalactic Database (NED) which is operated by the Jet Propulsion Laboratory, California Institute of Technology under contract with NASA. This research used the facilities of the Canadian Astronomy Data Centre operated by the National Research Council of Canada with the support of the Canadian Space Agency.


\bibliographystyle{mnras}
\bibliography{mybibliography}


\appendix
\section{Ancillary Data}\label{ancillary}
We provide here a brief description of the archival radio and optical
images referred to in this paper.

\subsection{VLA}
\label{VLAdata}

\subsubsection{Observations and data reduction}\label{VLAobs}
In order to compare the core and jet emission detected at 230\,GHz
with that visible at lower frequencies, we extracted VLA data at 4.9,
8.5 or 14.9\,GHz for our target sources from the NRAO archive. These
frequencies were chosen to give a reasonable compromise between high
resolution (to match our ALMA images as closely as possible) and
sensitivity to extended structure. The datasets are inevitably
heterogeneous, with wide ranges of resolution, uv coverage and
integration time. In particular, the only data available for NGC\,612
have very low spatial resolution. We calibrated the archival data using
standard methods, with the flux-density scale set using observations
of 3C\,48 or 3C\,286.  Polarization leakage was calibrated for
NGC\,612 and IC\,4296 and the absolute ${\bf E}$-vector polarization
position angle was set for these sources using observations of
3C\,286.  For NGC\,3557, we combined data from three VLA
configurations and for NGC\,7075 and ESO443$-$G024 we show images with
two combinations of frequency and resolution; in all other cases
useful data were only available for a single configuration and
frequency. All of the datasets were self-calibrated in phase and
amplitude and the final images were made using single or multi-scale
{\sc clean} depending on the complexity of the brightness
distribution.  The datasets and image parameters are listed in
Table~\ref{tab:VLA}.  We display the linear polarization as vectors
with lengths proportional to the degree of polarization, $p = P/I$,
where $P$ and $I$ are polarized and total intensity, respectively,
orientated along the local ${\bf E}$-vector direction.

\begin{figure}
  \includegraphics[width=0.45\textwidth]{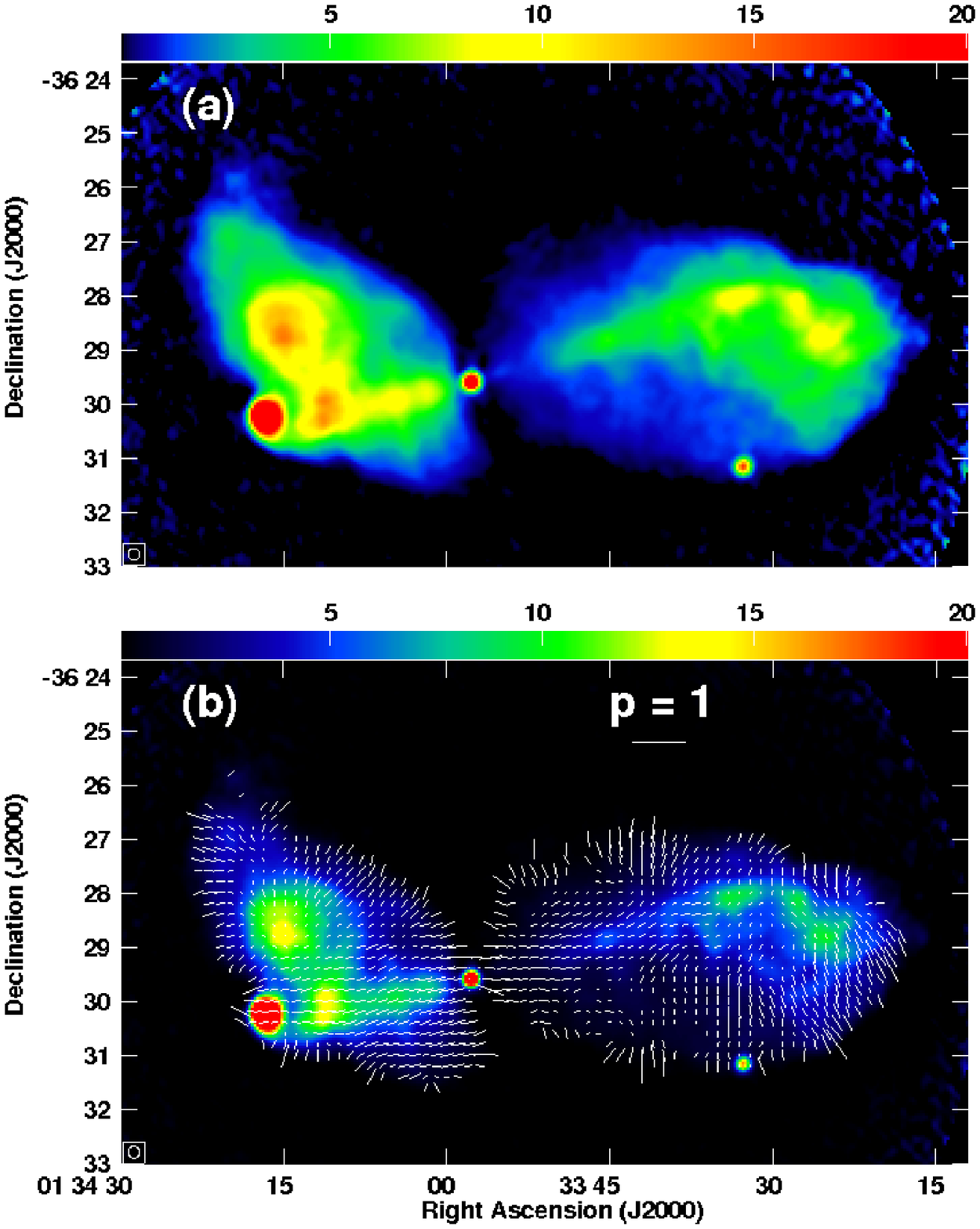}
  \caption{VLA images of NGC\,612 at 4.9\,GHz. (a) Total intensity in
    the range 0 -- 20\,mJy\,beam$^{-1}$. The circular pattern visible
    on the right-hand side of the plot is an artefact caused by the
    correction for primary-beam attenuation. (b) ${\bf E}$-vectors
    with lengths proportional to degree of polarization, $p = P/I$,
    superimposed on total intensity. The vector scale is indicated by
    the labelled bar.\label{fig:ngc612_appendix}}
\end{figure}

\begin{figure}
  \includegraphics[width=0.4\textwidth]{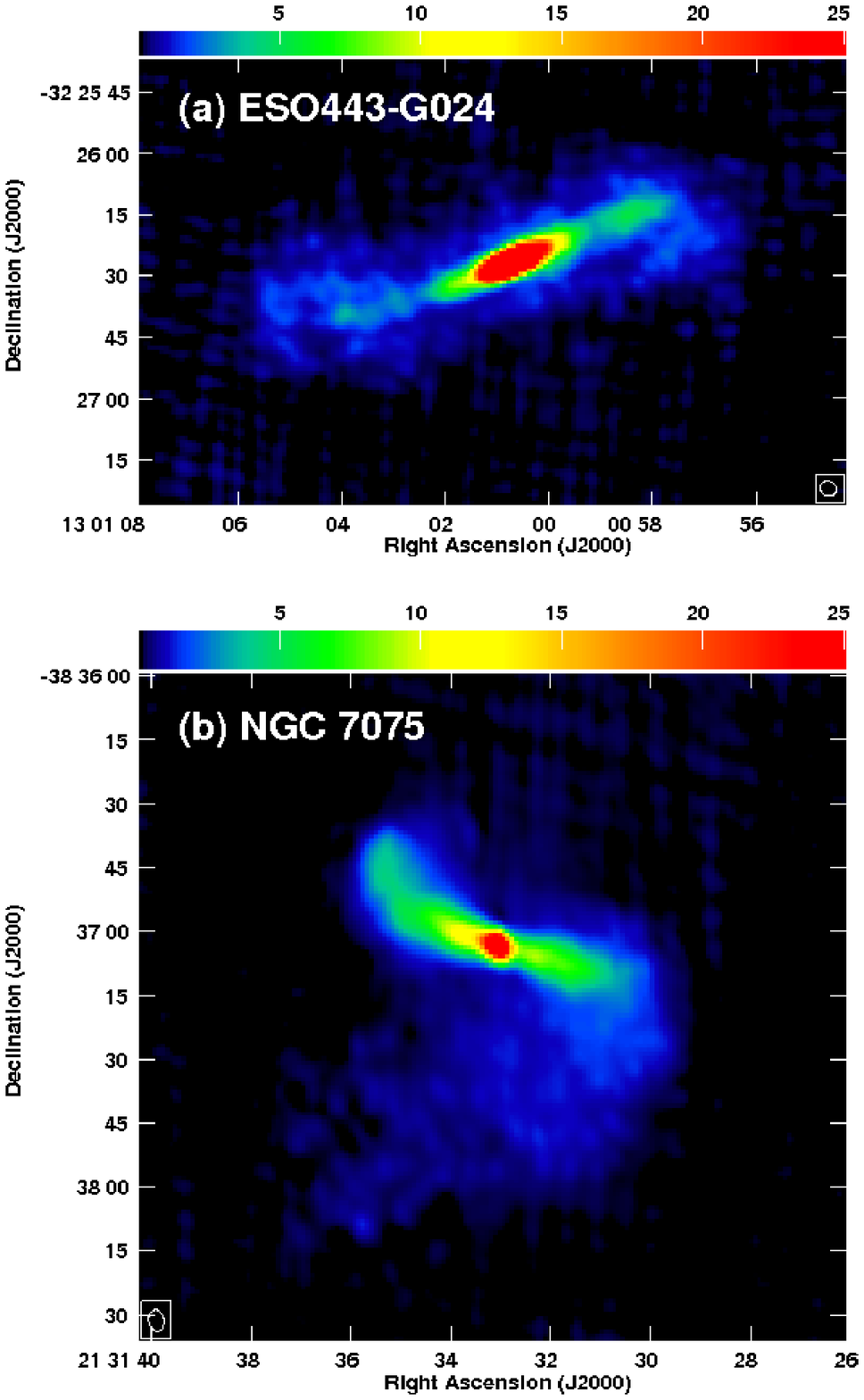}
  \caption{4.9-GHz VLA images of (a) ESO443-G024 and (b) NGC\,7075.\label{fig:vlamaps}}
\end{figure}

The highest-resolution images available are plotted together with the
corresponding 230-GHz images in Figure~1. Lower-resolution images of
four sources, ESO443$-$G024, IC\,4296, NGC\,612 and NGC\,7075, are
significantly deeper than any published to date and/or include linear
polarization for the first time. These are therefore shown below
(Fig.~\ref{fig:ngc612_appendix} -- \ref{fig:ic4296_appendix}).

\begin{table*}
  \caption{Archival VLA data used in this study.}
   \begin{tabular}{lll c c c c c c c c c c}
    \hline
    &&&&&&&&&&\\
    Name & Code & Date & t & Config & $\nu$ & $\Delta\nu$ &   $\theta$\textsubscript{maj}   & $\theta$\textsubscript{min}    &   PA  &  RMS           \\ 
         &      &      & (s) &        & (MHz)  & (MHz)          &\multicolumn{2}{c}{(arcsec)} & (deg) & ($\mu$Jy\,beam$^{-1}$) \\
    (1) & (2) & (3) & (4) & (5) & (6) & (7) & (8) & (9) & (10) & (11) \\
    \hline
    &&&&&&&&&&\\
    IC\,1531       & AH640 & 1998 06 19 &    40 & A   & 8460.1 & 100.0 & 0.58 & 0.23 & $-$3.4  & 210 \\
    NGC\,612       & AK135 & 1985 10 29 & 12700 & C/D & 4860.1 & 100.0 & 13.00 & 13.00 & $-$   &  29 \\
    PKS\,0718$-$34 & AL508 & 1999 10 18 &    45 & A/B & 8460.1 & 100.0 & 2.29 & 0.58 &    14.9 & 220 \\
    NGC\,3100      & AD270 & 1991 07 21 &   260 & A   & 4860.1 & 100.0 & 0.90 & 0.32 & $-$6.9  & 140 \\
    NGC\,3557      & AB289 & 1984 12 14 &  1640 & A   & 4876.1 &  50.0 & 1.38 & 0.35 & $-$12.4 & 53 \\
                   & AW136 & 1985 06 22 &   340 & B/C & 4860.1 & 100.0 &      &      &         &    \\
                   & AB377 & 1987 02 16 & 14120 & C/D & 4860.1 & 100.0 &      &      &         &    \\
    ESO443$-$G024  & AJ141 & 1986 10 04 &   169 & B/C & 4860.1 & 100.0 & 4.13 & 3.62 & 65.9    & 200 \\
                   & AM141 & 1985 02 18 &   460 & A   &14938.7 & 100.0 & 0.60 & 0.60 & $-$     & 970 \\
    IC\,4296       & AE016 & 1982 07 09 & 20200 & A/B & 4872.6 &  25.0 & 1.58 & 0.93 & $-$46.8 & 55  \\
    NGC\,7075      & AH640 & 1998 05 18 &  530  & A   & 8460.1 & 100.0 & 0.79 & 0.22 & $-$2.1  & 270 \\
                   & AG478 & 1996 01 30 &    35 & B/C & 4860.1 & 100.0 & 5.06 & 3.59 &  9.7    & 110 \\
    IC\,1459       & AG674 & 2004 09 07 &   240 & A   & 8460.1 & 100.0 & 0.80 & 0.17 & 22.2    & 210 \\
    &&&&&&&&&&\\
    \hline
\end{tabular} 
\parbox[t]{1\textwidth}{ \textit{Notes.} $-$ Columns: (1) Source name.(2)
    VLA proposal code. (3) Date of observation. (4) On-source
    integration time (scaled to 27 antennas). (5) VLA
    configuration (A/B, B/C and C/D are hybrid configurations with a
    long N arm).(6) Centre reference frequency. (7) Total bandwidth. (8) Beam major axis. (9) Beam minor
    axis. (10) Beam position angle. (11) Rms image noise
    level for $I$. \label{tab:VLA}}   
\end{table*}

\subsubsection{NGC\,612}
\label{NGC612VLA}

The image of NGC\,612 shown in Figure~\ref{fig:ngc612_appendix} was
made from the same dataset as that analysed by \citet{Morganti93}, but
has a significantly lower rms noise level. In addition, linear
polarization has been calibrated and the image has been corrected for
attenuation by the primary beam. The effects of Faraday rotation on
the degree and direction of polarization are 4.9\,GHz are very small
\citep{Kaczmarek18}. The apparent magnetic field direction (orthogonal
to the plotted vectors) is circumferential in both lobes and
transverse in the jets.

\subsubsection{ESO443$-$G024 and NGC\,7075}
\label{ESO443_NGC7075VLA}

The 4.9-GHz images of ESO443-G024 and NGC\,7075
(Fig.~\ref{fig:vlamaps}) show the overall structures of the sources,
both of which have lobed twin-jet FR\,I morphologies. The jet
structure of ESO443-G024 is symmetrical on large scales
(Fig.~\ref{fig:vlamaps}a); side-to-side asymmetry on scales of a few
arcseconds is evident from Fig.~1(f), indicating that
the NW jet is approaching.  For NGC\,7075, the base of the NE jet
appears brighter even at low resolution (Fig.~\ref{fig:vlamaps}b) and
only the NE (approaching) side is visible in the high-resolution VLA
and ALMA images (Fig.~1h).
  
\subsubsection{IC\,4296}\label{IC4296VLA}

\begin{figure*}
  \includegraphics[scale=0.6]{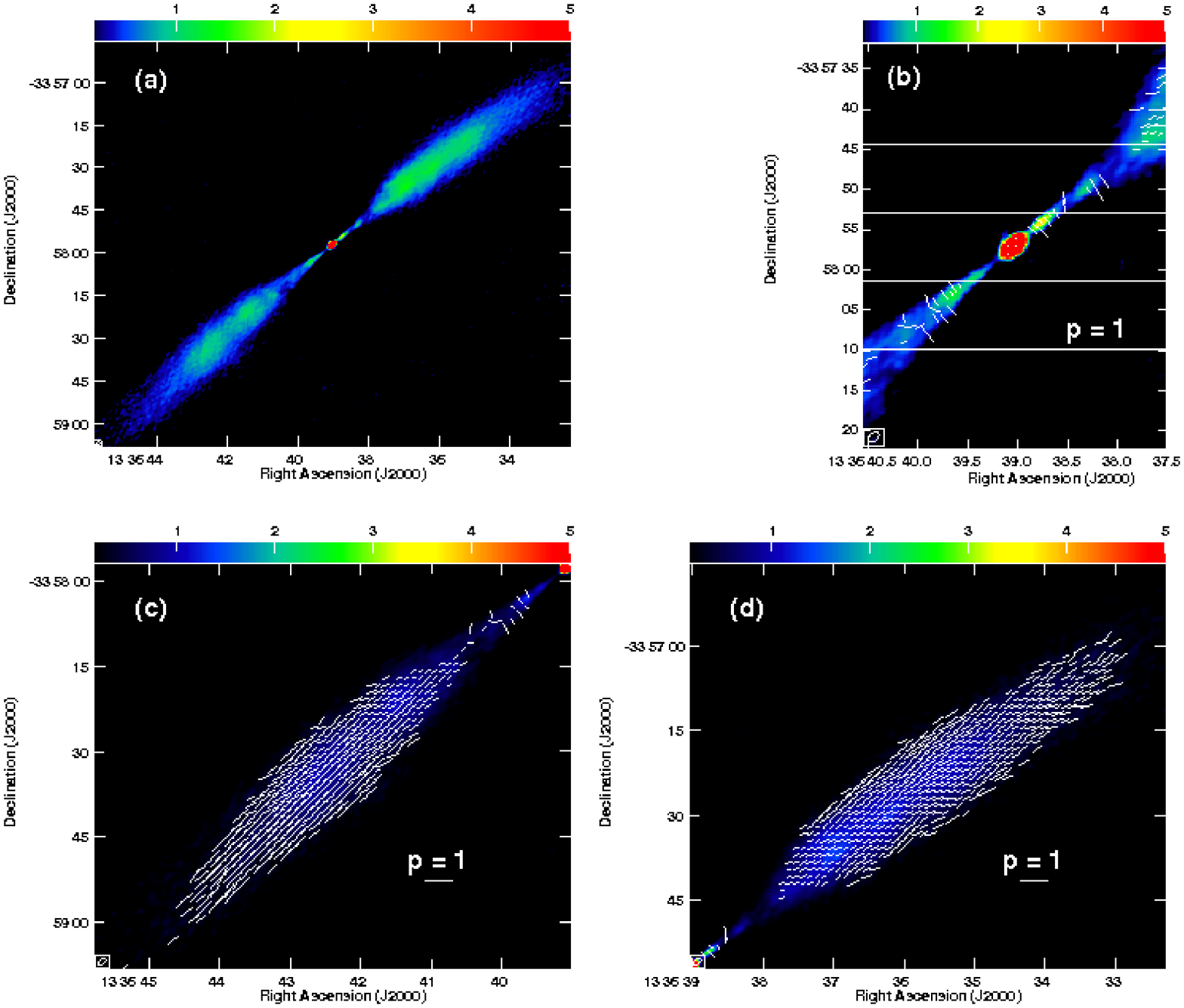}
  \caption{VLA images of IC\,4296 at 4.9\,GHz. (a) Total intensity. (b)
  ${\bf E}$-vectors with lengths proportional to degree of
  polarization, $p = P/I$, superimposed on total intensity for the
  inner jets. (c) and (d) As (b), but for larger regions around the SW
  and NE jets, respectively. The vector scales in panels (b) -- (d)
  are indicated by the labelled bars.\label{fig:ic4296_appendix}}
\end{figure*}

This source was studied in detail by \citet{Killeen86}, but they did
not present images for their highest-resolution 4.9-GHz dataset, which
we have therefore re-reduced. The resulting total-intensity images,
shown in Fig.~\ref{fig:ic4296_appendix}, have a resolution of $1.58 \times
0.93$~arcsec$^2$ and show the jet bases in detail.  The jets are
unusually symmetrical and are likely to be close to the plane of the
sky.  The Faraday rotations estimated for the jets by
\citet{Killeen86} are in the range $-40$ -- 0\,rad\,m$^{-2}$, implying
that the magnitude of the rotation of the observed ${\bf E}$-vectors
from their intrinsic directions is $\la 8^\circ$.  The apparent
magnetic field within $\approx$20~arcsec of the core
(Fig.~\ref{fig:ic4296_appendix}b) is therefore aligned with the axis on both
sides of the nucleus.  Figs~\ref{fig:ic4296_appendix}(c) and (d) show
polarization vectors for the SE and NW jets, respectively. The
magnetic-field structure farther from the core than the flaring points
at $\approx$20~arcsec is remarkably uniform, with an apparent field
direction orthogonal to the jet axis and $p \approx 0.3 -
0.5$. IC\,4296 and 3C\,449 \citep{Feretti99} have remarkably
symmetrical jet intensity and polarization distributions, consistent
with their axes being close to the plane of the sky
\citep[e.g.][]{Laing14}.

\subsection{HST}\label{HST}
To investigate the dust-CO connection, we downloaded HST images (available only for NGC\,3557 and IC\,4296) from the Hubble Legacy Archive at the Canadian Astronomy Data Centre (CADC)\footnote{http://www.cadc-ccda.hia-iha.nrc-cnrc.gc.ca/en/hsthla/}. This provides enhanced HST image products with an absolute astrometric accuracy $\approx$0.3\,arcsec.

Both galaxies were observed with the HST Wide-Field and Planetary Camera 2 (WFPC2) through the F555W optical filter ($\lambda\approx$472$-$595~nm). The final image products have a pixel scale of 0.1\,arcsec\,pixel$^{-1}$ and an image FWHM of 0.08\,arcsec.
NGC\,3557 was observed on 1997 March 10 with a total on-source time of 1900~s (proposal ID: 6587), IC\,4296 on 1997 July 16 with a total exposure time of 1700~s (proposal ID: 6587).  Both images are documented by \citet{Lauer05} and were subsequently used by \citet{Balmaverde06}.

The HST images of NGC\,3557 and IC\,4296 are discussed in Section~6.4 and presented in Figure~11, where they are overlaid on CO integrated intensity maps. 

\subsection{Other optical images}\label{Other_optical}
Archival optical images from ground-based telescopes were also available for NGC\,612 and NGC\,3100 (Fig.~12).
 
We retrieved the NGC\,612 optical image from NED\footnote{https://ned.ipac.caltech.edu}. It was observed with the UK Schmidt Telescope on 1977 September 18 at  $\lambda\approx468$~nm, with a total exposure time of 4500~s. The pixel scale is 1.7\,arcsec\,pixel$^{-1}$.  The B$-$I image was derived by \citet{Veron01} from CCD images taken with EFOSC on the ESO 3.6m Telescope.

The optical images of NGC\,3100 were retrieved from the Carnegie-Irvine Galaxy Survey (CGS) database\footnote{https://cgs.obs.carnegiescience.edu/CGS/Home.html}. NGC\,3100 was observed with the Tek5 CCD camera of the du Pont 2.5-meter telescope at Las Campanas Observatory on 2004 April 19. The observations were made using four filters (Harris B, V, R and I; $\lambda$ from 300 to 1100 nm). The exposure times were 120 -- 360~s in each filter, with resolutions ranging from 0.77\,arcsec (B) to 0.62\,arcsec (I).

The multi-wavelength set of optical images available for NGC\,3100 were used to produce a B-I colour map (bottom-right inset of Fig.~12a). 
Following \citet{deKoff00} and \citet{deRuiter02}, we first scaled the B image (where the dust absorption is most prominent) for the mean magnitude value measured in the I image (where the dust absorption is smallest), and then divided the B by the I image. In the resulting image we assumed that dust absorption features are present where the pixel values are $\leq$0.85 (i.e. where at least 15\% of the emission is absorbed). In regions without absorption, the pixel values are close to unity.

\bsp	
\label{lastpage}
\end{document}